\newcommand{\DoPrePrint}{0} % 0 for 2-column submission/review format; 1 for double-spaced, line-numbered preprint
\ifnum\DoPrePrint=1
  \RequirePackage{lineno} 
  \documentclass[aps,prl,preprint,showpacs,superscriptaddress,nofootinbib,floatfix,linenumbers]{revtex4}

  \usepackage{lineno}
\else
  \documentclass[aps,prd,twocolumn,showpacs,superscriptaddress,nofootinbib]{revtex4}
\fi

\usepackage{amsmath}   % need for subequations
\usepackage{graphicx}  % for figures
\usepackage[caption=false,labelformat=simple]{subfig}
\usepackage{xspace}
\usepackage{units}
\usepackage{gensymb}
\usepackage{hyperref}

\newcommand{\minerva}{MINERvA\xspace}
\newcommand{\minos}{MINOS\xspace}
\newcommand{\miniboone}{MiniBooNE\xspace}

\newcommand{\numu}{\ensuremath{\nu_{\mu}}\xspace}

\newcommand{\cconepiplus}{CC1\ensuremath{\pi^{+}}\xspace}
\newcommand{\cconepi}{CC1\ensuremath{\pi^{\pm}}\xspace}
\newcommand{\ccpi}{CCN\ensuremath{\pi^{\pm}}\xspace}
\newcommand{\Tpi}{\ensuremath{T_{\pi}}\xspace}
\newcommand{\thetapi}{\ensuremath{\theta_{\pi}}\xspace}

\newcommand{\sizecheck}{0} % 0 to do nothing; 1 to check size
\newcommand{\PRLsupp}{0}   % 0 to leave the appendix at the end of the paper; 1 to put the appendix in a separate supplement
\ifnum\PRLsupp=0
  
\else
\fi

\newif\ifpdf
\ifx\pdfoutput\undefined
   \pdffalse
\else
   \pdfoutput=1
   \pdftrue
\fi
\ifpdf
   \usepackage{graphicx}
   \usepackage{epstopdf}
\else
   \usepackage{graphicx}
\fi

\begin{document}

\ifnum\DoPrePrint=1
\linenumbers
\fi
\title{Charged pion production in $\nu_\mu$ interactions \\
on hydrocarbon at $\langle E_{\nu}\rangle$= 4.0~GeV}

%% THIS IS THE AUTHOR LIST FROM THE CCQE ANTI-NEUTRINO PAPER

%% MANUAL PARTS OF AUTHOR LIST

%% (1) need to add ``\thanks{\deceased}'' after DeMaat, Gobbi, Tzanakos
\newcommand{\deceased}{Deceased}

\newcommand{\wroclaw}{Institute of Theoretical Physics, Wroc\l aw University, Wroc\l aw, Poland}    
%% \author{J.T.~Sobczyk}                     \affiliation{\wroclaw}  \affiliation{\FNAL}

%% (2b) Also technical authors non in Glaucus who did not sign the communications paper.  Basically the W&M crowd are the only ones who have requested this

%% AUTOMATIC LIST (EDITED AS ABOVE)
%% List of institution addresses, in command form.
\newcommand{\Rutgers}{Rutgers, The State University of New Jersey, Piscataway, NJ 08854, USA}
\newcommand{\Hampton}{Hampton University, Dept. of Physics, Hampton, VA 23668, USA}
\newcommand{\Dortmund}{Institute of Physics, Dortmund University, 44221, Germany }
\newcommand{\Otterbein}{Department of Physics, Otterbein University, 1 South Grove Street, Westerville, OH 43081, USA}
\newcommand{\JMU}{James Madison University, Harrisonburg, VA 22807, USA}
\newcommand{\Florida}{University of Florida, Department of Physics, Gainesville, FL 32611, USA}
\newcommand{\UCIrvine}{Department of Physics and Astronomy, University of California, Irvine, Irvine, CA 92697-4575, USA}
\newcommand{\CBPF}{Centro Brasileiro de Pesquisas F\'{i}sicas, Rua Dr. Xavier Sigaud 150, Urca, Rio de Janeiro, Rio de Janeiro, 22290-180, Brazil}
\newcommand{\PUCP}{Secci\'{o}n F\'{i}sica, Departamento de Ciencias, Pontificia Universidad Cat\'{o}lica del Per\'{u}, Apartado 1761, Lima, Per\'{u}}
\newcommand{\INRM}{Institute for Nuclear Research of the Russian Academy of Sciences, 117312 Moscow, Russia}
\newcommand{\Jlab}{Jefferson Lab, 12000 Jefferson Avenue, Newport News, VA 23606, USA}
\newcommand{\Pittsburgh}{Department of Physics and Astronomy, University of Pittsburgh, Pittsburgh, PA 15260, USA}
\newcommand{\Guanajuato}{Campus Le\'{o}n y Campus Guanajuato, Universidad de Guanajuato, Lascurain de Retana No. 5, Colonia Centro, Guanajuato 36000, Guanajuato M\'{e}xico.}
\newcommand{\Athens}{Department of Physics, University of Athens, GR-15771 Athens, Greece}
\newcommand{\Tufts}{Physics Department, Tufts University, Medford, MA 02155, USA}
\newcommand{\WM}{Department of Physics, College of William \& Mary, Williamsburg, VA 23187, USA}
\newcommand{\FNAL}{Fermi National Accelerator Laboratory, Batavia, IL 60510, USA}
\newcommand{\Purdue}{Department of Chemistry and Physics, Purdue University Calumet, Hammond, IN 46323, USA}
\newcommand{\MCLA}{Massachusetts College of Liberal Arts, 375 Church Street, North Adams, MA 01247, USA}
\newcommand{\UMD}{Department of Physics, University of Minnesota -- Duluth, Duluth, MN 55812, USA}
\newcommand{\Northwestern}{Northwestern University, Evanston, IL 60208, USA}
\newcommand{\UNI}{Universidad Nacional de Ingenier\'{i}a, Apartado 31139, Lima, Per\'{u}}
\newcommand{\Rochester}{University of Rochester, Rochester, NY 14627, USA}
\newcommand{\Austin}{Department of Physics, University of Texas, 1 University Station, Austin, TX 78712, USA}
\newcommand{\USM}{Departamento de F\'{i}sica, Universidad T\'{e}cnica Federico Santa Mar\'{i}a, Avenida Espa\~{n}a 1680 Casilla 110-V, Valpara\'{i}so, Chile}
\newcommand{\Geneva}{University of Geneva, 1211 Geneva 4, Switzerland}
\newcommand{\Chicago}{Enrico Fermi Institute, University of Chicago, Chicago, IL 60637, USA}
\newcommand{\hired}{}
\newcommand{\OregonState}{Department of Physics, Oregon State University, Corvallis, OR 97331, USA}
\newcommand{\bmeThanks}{now at SLAC National Accelerator Laboratory, Stanford, CA 94309, USA}
\newcommand{\higueraThanks}{University of Houston, Houston, TX 77204, USA}
\newcommand{\LazaThanks}{also at Department of Physics, University of Antananarivo, Madagascar}
\newcommand{\ticeThanks}{now at Argonne National Laboratory, Argonne, IL 60439, USA }
\newcommand{\twaltonThanks}{now at Fermi National Accelerator Laboratory, Batavia, IL 60510, USA}
\newcommand{\janThanks}{also at Institute of Theoretical Physics, Wroc\l aw University, Wroc\l aw, Poland}

% 81 total signatories.
\author{B.~Eberly}\altaffiliation{\bmeThanks}     \affiliation{\Pittsburgh}
\author{L.~Aliaga}                        \affiliation{\WM}
\author{O.~Altinok}                       \affiliation{\Tufts}
\author{M.G.~Barrios~Sazo}                \affiliation{\Guanajuato}
\author{L.~Bellantoni}                    \affiliation{\FNAL}
\author{M.~Betancourt}                    \affiliation{\FNAL}
\author{A.~Bodek}                         \affiliation{\Rochester}
\author{A.~Bravar}                        \affiliation{\Geneva}
\author{H.~Budd}                          \affiliation{\Rochester}
\author{M.~J.~Bustamante~}                \affiliation{\PUCP}
\author{A.~Butkevich}                     \affiliation{\INRM}
\author{D.A.~Martinez~Caicedo}            \affiliation{\CBPF}  \affiliation{\FNAL}
\author{M.F.~Carneiro}                    \affiliation{\CBPF}
\author{M.E.~Christy}                     \affiliation{\Hampton}
\author{J.~Chvojka}                       \affiliation{\Rochester}
\author{H.~da~Motta}                      \affiliation{\CBPF}
\author{M.~Datta}                         \affiliation{\Hampton}
\author{J.~Devan}                         \affiliation{\WM}
\author{S.A.~Dytman}                      \affiliation{\Pittsburgh}
\author{G.A.~D\'{i}az~}                   \affiliation{\Rochester}  \affiliation{\PUCP}
\author{J.~Felix}                         \affiliation{\Guanajuato}
\author{L.~Fields}                        \affiliation{\Northwestern}
\author{R.~Fine}                          \affiliation{\Rochester}
\author{G.A.~Fiorentini}                  \affiliation{\CBPF}
\author{A.M.~Gago}                        \affiliation{\PUCP}
\author{R.Galindo}                        \affiliation{\USM}
\author{H.~Gallagher}                     \affiliation{\Tufts}
\author{T.~Golan}                         \affiliation{\Rochester}  \affiliation{\FNAL}
\author{R.~Gran}                          \affiliation{\UMD}
\author{D.A.~Harris}                      \affiliation{\FNAL}
\author{A.~Higuera}\altaffiliation{\higueraThanks}  \affiliation{\Rochester}  \affiliation{\Guanajuato}
\author{K.~Hurtado}                       \affiliation{\CBPF}  \affiliation{\UNI}
\author{T.~Kafka}                         \affiliation{\Tufts}
\author{J.~Kleykamp}                      \affiliation{\Rochester}
\author{M.~Kordosky}                      \affiliation{\WM}
\author{T.~Le}                            \affiliation{\Tufts}  \affiliation{\Rutgers}
\author{E.~Maher}                         \affiliation{\MCLA}
\author{S.~Manly}                         \affiliation{\Rochester}
\author{W.A.~Mann}                        \affiliation{\Tufts}
\author{C.M.~Marshall}                    \affiliation{\Rochester}
\author{K.S.~McFarland}                   \affiliation{\Rochester}  \affiliation{\FNAL}
\author{C.L.~McGivern}                    \affiliation{\Pittsburgh}
\author{A.M.~McGowan}                     \affiliation{\Rochester}
\author{B.~Messerly}                      \affiliation{\Pittsburgh}
\author{J.~Miller}                        \affiliation{\USM}
\author{A.~Mislivec}                      \affiliation{\Rochester}
\author{J.G.~Morf\'{i}n}                  \affiliation{\FNAL}
\author{J.~Mousseau}                      \affiliation{\Florida}
\author{T.~Muhlbeier}                     \affiliation{\CBPF}
\author{D.~Naples}                        \affiliation{\Pittsburgh}
\author{J.K.~Nelson}                      \affiliation{\WM}
\author{A.~Norrick}                       \affiliation{\WM}
\author{J.~Osta}                          \affiliation{\FNAL}
\author{J.L.~Palomino}                    \affiliation{\CBPF}
\author{V.~Paolone}                       \affiliation{\Pittsburgh}
\author{J.~Park}                          \affiliation{\Rochester}
\author{C.E.~Patrick}                     \affiliation{\Northwestern}
\author{G.N.~Perdue}                      \affiliation{\FNAL}  \affiliation{\Rochester}
\author{L.~Rakotondravohitra}\altaffiliation{\LazaThanks}  \affiliation{\FNAL}
\author{M.A.~Ramirez}                     \affiliation{\Guanajuato}
\author{R.D.~Ransome}                     \affiliation{\Rutgers}
\author{H.~Ray}                           \affiliation{\Florida}
\author{L.~Ren}                           \affiliation{\Pittsburgh}
\author{P.A.~Rodrigues}                   \affiliation{\Rochester}
\author{D.~Ruterbories}                   \affiliation{\Rochester}
\author{G.~Salazar}                       \affiliation{\UNI}
\author{H.~Schellman}                     \affiliation{\OregonState}  \affiliation{\Northwestern}
\author{D.W.~Schmitz}                     \affiliation{\Chicago}  \affiliation{\FNAL}
\author{C.~Simon}                         \affiliation{\UCIrvine}
\author{J.T.~Sobczyk}\altaffiliation{\janThanks}                     \affiliation{\FNAL}
\author{C.J.~Solano~Salinas}              \affiliation{\UNI}
\author{N.~Tagg}                          \affiliation{\Otterbein}
\author{B.G.~Tice}\altaffiliation{\ticeThanks}    \affiliation{\Rutgers}
\author{E.~Valencia}                      \affiliation{\Guanajuato}
\author{T.~Walton}\altaffiliation{\twaltonThanks}  \affiliation{\Hampton}
\author{J.~Wolcott}                       \affiliation{\Rochester}
\author{M.Wospakrik}                      \affiliation{\Florida}
\author{G.~Zavala}                        \affiliation{\Guanajuato}
\author{A.~Zegarra}                       \affiliation{\UNI}
\author{D.~Zhang}                         \affiliation{\WM}
\author{B.P.Ziemer}                       \affiliation{\UCIrvine}

%
%% END AUTOMATIC PART
\collaboration{ \minerva  Collaboration}\ \noaffiliation

\date{\today}

%nubar CCQE paper \pacs{13.15.+g,25.30.Pt,21.10.-k}
\begin{abstract}

Charged pion production via charged-current $\nu_{\mu}$ interactions on plastic scintillator (CH)
is studied using the MINERvA detector exposed to the NuMI wideband neutrino beam at Fermilab. 
Events with hadronic invariant mass $W<1.4$ GeV and $W<1.8$ GeV are selected in separate analyses:   
the lower $W$ cut isolates single pion production, which is expected to occur primarily through the $\Delta(1232)$ resonance, while  
results from the higher cut include the effects of higher resonances.
Cross sections as functions of pion angle and kinetic energy are compared to predictions from theoretical
calculations and generator-based models for neutrinos ranging in energy from 1.5--10~GeV. The data are best described by calculations
which include significant contributions from pion intranuclear rescattering.  These measurements constrain the primary interaction rate and the role of final state 
interactions in pion production, both of which need to be well understood by neutrino oscillation experiments. 

\end{abstract}
\ifnum\sizecheck=0  
\maketitle
\fi

\section{Introduction}
\label{sec:intro}
Charged-current pion production by few-GeV neutrinos interacting 
with nuclei (e.g. carbon, oxygen, and argon) is an important process 
for current and future long baseline neutrino oscillation 
experiments~\cite{Abe:2011ks,Ayres:2004js,LBNE}.  
Recent measurements highlight the important role that the nuclear medium plays
in the production and propagation of hadrons produced in neutrino-nucleus
interactions~\cite{MINERVAnuprl,MINERVAnubarprl,mboone-ccqe,mboone-ccqe-nubar}.
These experiments find cross section distortions which are absent in scattering from free
nucleons and affect both event rates and final state kinematics.  
These effects impact oscillation experiments, such as T2K~\cite{t2k-theta13} 
and MiniBooNE~\cite{AguilarArevalo:2010wv}, that rely on the charged-current quasielastic (CCQE)
interaction on bound neutrons, $\nu_\ell n \rightarrow \ell^{-} p$, 
to reconstruct the neutrino energy.  Although this is a relatively 
well-understood reaction with simple kinematics, the reconstruction
and interpretation of events that appear quasielastic 
are complicated by the presence of the nuclear medium.
For example, if a charged-current interaction produces a single $\pi^{+}$ (\cconepiplus),
e.g., $\nu_\ell N(p) \rightarrow \ell^{-} p \pi^{+}$, and the pion is 
absorbed by the target nucleus in a Final State Interaction (FSI), 
the event will mimic the quasielastic topology.
In such a case, the reconstructed neutrino energy may
be significantly underestimated~\cite{Lalakulich:2012hs} and, in the absence of an accurate FSI model, this will lead to a 
bias in the measured oscillation parameters.  Therefore, both
pion production and the effect of the nuclear environment on that 
production must be accurately determined.  

In addition to being absorbed,
pions may undergo elastic, inelastic, or charge-exchange scattering before exiting the nucleus.  Neutrino experiments 
model these processes with Monte Carlo event generators that use particle cascade algorithms constrained by 
cross section measurements of pion
absorption and scattering on various target nuclei.
This technique assumes that interactions of pions created within a nucleus are identical to those of accelerator beam pions, an
assumption which can be probed by measurements of pion production
in electron- and neutrino-scattering experiments.
The only existing electron-scattering experiment on heavy nuclei~\cite{Qian:2009aa} examined 
the ``color transparency'' of pion production, but was done at 
higher energies than those that are relevant to neutrino oscillation experiments;   
hadronic invariant masses (pion kinetic energies) accessed were 
greater than 2.1~GeV (2~GeV).

The earliest neutrino \cconepiplus measurements used hydrogen or deuterium
targets~\cite{Bell:1978fnal,Radecky:1981fn,Kitagaki:1986ct,Allen:1986,WA25:1990} or reported neutrino-nucleon cross 
sections extracted from nuclear target data by model-dependent corrections~\cite{Block:1964,Budagov:1969,SKAT:1989}.  
These data, particularly the ANL~\cite{Radecky:1981fn} and BNL~\cite{Kitagaki:1986ct} data, are used to constrain the 
neutrino-nucleon pion production models contained in event generators, but these constraints are fairly weak because the ANL and BNL 
measurements differ by up to $\sim$40\% in normalization.  A recent re-analysis of the two experiments prefers the 
ANL measurement~\cite{newphil}.

There are a few measurements of $\nu_\mu$ \cconepiplus on nuclear targets, which provide insight into the nuclear effects 
important to neutrino oscillation experiments.  The K2K~\cite{K2K_piprod} and MiniBooNE~\cite{miniboone_piratio} 
collaborations measured the \cconepiplus to CCQE 
cross section ratio on carbon and mineral oil (CH$_{2}$) targets, respectively.  MiniBooNE 
also reported an absolute cross section measurement of \cconepiplus
on a nuclear target (CH$_{2}$) for $E_{\nu}\sim 1$~GeV~\cite{miniboone_piprod}.  This measurement is primarily 
sensitive to pions with kinetic energies from 20 to 400~MeV produced by $\Delta(1232)$ decays.
The kinetic energy spectrum of charged pions reported by MiniBooNE
does not show the suppression of pions predicted by beam-based models of 
FSI~\cite{gibuu-pi,valencia-pi,Rodrigues:2014jfa}, particularly around 160 MeV 
where the total pion-carbon cross section peaks and pion absorption is greatest.  At present, 
oscillation experiments must account for this discrepancy by assigning large 
systematic errors on the size of pion FSI~\cite{t2k_osc_long}.

The analyses presented here measure flux-integrated differential cross sections in pion 
kinetic energy \Tpi and pion angle with respect to the neutrino direction \thetapi.  The signal is defined to be a 
charged-current $\nu_\mu$ interaction in the \minerva tracking detector (mostly CH).  The \cconepi (\ccpi) measurement signal definition 
requires that exactly one (at least one) charged pion exits the target nucleus.  There is no restriction on neutral pions or other mesons.  The 
\cconepi (\ccpi) signal is also restricted to $1.5\le E_\nu \le \unit[10.0]{GeV}$ and hadronic invariant mass $W<1.4$ ($<1.8$) GeV.  Charged-current 
coherent pion production is included in the signal definitions. 

These are the first such measurements on a nuclear target in the few-GeV energy range that is important for the NOvA~\cite{Ayres:2004js} and DUNE~\cite{LBNE} oscillation 
experiments. The \cconepi
measurement is dominated by the excitation of the $\Delta(1232)$ P$_{33}$ resonance, which facilitates comparison 
to theoretical calculations, neutrino event generators, and the MiniBooNE measurement.  The \ccpi measurement, of which the \cconepi events 
are a subset, is complementary since it samples about six resonances and additional nonresonant processes.

The remainder of this paper is organized as follows.  Section~\ref{sec:expt} describes the \minerva experiment and the NuMI beam line.  
Section~\ref{sec:expt_sim} discusses the simulations used to analyze data.  The event reconstruction, including 
track reconstruction, particle identification, and the hadronic recoil energy measurement, is described in Section~\ref{sec:reco}.  The event selection 
criteria for both analyses are provided in Section~\ref{sec:ev_rec}.  Section~\ref{sec:xs} describes the procedure used to extract cross sections 
from the selected events.  Finally, Section~\ref{sec:results} presents 
and discusses the measured cross sections, and Section~\ref{sec:summary} summarizes this paper.

\section{\minerva Experiment}
\label{sec:expt}
The \minerva experiment combines a fine-grained tracking detector with the high-intensity 
NuMI beam line~\cite{Anderson:1998zza} and the MINOS near detector~\cite{Michael:2008bc}.  
The neutrino beam is created by directing 120 GeV protons onto a graphite target, producing charged 
particles (mostly pions and kaons) which are focused into a 
beam by two magnetic horns.  Downstream of the horns, most of the pions and kaons decay within 
the 675 m helium-filled decay pipe to produce neutrinos.  Approximately 97\% of the muon neutrinos that enter \minerva 
are produced by pion decay, with the remainder produced from kaon decay.

The \minerva detector consists of a central tracking volume 
preceded by nuclear targets, which are not used in this analysis, and surrounded by electromagnetic 
and hadronic calorimeters.
In the tracking volume, triangular polystyrene scintillator strips with a 1.7~cm strip-to-strip pitch are arranged into planes
arrayed perpendicularly to the horizontal axis, which is inclined by 3.4$^\circ$
relative to the beam direction.  
Three plane orientations, at $0^\circ$ and $\pm 60^\circ$ relative to the vertical axis, enable 
unambiguous 3-dimensional reconstruction of the neutrino
interaction point and charged particle tracks.  Each scintillator strip contains a wavelength-shifting fiber 
that is read out by a multi-anode photomultiplier tube.  The 3.0 ns timing resolution of the readout electronics is 
adequate for separating multiple interactions within a single beam spill.   
The MINOS near detector, located 2~m downstream of the \minerva detector, 
is used to reconstruct muon momentum and charge.  More information on the design, calibration, and performance of 
the \minerva detector, including the elemental composition of the tracking volume, is provided in Ref.~\cite{minerva_nim}.

The data for this measurement were collected 
between March 2010 and April 2012 and correspond to an integrated $3.04\times 10^{20}$ protons on 
target (POT).  For these data the horn current was configured to produce 
a muon neutrino beam, and the MINOS detector's magnet polarity was set to focus negative muons.   

\section{Experiment Simulation}
\label{sec:expt_sim}
The neutrino beam is simulated by a GEANT4-based model~\cite{Agostinelli2003250,1610988} that is tuned to agree with 
hadron production measurements on carbon~\cite{Alt:2006fr, Lebedev:2007zz} by the procedure described in Ref.~\cite{MINERVAnubarprl}.  Uncertainty on
the neutrino flux is determined by the precision in these measurements, uncertainties in the beam line
focusing system and alignment~\cite{Pavlovic:2008zz}, and comparisons between different hadron production models in regions not
covered by the hadron production data referenced above.  The integrated neutrino flux over the range 
$1.5\le E_\nu \le 10$~GeV is estimated to be $\unit[2.77 \times 10^{-8}]{cm^{-2}/POT}$.  Table~\ref{tab:nu_flux} lists the flux as a function of energy.

\begingroup
\squeezetable
\begin{table*}[t]
\centering
\begin{tabular}{l|ccccccccccccc}
\hline \hline
$E_\nu$ (GeV) & 
$1.5 - 2$ &
$2 - 2.5$ &
$2.5 - 3$ &
$3 - 3.5$ &
$3.5 - 4$ &
$4 - 4.5$ &
$4.5 - 5$ &
$5 - 5.5$ \\
Flux ($\nu_{\mu}$/cm$^2$/POT ($\times 10^{-8}$)) &
$0.291	$ &
$0.387	$&
$0.476	$&
$0.502   $&
$0.402	$&
$0.242	$&
$0.131	$&
$0.077	$ \\
\hline
$E_\nu$ (GeV) & 
$5.5 - 6$ &
$6 - 6.5$ &
$6.5 - 7$ &
$7 - 7.5$ &
$7.5 - 8$ &
$8 - 8.5$ &
$8.5 - 9$ &
$9 - 9.5$ &
$9.5 - 10$ \\
Flux ($\nu_{\mu}$/cm$^2$/POT ($\times 10^{-8}$)) &
$0.053   $&
$0.041	$&
$0.035	$&
$0.030	$&
$0.026	$&
$0.023	$&
$0.021	$&
$0.019	$&
$0.017	$ \\ \hline \hline
\end{tabular}
\caption{The $\nu_\mu$ flux per POT for the data included in this analysis.}
\label{tab:nu_flux}
\end{table*}
\endgroup

The \minerva detector response is also simulated by a GEANT4-based model.  The muon energy loss scale of the detector
is known to within 2\% by requiring agreement between data and simulation of both the photon statistics and the reconstructed
energy deposited by momentum-analyzed throughgoing muons. Calorimetric corrections used to reconstruct the
energy of hadronic showers are determined from the simulation by the procedure described in Ref.~\cite{minerva_nim}.  The 
uncertainties on the hadron interaction models in GEANT4 are determined to be $\sim$10\% by external 
data~\cite{Ashery:1981tq,Allardyce:1973ce,Saunders:1996ic,Lee:2002eq}.  The tracking efficiency and 
energy response of single hadrons, as well as the scintillation Birks constant, are determined from measurements made 
with a scaled-down replica of the MINERvA detector 
in a low energy hadron test beam~\cite{mnv_testbeam}. The response of the MINOS near detector to muons is determined by a tuned
GEANT-based
simulation~\cite{Michael:2008bc}.

Neutrino interactions are simulated using the GENIE 2.6.2 neutrino event generator~\cite{Andreopoulos201087}.  Non-coherent interactions 
are treated as neutrino-nucleon scattering within a relativistic Fermi gas.  The nucleon momentum distribution is 
modified with a high-energy tail to account for nucleon-nucleon interactions, but interactions with correlated 
nucleon pairs are not included in the simulation.  Pauli-blocking is applied to quasielastic and elastic scattering, but not to 
resonance baryon production.  The structure functions in the deep inelastic scattering (DIS) model are modified to reproduce 
the shadowing, anti-shadowing, and EMC effects observed in charged-lepton nuclear scattering data.

Almost all pion-production events observed in \minerva are due to baryon resonance production, nonresonant pion production (including DIS), 
and coherent pion production.  For baryon resonance production at $W<1.7$~GeV, the formalism of 
Rein-Sehgal~\cite{Rein:1980wg} is used with modern resonance properties~\cite{pdg} and an axial mass $M_{A}=1.12\pm0.22$~GeV.  
However, GENIE differs from Rein-Sehgal in a couple ways. Resonance interference and lepton mass terms in the 
cross section calculation are not included. 
Most signficantly, the angular spectrum of the $\Delta$ decay is nominally isotropic in GENIE; this analysis instead reweights GENIE such 
that the $\Delta$ decay angular anisotropy is half that predicted by 
Rein-Sehgal.  Excursions from isotropic to the full Rein-Sehgal anisotropic prediction are included as a systematic uncertainty.
Nonresonant pion production is simulated using the Bodek-Yang 
model~\cite {Bodek:2004pc} and is constrained below $W=1.7$~GeV by 
neutrino-deuterium bubble chamber data. Coherent pion production is described according to the 
model of Rein-Sehgal modified with lepton mass terms~\cite{Rein:2007}.  Uncertainties on the components of the neutrino-interaction model are 
provided by GENIE. 

Pion and nucleon FSI processes are modeled in GENIE using an effective
intranuclear cascade model~\cite{Dytman:2011zz}, called the ``hA'' model, that simulates the full
cascade as a single interaction and tunes the overall interaction rate
to hadron-nucleus total reaction cross section data.  For 
light nuclei such as carbon, a single interaction happens for a large
fraction of the events.  The final state particle multiplicity and kinematic 
distributions are also tuned to data.  
This model has good agreement with a wide range 
of data from hadron-nucleus scattering experiments for many 
targets. 
Uncertainties in the FSI model are evaluated by varying its parameters
within measured uncertainties~\cite{Lee:2002eq,Ashery:1981tq}.

\section{Event Reconstruction}
\label{sec:reco}
Track reconstruction and the calorimetric energy measurement are the most important components of \ccpi event 
reconstruction.  The reconstruction techniques are fully described in Refs.~\cite{minerva_nim,pittir20853}; the 
most important details are presented here.  Before any reconstruction is performed, the calibrated energy 
deposits within the scintillator strips are grouped into objects called clusters according to timing and spatial proximity.  

\subsection{Track Reconstruction}

Charged particle tracks are reconstructed by applying two pattern recognition algorithms to the clusters found within the tracking volume and downstream calorimeters. 
The first algorithm finds lines separately in each of the three plane orientations (views), then 
attempts to merge one line from each view into a three dimensional track.  Once all accepted three-view 
combinations are found, additional tracks are made from compatible two-view 
combinations if there are overlapping clusters within the unused view.  All tracks are fit 
with a Kalman filter that includes multiple scattering.  The tracks found by 
this algorithm are limited to a polar angle $<70^{\circ}$ 
and must traverse at least nine scintillator planes, 
which corresponds to a \Tpi threshold of about 80~MeV.

In order to lower the pion energy tracking threshold, a second track pattern recognition 
algorithm is employed.  First, all possible combinations of four clusters located within consecutive scintillator planes are formed into track seeds.  
Next, two seeds are merged into a longer seed if they share at least one 
cluster, have similar polar angles, fit well to a straight line, and pass a Kalman filter fit.  Merged seeds may 
be merged with additional seeds, and merging continues until all possible merges are exhausted.  All merged seeds are 
retained as reconstructed tracks.  This algorithm is unable to find tracks with 
a polar angle $>55^{\circ}$, but can reconstruct particles that traverse as few as five scintillator planes, 
resulting in a \Tpi threshold of 50~MeV.  

\begin{figure}[tp]
\centering
\ifnum\PRLsupp=0
  \includegraphics[width=\columnwidth]{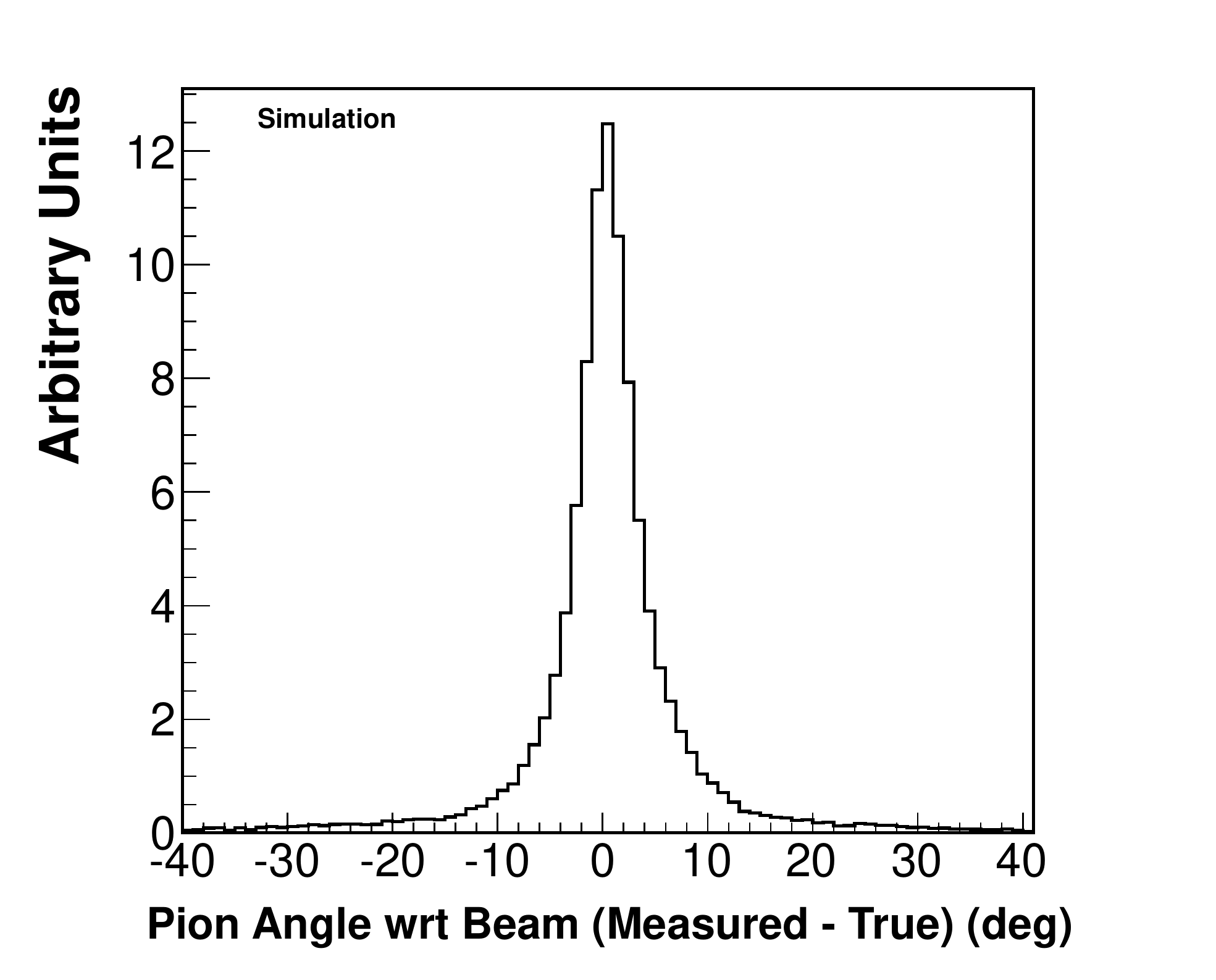}
\else
 \includegraphics[width=0.99\columnwidth]{figures/resid_pi_theta} % shrink to fit PL 
\fi
    \vspace{-7pt}
\caption{The resolution of the pion angle with respect to the neutrino beam.  Only pions 
from events selected by the \ccpi event selection are included.  The full width at half maximum is 5$^\circ$.}
    \vspace{-10pt}
\label{fig:pi_ang_resid}
\end{figure}

The combined efficiency of the two track pattern recognition 
algorithms to find tracks for pions with $\Tpi>50$~MeV in simulated \ccpi events with $W<1.8$~GeV is 42\%.  The primary 
reasons for pion tracking inefficiency are secondary interactions in the detector and activity in high-multiplicity events that 
obscures the pion.  Figure~\ref{fig:pi_ang_resid} shows the angular resolution of
pion tracks in \ccpi events selected by the event selection described in Section~\ref{sec:ev_rec}.  

Neutrino event candidates are reconstructed by finding the longest track in the event, then 
searching for additional tracks that share a common vertex with the longest track.  Kinked tracks, which are often the 
result of secondary interactions, are reconstructed by iteratively searching for additional tracks 
starting at the endpoint of the previously found tracks.  Tracks that exit the downstream end of 
the \minerva detector are matched to tracks in MINOS found by the independent MINOS reconstruction; if a match is found, 
it is identified as a muon 
track and the event is retained as a $\nu_\mu$ charged-current interaction candidate. Additional tracks that share a common vertex 
with the muon track are hadron track candidates.  The \minos match requirement is greater than 90\% 
efficient for muons with momenta greater than 1.5~GeV and angles with respect to the beam less than 20$^\circ$. 
The muon energy $E_\mu$ and charge reconstruction use the reconstructed track curvature and range in MINOS.  

\subsection{Charged Pion Identification}
All hadron track candidates that are fully contained within the \minerva detector are classified as pion-like or proton-like by 
a particle identification algorithm that fits the pattern of energy deposition along each track to the Bethe-Bloch 
formula under pion and proton hypotheses.  The fit is allowed to ignore the last cluster on the track or extend up to two planes 
beyond the end of the track without penalty, but is otherwise consistent with the range of the track.  This is done to account 
for mis-reconstruction of the track end position.  Contamination from 
overlapping vertex activity biases pion track fits towards the proton hypothesis; this is avoided by finding the 
portion of the track with an energy profile that is consistent with multiple overlapping particles and not including it in the fit. 

\begin{figure}[tp]
\centering
\ifnum\PRLsupp=0
  \includegraphics[width=\columnwidth]{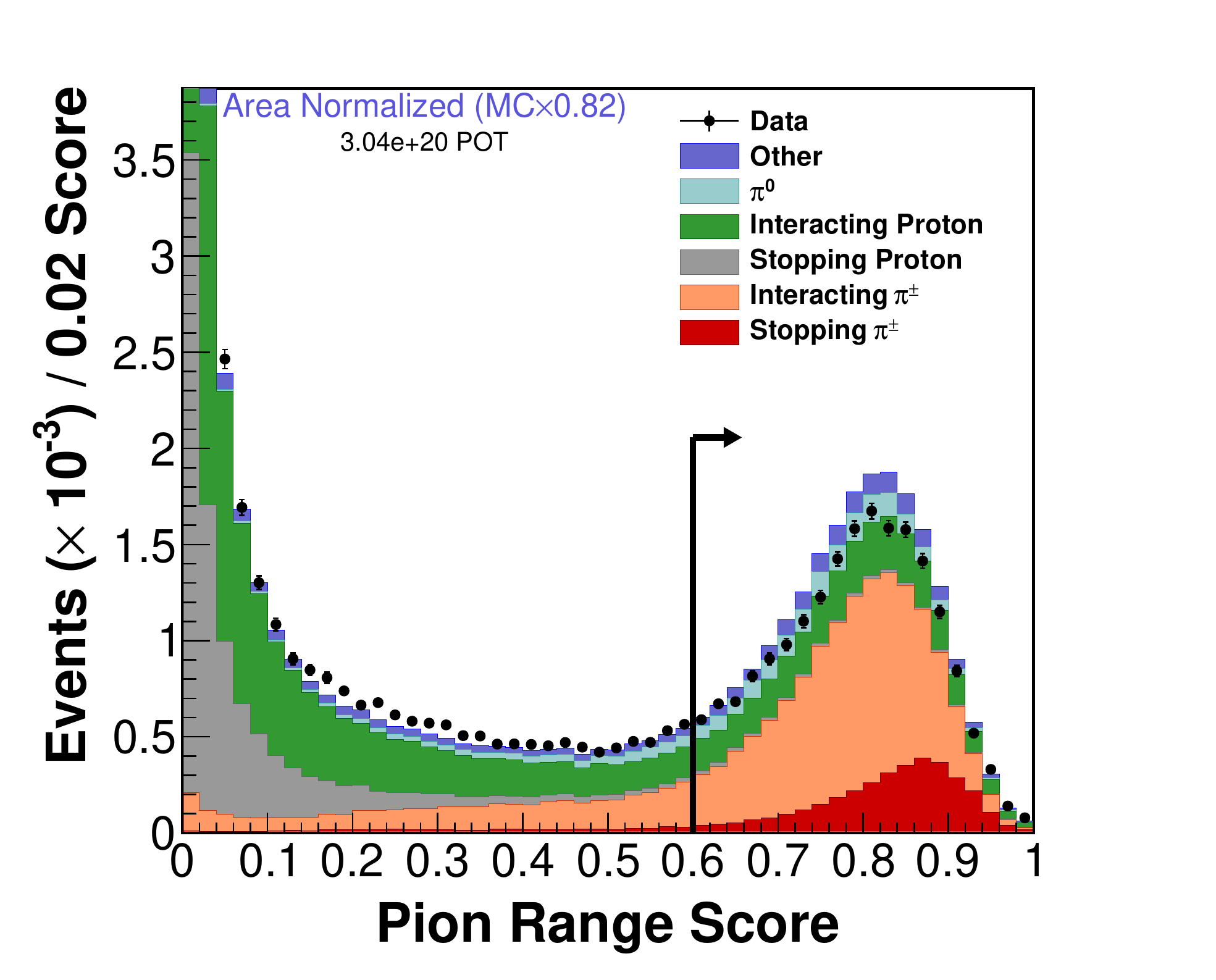}
\else
 \includegraphics[width=0.99\columnwidth]{figures/pi_dedx_score} % shrink to fit PL 
\fi
    \vspace{-7pt}
\caption{A data-simulation comparison of the pion range score $s_{\pi}$.  All \ccpi event selections are applied except for the Michel electron 
requirement.  The simulation is multiplied by a factor of 0.82 to match the area of the data.  A stopping particle is defined as one 
which is fully contained in the \minerva detector without experiencing a secondary interaction. }
    \vspace{-10pt}
\label{fig:pi_score}
\end{figure}

\begin{figure}[tp]
\centering
\ifnum\PRLsupp=0
  \includegraphics[width=\columnwidth]{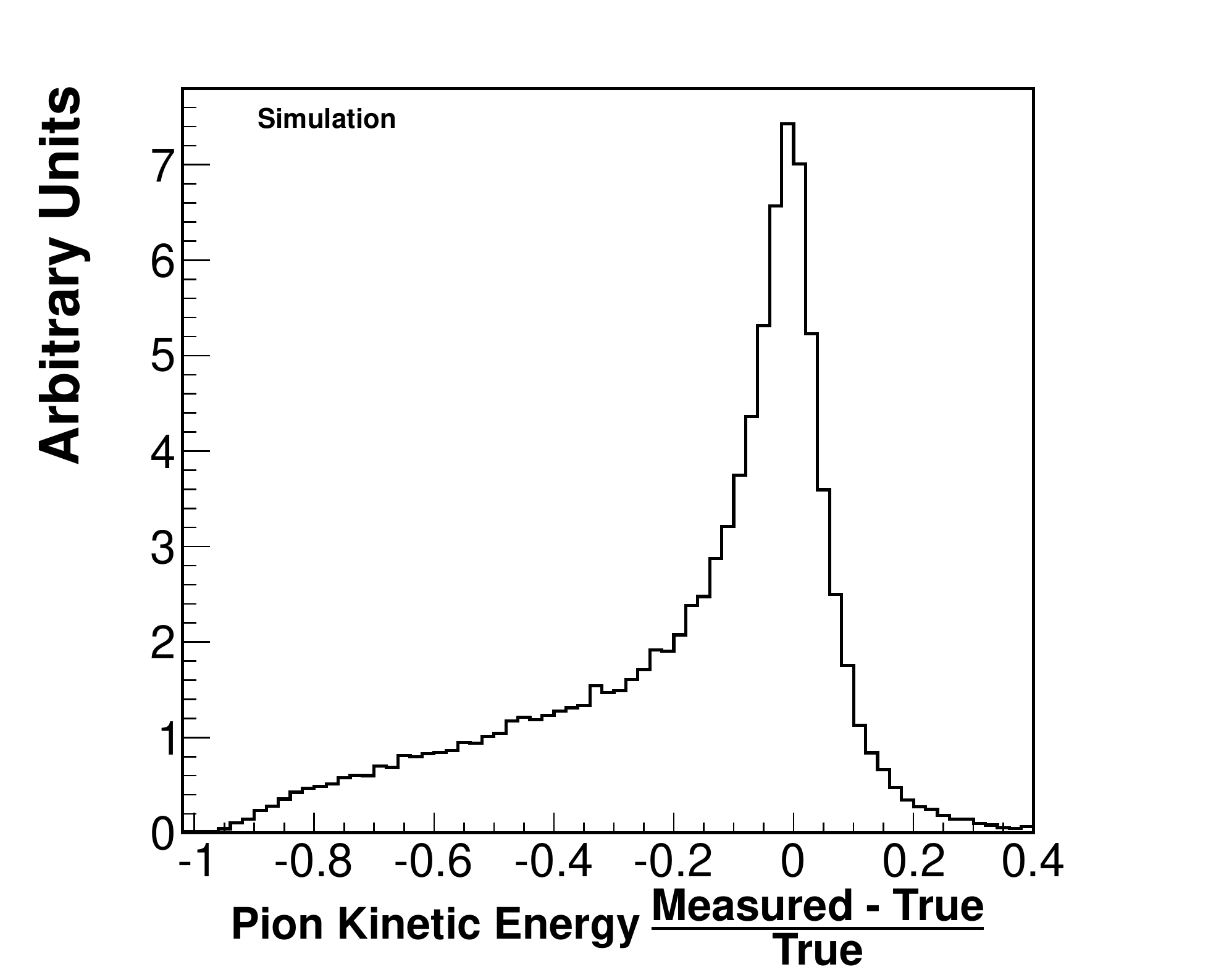}
\else
 \includegraphics[width=0.99\columnwidth]{figures/resid_pi_ke} % shrink to fit PL 
\fi
    \vspace{-7pt}
\caption{The \Tpi resolution.  Only pions 
from events selected by the \ccpi event selection are included.  The lowside tail consists of inelastic secondary interactions 
with a charged pion in the final state.}
    \vspace{-10pt}
\label{fig:pi_ke_resid}
\end{figure}

The pion range score $s_{\pi}$ is calculated from the $\chi^{2}$ of the best fit under each hypothesis by the equation

\begin{equation}\label{eq:dEdx_score}
s_{\pi} = 1 - \frac{\chi_{\pi,DOF}^{2}}{\sqrt{\chi_{\pi,DOF}^{4} + \chi_{p,DOF}^{4}}},
\end{equation}
where  $\chi_{\pi,DOF}^{2}$ is the pion best fit $\chi^{2}$ per degree of freedom and $\chi_{p,DOF}^{2}$ is the proton 
best fit $\chi^{2}$ per degree of freedom.  Figure~\ref{fig:pi_score} presents the $s_{\pi}$ distribution of hadronic track 
candidates in events passing the muon and calorimetric \ccpi selections described in Section~\ref{sec:ev_rec}.  Tracks with 
$s_{\pi}>0.6$ are identified as charged pion candidates.  The kinetic energy of the 
best pion fit determines the reconstructed \Tpi, which can be as low as 35~MeV when the best fit does not include the last cluster 
on the track.  The \Tpi resolution is shown in Fig.~\ref{fig:pi_ke_resid}. 

\begin{figure}[tp]
\centering
\ifnum\PRLsupp=0
  \includegraphics[width=\columnwidth]{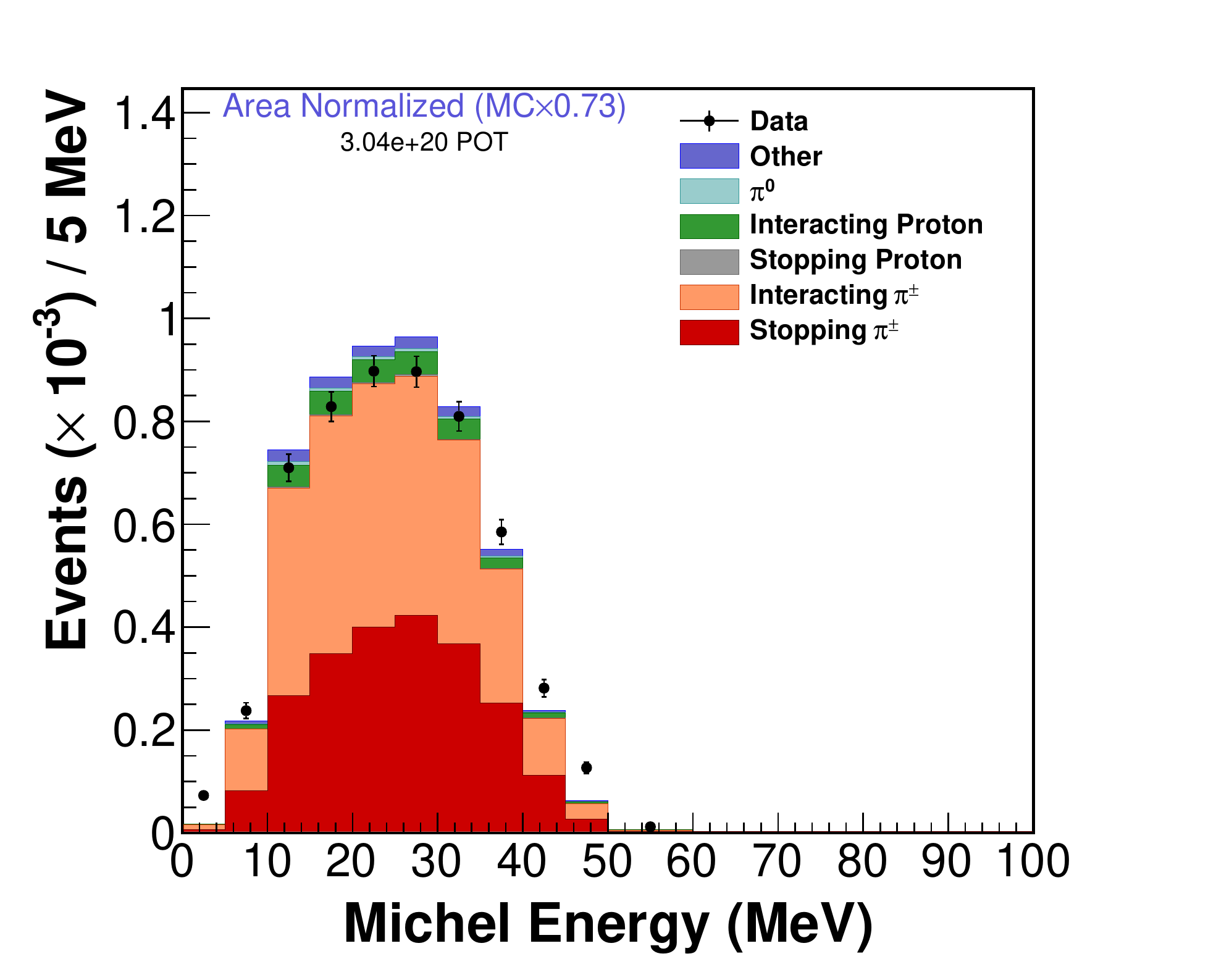}
\else
 \includegraphics[width=0.99\columnwidth]{figures/michel_e_an} % shrink to fit PL 
\fi
    \vspace{-7pt}
\caption{The visible energy distribution of Michel candidates selected by the \ccpi analysis.  The simulation is scaled by a factor of 0.73 to match 
the area of the data.}
    \vspace{-10pt}
\label{fig:michel}
\end{figure}

Charged pions are also identified by the Michel electron from the $\pi\rightarrow \mu \rightarrow e$ decay chain.  Michel 
candidates are found by searching for delayed energy deposits in each view within a $35 \times 25$~cm$^{2}$ (transverse $\times$ 
longitudinal) box centered on the end position of each hadron track.  The large search box accounts for track mis-reconstruction 
and the potential size of the Michel shower, but will often include energy from unrelated neutrino-induced activity 
that occurs later in the beam spill.  To avoid this, the total visible energy of the Michel candidate must be less than 55~MeV and 
the total number of scintillator strips cannot exceed 35.  These restrictions are motivated by the well-understood kinematics 
of muon decay.  Figure~\ref{fig:michel} shows a comparison of the reconstructed Michel 
visible energy spectrum in data and simulation for pion candidates with $s_{\pi}>0.6$ in \ccpi candidate events.  The means of the data 
and simulation are consistent within the 3\% uncertainty on the detector energy response to electromagnetic 
particles.  Michel candidates are associated with an at-rest $\pi^{+}$ with an efficiency of 80\%, as validated
in data with stopped muons from upstream neutrino interactions.

\subsection{Neutrino Energy Reconstruction} 
The neutrino energy in \ccpi events can be reconstructed kinematically using the reconstructed four-momentum of the muon and pion, 
but this requires the assumptions that there is only one nucleon in the final state and that the pion did not experience FSI. Instead, 
this analysis employs a calorimetric energy reconstruction that utilizes the final-state recoil energy $E_\mathrm{recoil}$ 

\begin{equation}\label{eq:e_recoil_def}
E_\mathrm{recoil} \equiv E_\nu - E_\mu,
\end{equation}
which is reconstructed as the calorimetrically-weighted sum of the visible energy not associated with the muon track, i.e.

\begin{equation}\label{eq:e_recoil_rec}
E_\mathrm{recoil} = \beta\left(\alpha\sum\limits_{i} C_{i}E_{i}\right),
\end{equation}
where $E_{i}$ is the non-muon reconstructed energy in subdetector $i$ (the tracking detector, downstream electromagnetic calorimeter, downstream hadronic
calorimeter, and the outer calorimeter), $C_{i}$ is a calorimetric constant determined by the fraction of 
passive material in subdetector $i$, and $\alpha$ and $\beta(E)$ are model-dependent parameters, tuned to the true $E_\mathrm{recoil}$ using a 
simulated charged-current interaction sample, that account for undetected energy from neutral and exiting particles.  

\begin{figure}[tp]
\centering
\ifnum\PRLsupp=0
  \includegraphics[width=\columnwidth]{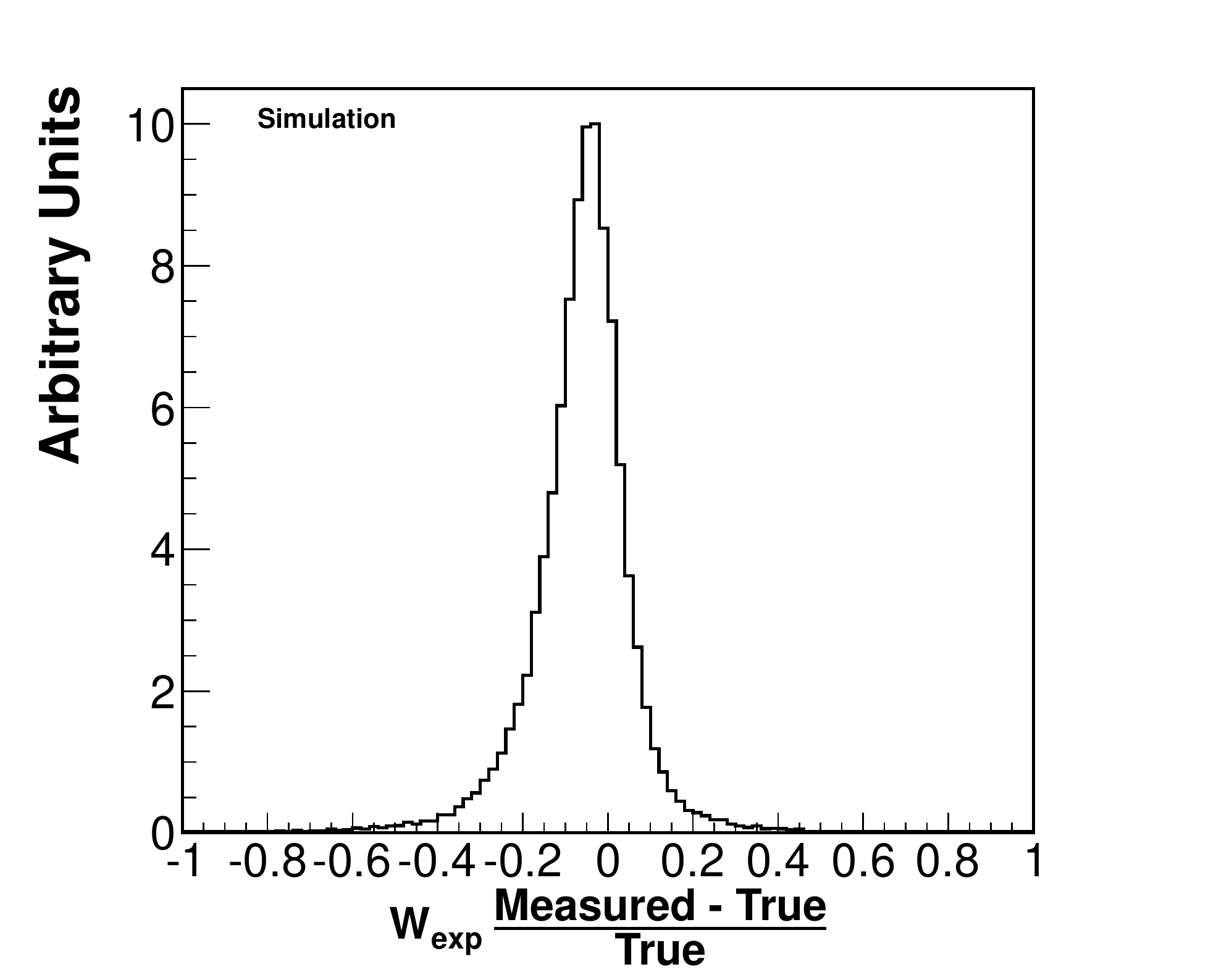}
\else
 \includegraphics[width=0.99\columnwidth]{figures/resid_W} % shrink to fit PL 
\fi
    \vspace{-7pt}
\caption{The $W_{exp}$ resolution, in which true refers to $W_{exp}$ calculated by using 
true quantities in \eqref{eq:Wexp_rec}.  Only 
events selected by the \ccpi event selection are included.  The full width at half maximum is 17\%}
    \vspace{-10pt}
\label{fig:Wexp_resid}
\end{figure}

Neutrino energy and other kinematic quantities are calculated from $E_\mathrm{recoil}$ and the reconstructed muon four-momentum using 
the following equations:
\begin{eqnarray}
E_\nu &=& E_\mu + E_\mathrm{recoil},\\
Q^2 &=& 2E_\nu(E_\mu-|\vec{p}_\mu|\cos(\theta_\mu))-m_\mu^2,\\
W_{exp}^2 &=& M_p^2 - Q^2 + 2M_pE_\mathrm{recoil}. \label{eq:Wexp_rec}
\end{eqnarray}
Here, $M_p (m_\mu$) is the proton (muon) mass, $p_\mu$ and $\theta_\mu$ are
the reconstructed momentum and angle of the muon with respect to the beam, and the $W_{exp}$ is $W$ calculated with the assumption of a single free 
target nucleon at rest.  Of the three quantities above, $W_{exp}$ is most important for this analysis.  The $W_{exp}$ resolution 
in selected \ccpi events are shown in Fig.~\ref{fig:Wexp_resid}.
The resolution on $W_{exp}$ is about 8.5\% with a bias of $-5\%$.  The bias is the result of using an inclusive sample of 
charged-current events, which have a higher average 
multiplicity than \ccpi events restricted to $W<1.8$~GeV, to tune $E_\mathrm{recoil}$.

\section{Event Selection} 
\label{sec:ev_rec}
Reconstructed \cconepi (\ccpi) events are required to contain one negatively-charged muon track and exactly one (at least one) charged-pion track 
joined at a common vertex.  The event vertex is 
restricted to occur within the central 110 planes of the scintillator tracking region and 
at least \unit[22]{cm} from any edge of the planes. These requirements define a 
fiducial region with a mass of \unit[5.57]{metric tons}, containing $(3.54 \pm 0.05) \times 10^{30}$ nucleons.

Charged pion tracks are identified by a containment requirement and two particle identification selections.  Each pion track 
is required to begin at the event vertex and stop in either the tracking or electromagnetic calorimeter regions 
of \minerva, which restricts the maximum pion kinetic energy to 350~MeV.  The particle identification selections require that there exist at 
least one track with $s_\pi>0.6$ and  
an associated Michel electron candidate.  The Michel selection disfavors 
both negatively-charged pions, which tend to be captured on a nucleus before decaying, and pions that experience secondary interactions in the detector.  The
\cconepi (\ccpi) analysis requires exactly (at least) one reconstructed charged pion track.

The reconstructed $E_\nu$ is required to be between 1.5 and 10 GeV in both analyses.  The lower bound of this selection is 
made to match the \minos muon acceptance threshold.  The upper bound reduces 
flux uncertainties, which are largest above 10~GeV.  The \cconepi analysis 
selects events with $W_{exp}<1.4$ GeV, while the \ccpi analysis selects $W_{exp}<1.8$ GeV.

\begin{figure*}[ht]
  \centering
  \subfloat[] {
    \includegraphics[width=\columnwidth]{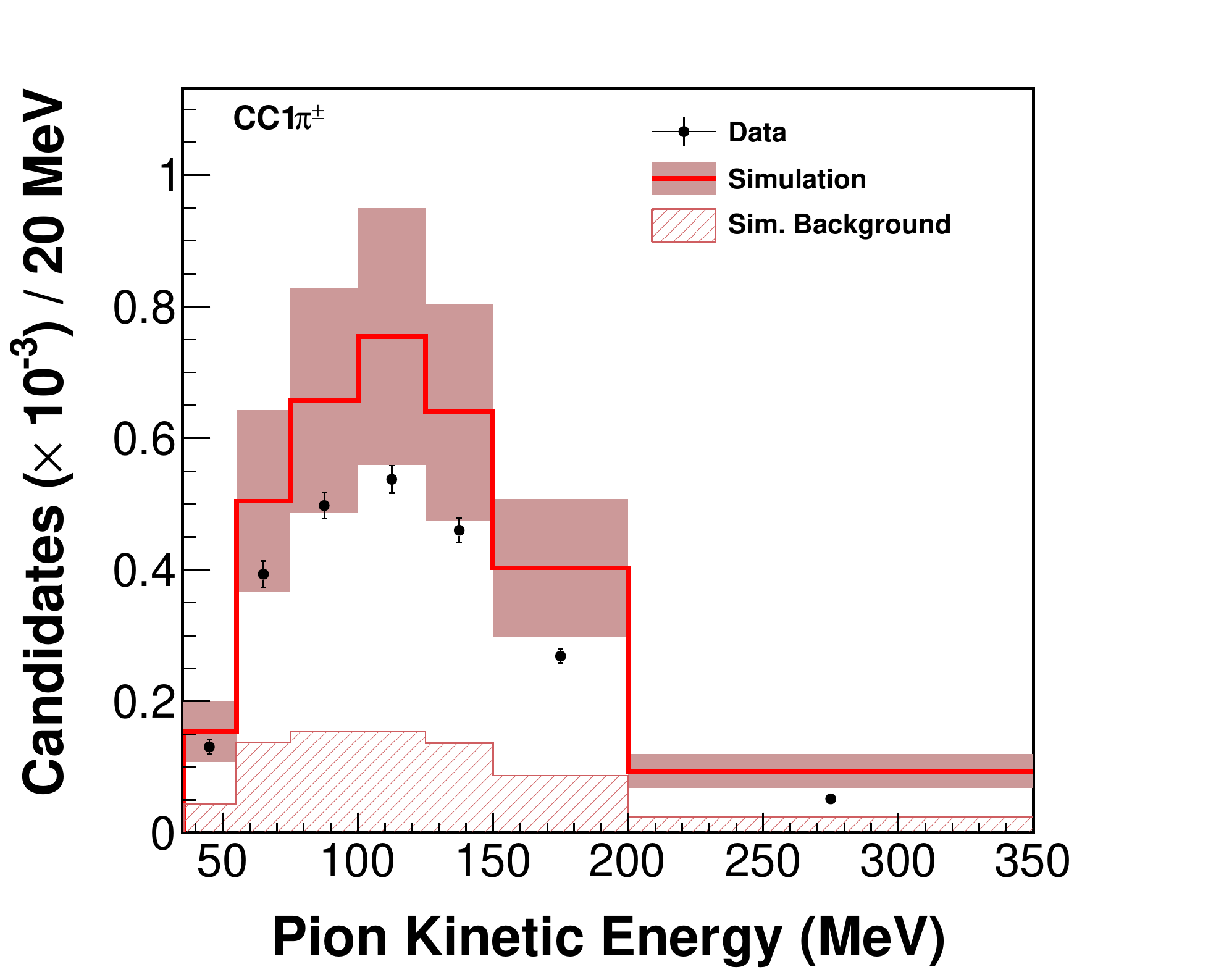}
  }
  \subfloat[] {
    \includegraphics[width=\columnwidth]{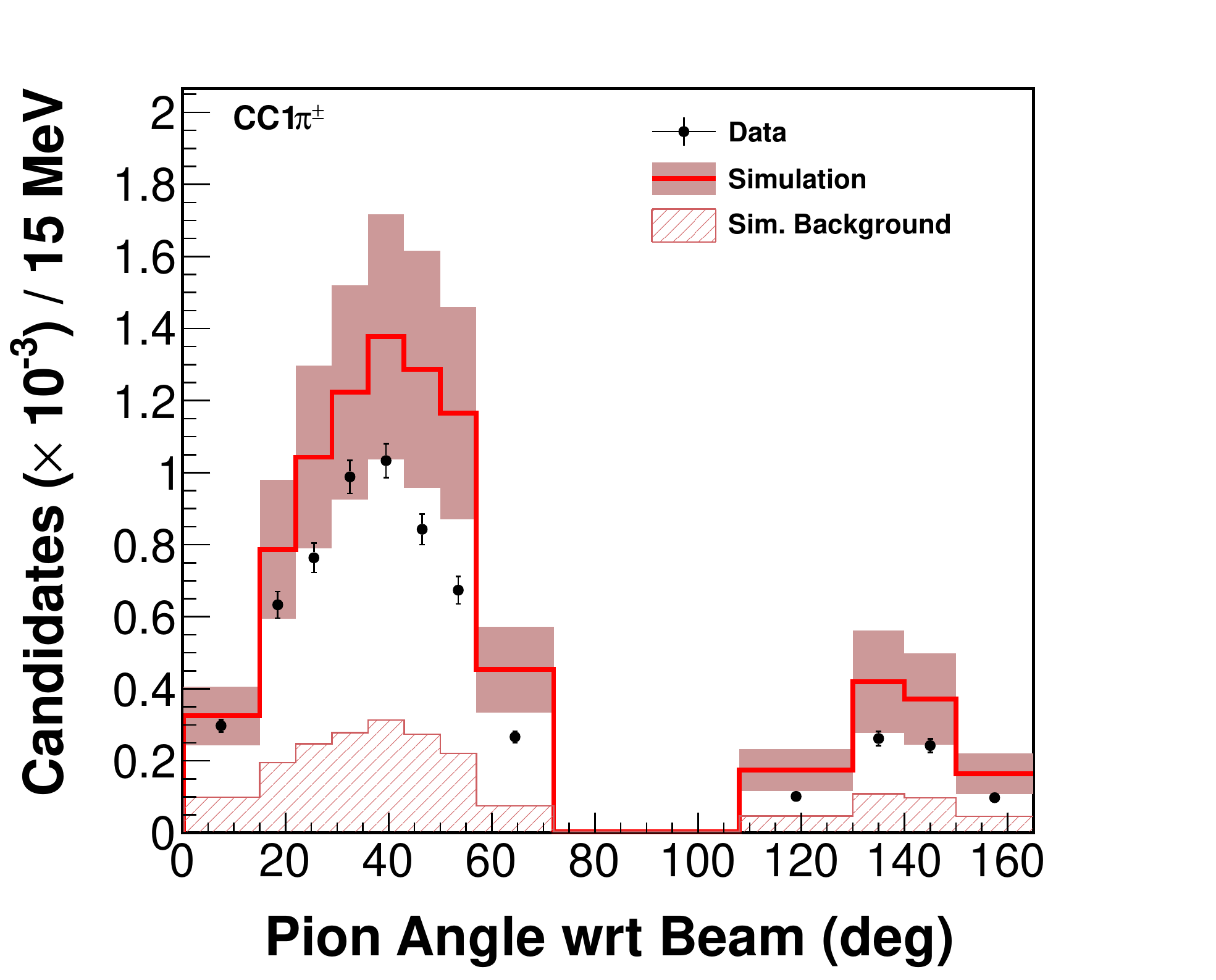}
  }
  \qquad
  \subfloat[] {
    \includegraphics[width=\columnwidth]{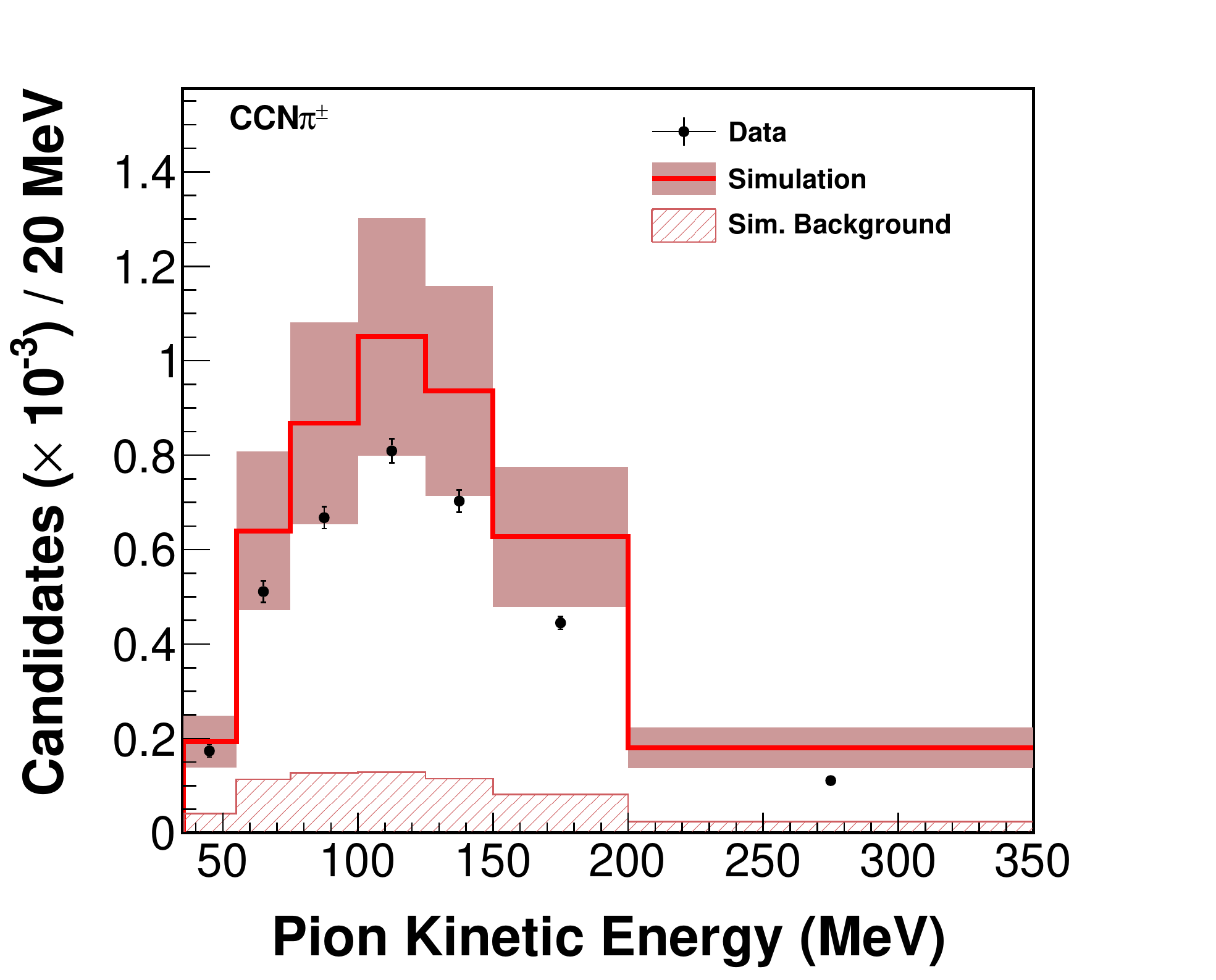}
  }
  \subfloat[] {
    \includegraphics[width=\columnwidth]{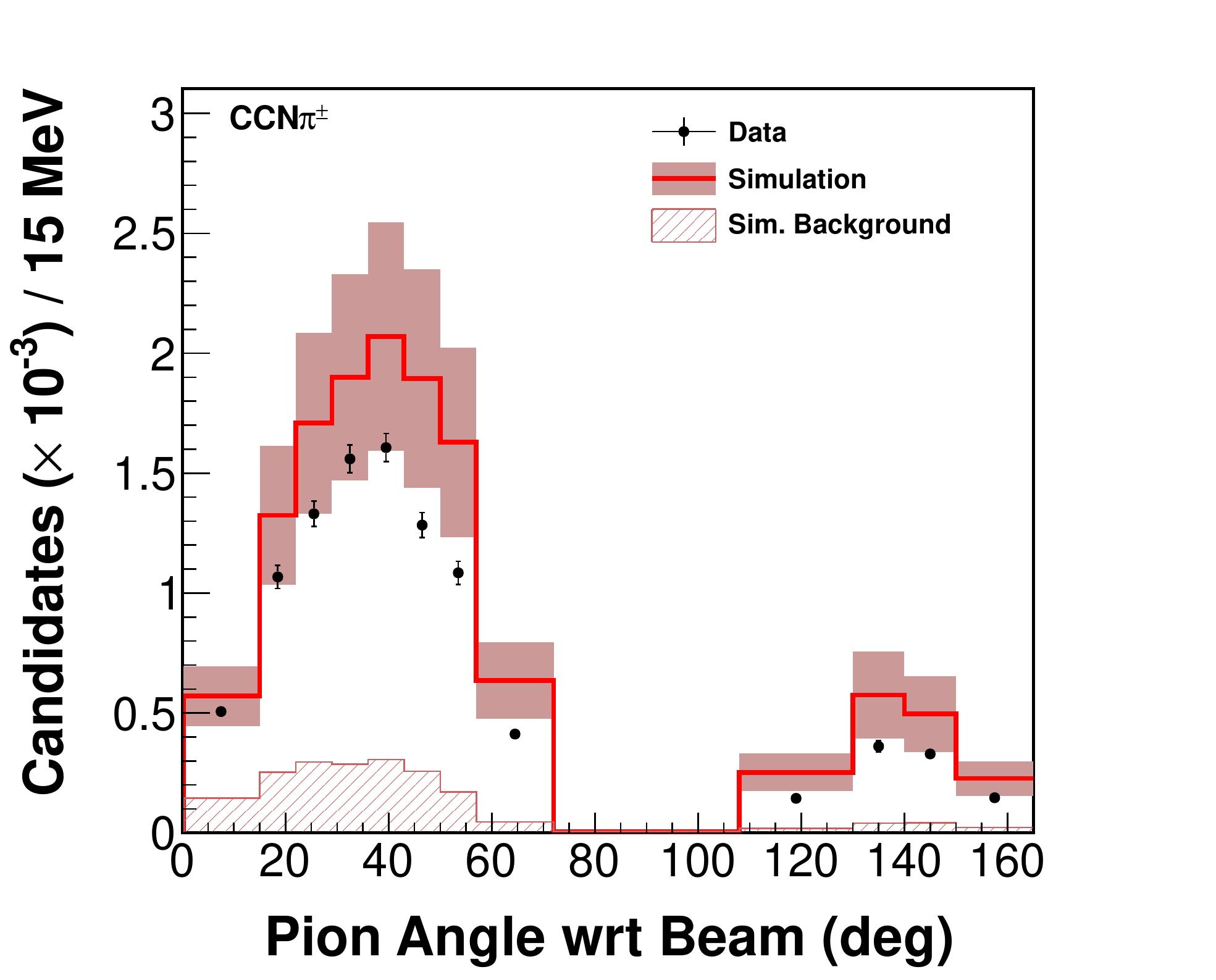}
  }
  \caption{Data-simulation comparisons of the \cconepi (top) and \ccpi (bottom) reconstructed pion kinetic energy and angle distributions after all event selections are applied.}
  \label{fig:reco_dist}
\end{figure*}

After all selections, 3474 (5410) events remain in the \cconepi (\ccpi) analysis.  Figure~\ref{fig:reco_dist} shows the selected \Tpi and 
\thetapi for both analyses.  There is a large normalization difference between simulation and data that is approximately the size 
of the total uncertainty in the prediction. This uncertainty is dominated by the uncertainty on the normalization of the resonance production cross section in 
neutrino-nucleon scattering, which GENIE determines by combining the ANL and BNL pion production data discussed in Section~\ref{sec:intro}.  
The cross section extraction procedure (see Section~\ref{sec:xs}) is designed to minimize the influence of this model uncertainty on the measured cross section distributions.

The selected pions are predicted to be 99.6\% 
(98.6\%) $\pi^+$ in the \cconepi (\ccpi) analysis because $\pi^-$ can only arise from FSI 
at low $W$ and are unlikely to survive the Michel electron requirement. The selection efficiency of 
charged pions between 35~MeV and 350~MeV in signal events is determined by simulation to be 4\% (3\%) in the \cconepi (\ccpi) analysis.  
The largest reductions in the selection efficiency are caused by the \minos-matched muon requirement, the pion track reconstruction 
inefficiency, and the Michel electron selection; the latter two are particularly affected by secondary pion scattering and absorption in the 
detector.  The signal purity of the \cconepi (\ccpi) event sample is 77\% (86\%).  Table~\ref{tab:bkgs} summarizes the background components 
in both analyses.  

\begingroup
\squeezetable
\begin{table}
\begin{tabular}{l|c|c}
  \hline\hline
  Background & \cconepi Sample (\%) & \ccpi Sample (\%) \\
  \hline
  $W > 1.4$ GeV ($1.8$ GeV) & 16.7 & 6.05 \\
  Particle Mis-ID & 4.12 & 6.67 \\
  Multiple Charged Pions & 1.61 & N/A \\
  $E_\nu > 10$ GeV & 0.45 & 0.84 \\
  Outside Fiducial Volume & 0.16 & 0.17 \\
  Not CC\numu & 0.13 &  0.18 \\
  \hline
  Total & 23.2 & 13.9 \\ 
  \hline\hline
\end{tabular}
  \caption{The predicted background components as percentages of the total selected sample after all event selections are applied.  
  The Particle Mis-ID background refers to events where another particle is misidentified as the reconstructed charged pion.}
  \label{tab:bkgs}
\end{table}
\endgroup

\section{Cross Section Extraction}
\label{sec:xs}
The \cconepi flux-integrated differential cross section per nucleon for kinematic variable $X$ (\Tpi and \thetapi in this analysis) in bin $i$ is

\begin{equation} \label{eq:diffxs}
\left(\frac{d\sigma}{dX}\right)_{i} = \frac{\sum_{j}U_{ij}\left(N_{j} - N_{j}^{bg}\right)}{\epsilon_{i}T\Phi\Delta_{i}},
\end{equation}
where $j$ is the index of a reconstructed $X$ bin, $U_{ij}$ is an unfolding function that calculates the contribution to true bin $i$ from reconstructed bin $j$, 
$N_{j}$ is the number of selected events, $N_{j}^{bg}$ is the predicted number of background events, $\epsilon_{i}$ is the signal charged-pion selection efficiency, 
$T$ is the number of nucleons in the fiducial volume, $\Phi$ is the \numu flux prediction integrated between 1.5 and 10 GeV, and $\Delta_{i}$ is the width of 
bin $i$.  The \ccpi analysis reports a slightly different observable because multiple-pion events are included in the signal:

\begin{equation} \label{eq:diffxs_ccpi}
\left(\frac{1}{T\Phi}\right)\left(\frac{dN_{\pi}}{dX}\right)_{i} = \frac{\sum_{j}U_{ij}\left(N_{\pi,j} - N_{\pi,j}^{bg}\right)}{\epsilon_{i}T\Phi\Delta_{i}}.
\end{equation}
Variable definitions are the same as those in \eqref{eq:diffxs}, except that $N_{\pi,j}$ and 
$N_{\pi,j}^{bg}$ are the number of selected charged pions and the predicted number of background charged pions in bin $j$, respectively.  
The integral of \eqref{eq:diffxs_ccpi} over $X$ yields the total number of 
charged pions $N_{\pi}$ divided by the integrated flux  and number of target nucleons.

\subsection{Background Subtraction}
\begin{figure*}[th]
\centering
\subfloat[] {
  \includegraphics[width=\columnwidth]{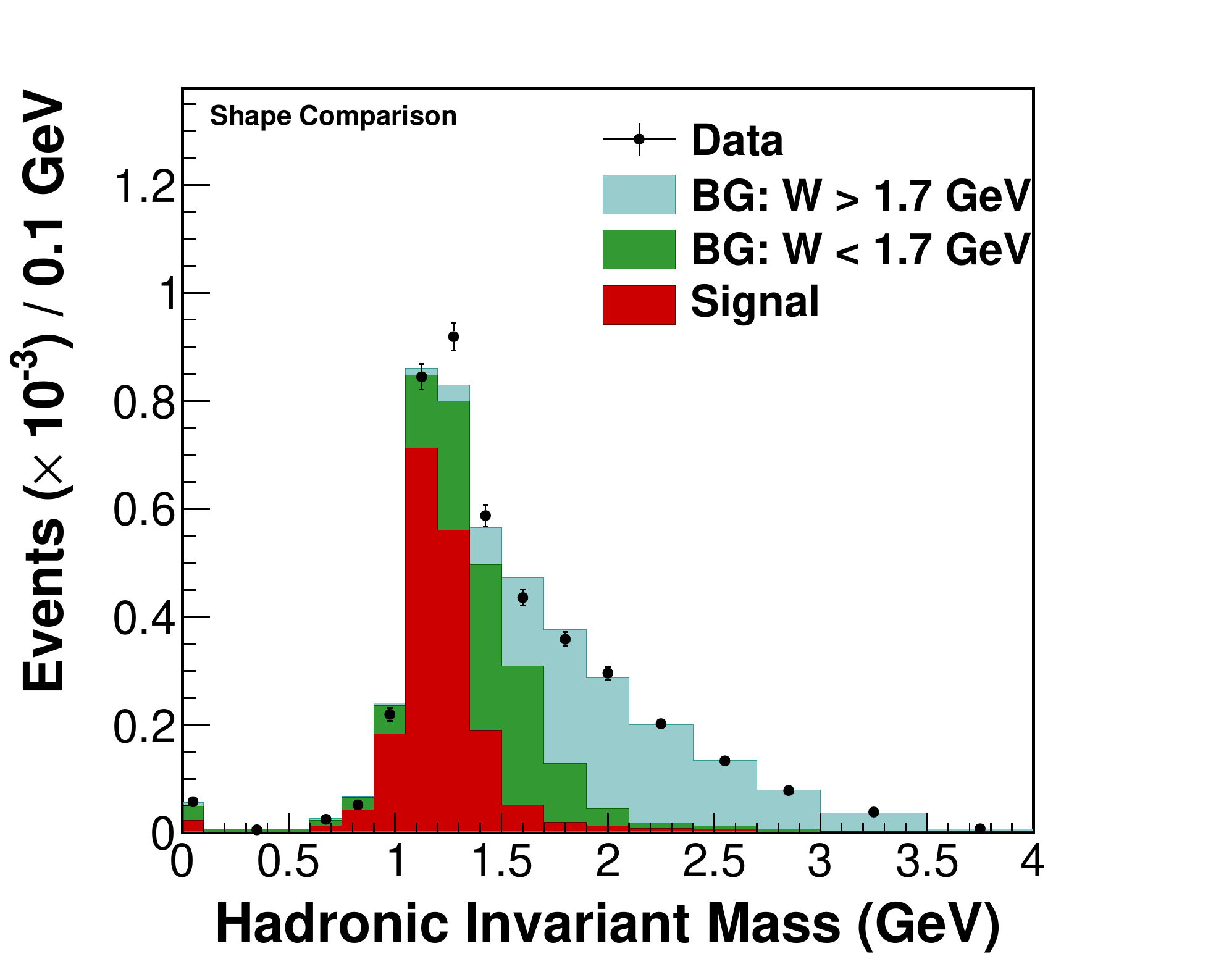}
}
\subfloat[] {
  \includegraphics[width=\columnwidth]{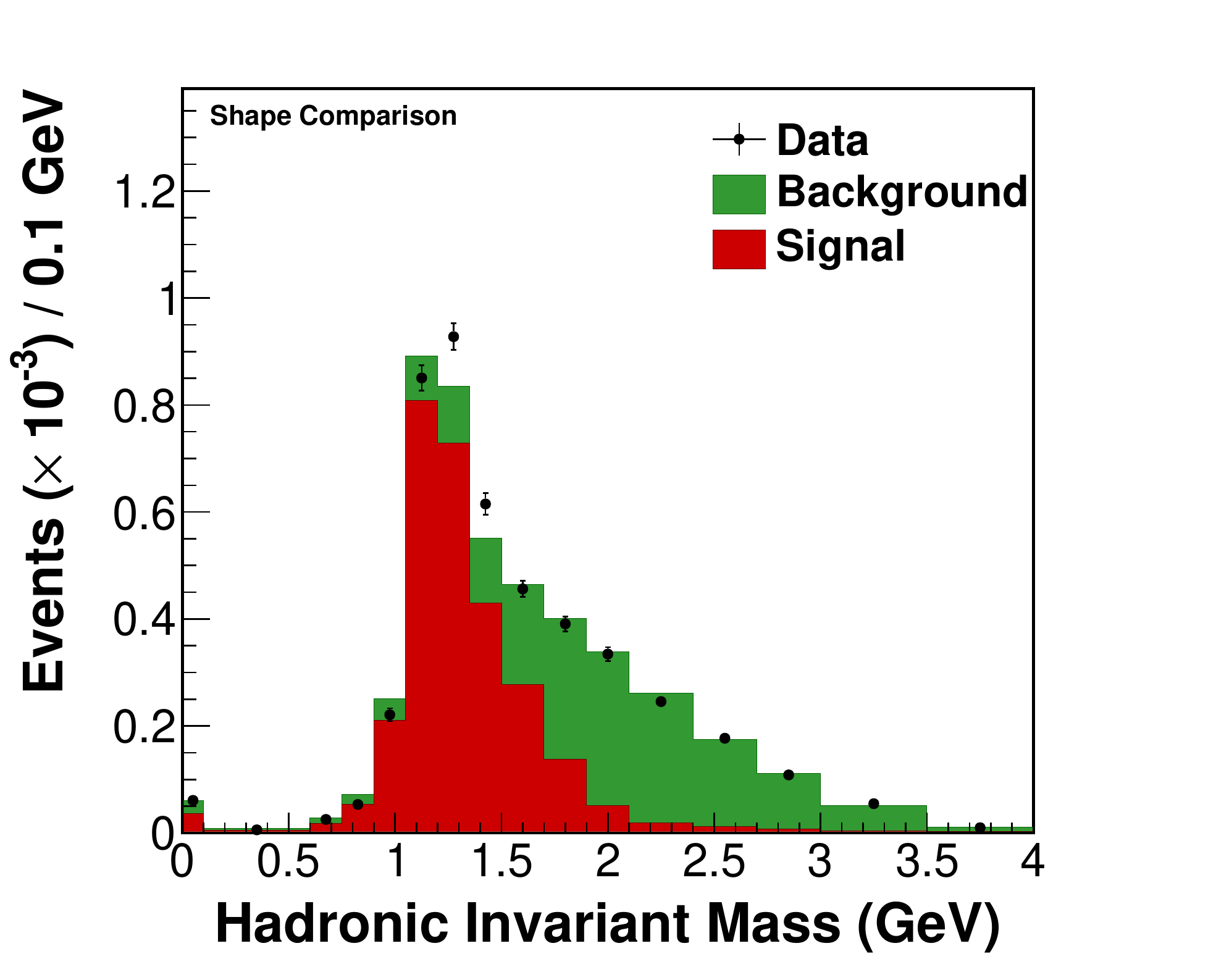}
}
    %\vspace{-7pt}
\caption{The \cconepi (left) and \ccpi (right) $W_{exp}$ distribution after fitting and reweighting the background (BG) and signal templates.  
The events below 100~MeV have a large amount of undetected hadronic energy and are not included in this analysis.}
    %\vspace{-10pt}
\label{fig:Wfit}
\end{figure*}

After event selection, the dominant 
background comes from pion production 
at higher $W$ and comprises 17\% (6\%) of the \cconepi (\ccpi) selected sample.  The total background is estimated using the reconstructed $W_{exp}$ distribution, 
in which each 
entry in the distribution is a charged pion candidate chosen by the event selection, excluding the cut on $W_{exp}$.  The simulated $W_{exp}$ distribution is 
divided into signal and background templates in bins of \Tpi and \thetapi; the \cconepi analysis further separates the background into two 
templates with true $W$ less than and greater than 1.7~GeV, which is the value at which GENIE turns off resonance production.  The normalizations of the 
signal and background templates are the fit parameters in maximum likelihood fits to the measured $W_{exp}$ distributions; each bin of \Tpi and \thetapi is fit 
independently.  The fits are restricted to $W_{exp}$ between 0.6~GeV and 2.4~GeV (3~GeV) in the \cconepi (\ccpi) analysis.  The $W_{exp}$ templates 
after fitting, 
integrated over all \Tpi and \thetapi, are shown in Fig.~\ref{fig:Wfit}. The detector calorimetric response uncertainty covers the $W_{exp}$ shape 
discrepancy that remains after the fit and is the dominant systematic uncertainty in the background estimate. 

The fit results are used to calculate weights that adjust the nominal predicted background.  In both analyses, the fit reduces the absolute background while increasing 
the prediction for the amount of background relative to the signal.  The fit procedure reduces the 
sensitivity of the background estimate to uncertainties in the simulation's cross section and FSI models, but increases sensitivity to uncertainties 
in the detector response and statistical fluctuations in the data.  The cumulative effect is positive, and the total uncertainty on the background prediction is 
reduced from 32\% to 24\% in the \cconepi analysis with a similar reduction in the \ccpi analysis.  More detail on the background subtraction procedure 
is provided in Ref.~\cite{pittir20853}.

\subsection{Unfolding}
The background-subtracted reconstructed \Tpi and \thetapi distributions are unfolded using a Bayesian 
procedure~\cite{D'Agostini:1994zf} with four iterations.  The unfolding migration matrix, which determines the probability that 
the true value of a quantity corresponds to a reconstructed value, is derived from simulation.  It is insensitive to FSI effects 
because the true values of \Tpi and \thetapi are calculated at the point where the pion exits the nucleus.  Also, 
the unfolding procedure is not sensitive to normalization uncertainties, including the large uncertainty in the resonance 
production cross section normalization.

The unfolding generally migrates events from low to high \Tpi bins, accounting for the tendency of the reconstruction to report a 
momentum that is too small for pions that interact inelastically in the detector.  The effect of unfolding on 
the \thetapi distribution is small except for the bin at 90$^{\circ}$, where the pion tracking efficiency is poor.  
Some of the pions in the neighboring bins are actually $\sim$90$^{\circ}$ pions that scatter close to the event vertex, 
such that the track reconstruction measures the scattered pion direction.  
The unfolding procedure relies on the GEANT4 interaction model to estimate this effect. 

\subsection{Efficiency Correction}
The efficiency and acceptance correction $\epsilon_{i}$ in Equations~\ref{eq:diffxs} and~\ref{eq:diffxs_ccpi} is calculated according to the
equation

\begin{equation}\label{eq:effdefinition}
\epsilon_{i} = \frac{w_{i}N_{\pi,i}^{S}}{N_{\pi,i}^{T}}, 
\end{equation}
where $N_{\pi,i}^{S}$ is the simulated number of signal pions retained by the event selection, $N_{\pi,i}^{T}$ is the total number of signal pions 
according to simulation, 
and $w_{i}$ corrects for discrepancies in the muon acceptance between data and the simulation.  Data is used to estimate 
other efficiencies, such as the acceptance of the Michel electron selection and the hadron reconstruction efficiencies, but these are found 
to match well to simulation and are included as a systematic uncertainty rather than as a correction.  The muon acceptance is 
compared in data and simulation by forming samples of exiting muon tracks in \minerva and MINOS that point toward the other detector, then 
searching for a matching track in the other detector.  The resulting corrections, which are all between 0.91 and 0.99, are measured separately for each 
data run period so that the time dependence of beam-intensity effects are accounted for.

\subsection{Systematic Uncertainties}
The cross section extraction procedure uses the simulation to estimate backgrounds, detector resolution and acceptance, 
selection efficiencies, and neutrino flux.  The systematic uncertainties on these quantities are evaluated by shifting each parameter in the simulation 
within its uncertainty $\sigma$ to produce a new simulated sample, referred to as an alternative simulation.  The cross sections are remeasured 
using each alternative simulation and a 
covariance matrix is formed from the results.  The covariance matrix for a single systematic uncertainty derived from $N$
alternative simulations is calculated as

\begin{equation}\label{eq:covmx}
C_{ij} = \frac{1}{N}\sum_{n}\left(x_{n,i} - u_{i}\right)\left(x_{n,j}-u_{j}\right),
\end{equation} 
where $i$ and $j$ indicate bins of the differential cross section and $x_{n,i}$ is the measurement of the differential cross section in bin $i$ using 
alternative simulation $n$.  The definition of $u_{i}$ changes according to the value of $N$.  If there is only one
alternative simulation, then $u_{i}$ is 
the value of the cross section measured from the nominal simulation.  Otherwise, $u_{i}$ is the mean of the measured 
cross section in all alternative simulations.  The total covariance matrix is the sum of $C_{ij}$ calculated for each systematic uncertainty.

The computational cost required to produce a new simulated sample for each systematic uncertainty is prohibitive.  Instead, this is 
effectively done for many uncertainties by reweighting the simulation or, in the case of detector resolution and energy scale uncertainties,
modifying the measured values 
event-by-event.  The effects of a few parameters, such as the effective nuclear size and quark hadronization time in GENIE, cannot be correctly estimated 
by either of these techniques.  In these cases, a new simulated sample is generated with the modified parameters.

Shape systematic uncertainties are reported for each measurement in order to mitigate certain large normalization uncertainties, such as the neutrino 
flux uncertainty.  The shape uncertainties are calculated by normalizing the cross section measurement in 
each alternative simulation so that the integrated cross sections measured in the alternative simulation and nominal simulation are equal.  The shape covariance 
matrix is calculated using the renormalized alternative simulation measurements.

Table~\ref{tab:syst-pi} lists the systematic 
uncertainties in the \cconepi analysis grouped according to the uncertainty source; the \ccpi analysis uncertainties are similar.  The total systematic 
uncertainty is between 16\% and 22\%, while the shape uncertainty ranges from from 3\% to 11\% per bin.  For comparison, the statistical uncertainties 
are approximately 3\% to 14\%.  The total uncertainties are generally systematics-limited, while the shape uncertainties are statistics-limited; the one 
notable exception to this trend is the kinetic energy measurement in the lowest bin (35--55~MeV), which is always statistics-limited.

\begingroup
\squeezetable
\begin{table}
\begin{tabular}{cccccccc}
\hline\hline
$T_{\pi}$ (MeV) & I & II & III & IV & V & Total \\
\hline
35 - 55 & 15 (9.7) & 9.7 (2.8) & 6.8 (2.9) & 8.5 (0.5) & 5.5 (2.2) & 22 (11) \\ 
55 - 75 & 12 (4.4) & 9.7 (3.3) & 8.5 (4.4) & 8.6 (0.4) & 4.8 (1.4) & 20 (7.2) \\ 
75 - 100 & 9.9 (4.6) & 8.9 (2.3) & 6.4 (2.8) & 9.0 (0.4) & 3.8 (0.6) & 18 (5.9) \\ 
100 - 125 & 10 (3.4) & 6.8 (1.7) & 4.9 (1.4) & 9.2 (0.7) & 3.0 (0.7) & 17 (4.2) \\ 
125 - 150 & 11 (3.0) & 6.7 (1.6) & 5.0 (1.5) & 8.9 (0.2) & 3.1 (0.4) & 17 (3.7) \\ 
150 - 200 & 11 (3.3) & 6.9 (2.2) & 3.1 (2.8) & 9.1 (0.4) & 2.7 (1.6) & 16 (5.1) \\ 
200 - 350 & 16 (7.2) & 8.5 (1.5) & 4.3 (3.1) & 9.2 (0.3) & 2.9 (1.2) & 21 (8.0) \\ 
\hline
\hline
$\theta_{\pi\nu}$ (degree) & I & II & III & IV & V & Total \\
\hline
0 - 15 & 11 (2.2) & 7.5 (6.7) & 11 (5.8) & 8.8 (0.6) & 4.9 (1.4) & 20 (9.3) \\ 
15 - 22 & 9.9 (2.3) & 9.2 (1.7) & 7.1 (2.3) & 9.2 (0.7) & 3.5 (0.4) & 18 (3.8) \\ 
22 - 29 & 10 (2.0) & 11 (1.8) & 4.4 (2.3) & 9.3 (0.5) & 3.3 (1.5) & 18 (3.9) \\ 
29 - 36 & 10 (1.9) & 12 (2.8) & 4.9 (2.2) & 9.1 (0.4) & 3.2 (1.6) & 19 (4.4) \\ 
36 - 43 & 11 (1.8) & 12 (3.1) & 5.6 (1.6) & 9.0 (0.2) & 3.3 (0.7) & 20 (4.0) \\ 
43 - 50 & 12 (2.0) & 12 (3.0) & 4.7 (1.5) & 9.4 (0.6) & 3.1 (0.8) & 20 (4.0) \\ 
50 - 57 & 12 (2.8) & 12 (3.1) & 3.9 (2.3) & 8.7 (0.6) & 4.7 (1.6) & 20 (5.1) \\ 
57 - 72 & 11 (1.5) & 10 (1.7) & 2.8 (4.3) & 8.6 (0.6) & 3.8 (0.6) & 18 (4.9) \\ 
72 - 108 & 11 (0.7) & 7.8 (1.8) & 6.1 (1.4) & 8.9 (0.2) & 4.4 (0.9) & 18 (2.5) \\ 
108 - 130 & 11 (2.3) & 6.4 (2.9) & 8.3 (4.1) & 9.2 (0.3) & 4.4 (0.6) & 19 (5.6) \\ 
130 - 140 & 9.7 (2.4) & 6.8 (2.6) & 7.7 (4.1) & 9.1 (0.2) & 4.3 (1.2) & 17 (5.5) \\ 
140 - 150 & 9.2 (2.9) & 7.3 (2.2) & 7.4 (3.9) & 9.0 (0.4) & 4.3 (0.6) & 17 (5.4) \\ 
150 - 165 & 9.7 (3.0) & 6.1 (3.2) & 5.6 (3.9) & 9.2 (0.5) & 5.4 (1.9) & 17 (6.2) \\ 
\hline\hline
\end{tabular}
\caption{Fractional systematic uncertainties (in per cent) on \cconepi $d\sigma/dT_{\pi}$ (top) and $d\sigma/d\theta_{\pi\nu}$ (bottom) 
associated with detector response (I), neutrino cross section model (II), nuclear effects including FSI (III), flux (IV), 
and other sources (V).  The absolute uncertainties are followed by shape uncertainties in parentheses.}
\label{tab:syst-pi}
\end{table}
\endgroup

The largest contribution to the total uncertainty comes from uncertainty in the detector response, particularly the average calorimetric response to 
events passing the analysis selections (6-11\%), to which the background-constraining fits are particularly sensitive.  The measurements at low pion 
kinetic energy are also very sensitive to the detector mass model uncertainty (7\% between 35 and 55 MeV) since this affects the pion track 
reconstruction threshold.  The total uncertainty also has large contributions from the neutrino-nucleon cross section model (6-12\%) and neutrino flux 
uncertainty ($\sim$9\%).  The primary uncertainty from the neutrino-nucleon cross section model comes from modeling the muon angular 
distribution in resonance production (7-12\%), which affects the estimated MINOS muon acceptance.  This uncertainty can be reduced to 4\% or less by
restricting the signal definition to muon angles less than $20^\circ$; the Appendix contains the results of this measurement variation.

The shape uncertainties are generally less sensitive to the systematic effects described above, especially to the neutrino flux (reduced to $<$1\%).  The exception to this 
is the measurement at \Tpi between 35 and 55 MeV, which retains sensitivity to the detector mass model.  Additionally, the shapes of the angular 
cross section measurements at forward angles are sensitive to the large uncertainties assumed in the $\Delta$ decay anisotropy model 
(see Section~\ref{sec:expt_sim}).

\section{Results}
\label{sec:results}
\subsection{Models}
The results of this measurement for the \cconepi and \ccpi
channels are presented in the following figures.  They are compared
with calculations from the theoretical work of Athar, Chaukin, and
Singh (ACS)~\cite{Athar} and the event generators GENIE~\cite{Andreopoulos201087}, NEUT~\cite{Hayato:2009zz}, 
and NuWro~\cite{Golan:2012wx}.  Predictions from the GiBUU model can be found in Refs.~\cite{gibuu-minerva, gibuu_mnv}.  Each prediction includes
models for the initial neutrino interaction, the nuclear structure
affecting the initial interaction, and
the FSI of the particles produced.  For resonance production, GENIE and NEUT use
the model of Rein and Sehgal~\cite{Rein:1980wg} without including resonance interference and with varying 
treatments of nuclear structure.  NuWro includes only the $\Delta$(1232) resonance, using the Adler model~\cite{nuwro_res,nuwro_res2}, 
and ACS uses the parameterization of Schreiner and Von Hippel~\cite{acs_res} and contains medium 
modifications to the $\Delta$ mass and decay width.  

\begingroup
\squeezetable
\begin{table*}[ht]
\begin{tabular}{l|l|l|l|l|l}
\hline\hline
Model & Nucleon Resonance & Nonresonance & Nucleon Momentum & $\Delta$ Modifications & FSI \\ 
\hline 
ACS~\cite{Athar} & Schreiner--Von Hippel~\cite{acs_res} & None & local relativistic & $\Delta$ mass and & attenuation only~\cite{VicenteVacas:1994} \\ 
    & & & Fermi gas & decay width & \\
\hline
GENIE~\cite{Andreopoulos201087} 2.6.2 & Rein--Sehgal~\cite{Rein:1980wg} & Bodek--Yang~\cite{Bodek:2004pc} with & global relativistic & no & effective cascade \\
    & without interference & extrapolation to lower $W$ & Fermi gas & & \\
\hline
NEUT~\cite{Hayato:2009zz} 5.3.3 & Rein--Sehgal & Rein--Sehgal & global relativistic & yes, via FSI model & Salcedo--Oset~\cite{Salcedo:1987md}, \\
    & without interference & & Fermi gas & & full cascade \\
\hline
NuWro~\cite{Golan:2012wx} & Adler~\cite{nuwro_res,nuwro_res2}, & Bodek--Yang with & global relativistic & yes, via FSI model & Salcedo--Oset, \\
    & $\Delta$(1232) only & extrapolation to lower $W$ & Fermi gas & & full cascade \\

\hline\hline
\end{tabular}
\caption{Summary of the models presented in this paper.}
\label{tab:models}
\end{table*}
\endgroup

NEUT takes the nonresonant interaction from Rein and Sehgal; ACS has no nonresonant mechanisms; 
GENIE and NuWro have similar approaches, using the model of 
Bodek and Yang~\cite{Bodek:2004pc} above the resonance region
and smoothly extrapolating it to lower $W$ to converge with 
the resonance model.  All models must choose between matching
the ANL~\cite{Radecky:1981fn} and BNL~\cite{Kitagaki:1986ct} data 
for charged-current pion production from 
nucleon targets because the BNL data is about 40\% larger than the 
ANL data for neutrino energies of $\sim$2 GeV.  While the
GENIE fit is midway between the two data sets, NEUT and NuWro fits
are closer to the ANL result.  

GENIE, NEUT, and NuWro use
a relativistic Fermi gas model for the nucleon momentum distribution, while
ACS uses a local Fermi gas model in which the Fermi momentum depends
on the radial distance from the center of the nucleus.  For FSI, NEUT and 
NuWro use the Salcedo--Oset model~\cite{Salcedo:1987md} in a cascade 
formalism which has nuclear medium corrections, while GENIE uses an effective cascade 
model which has similar agreement with pion-nucleus data.  ACS uses a model which includes 
pion attenuation, but not inelastic scattering which changes the pion 
energy and angle~\cite{VicenteVacas:1994}.  Thus, a calculation with 
excellent nuclear medium effects but incomplete FSI (ACS) is compared with 
calculations with simple nuclear structure and detailed FSI (GENIE, NuWro, NEUT).  
Table~\ref{tab:models} summarizes the models used by the predictions shown in this paper.

\subsection{\cconepi\ results}
\begin{figure}[h]
\centering
\subfloat[] {
    \includegraphics[width=\columnwidth]{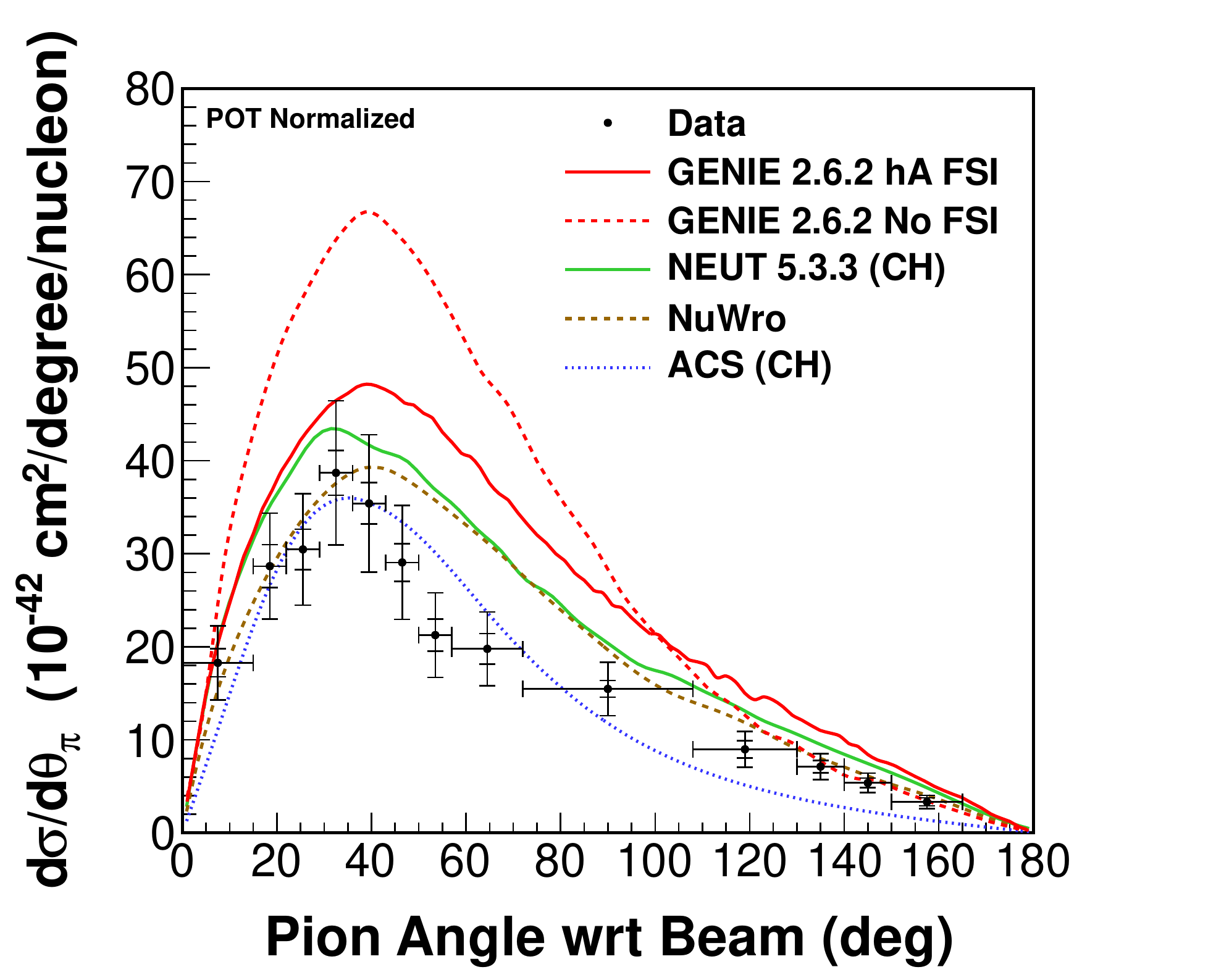}
}
\qquad
\subfloat[] {
    \includegraphics[width=\columnwidth]{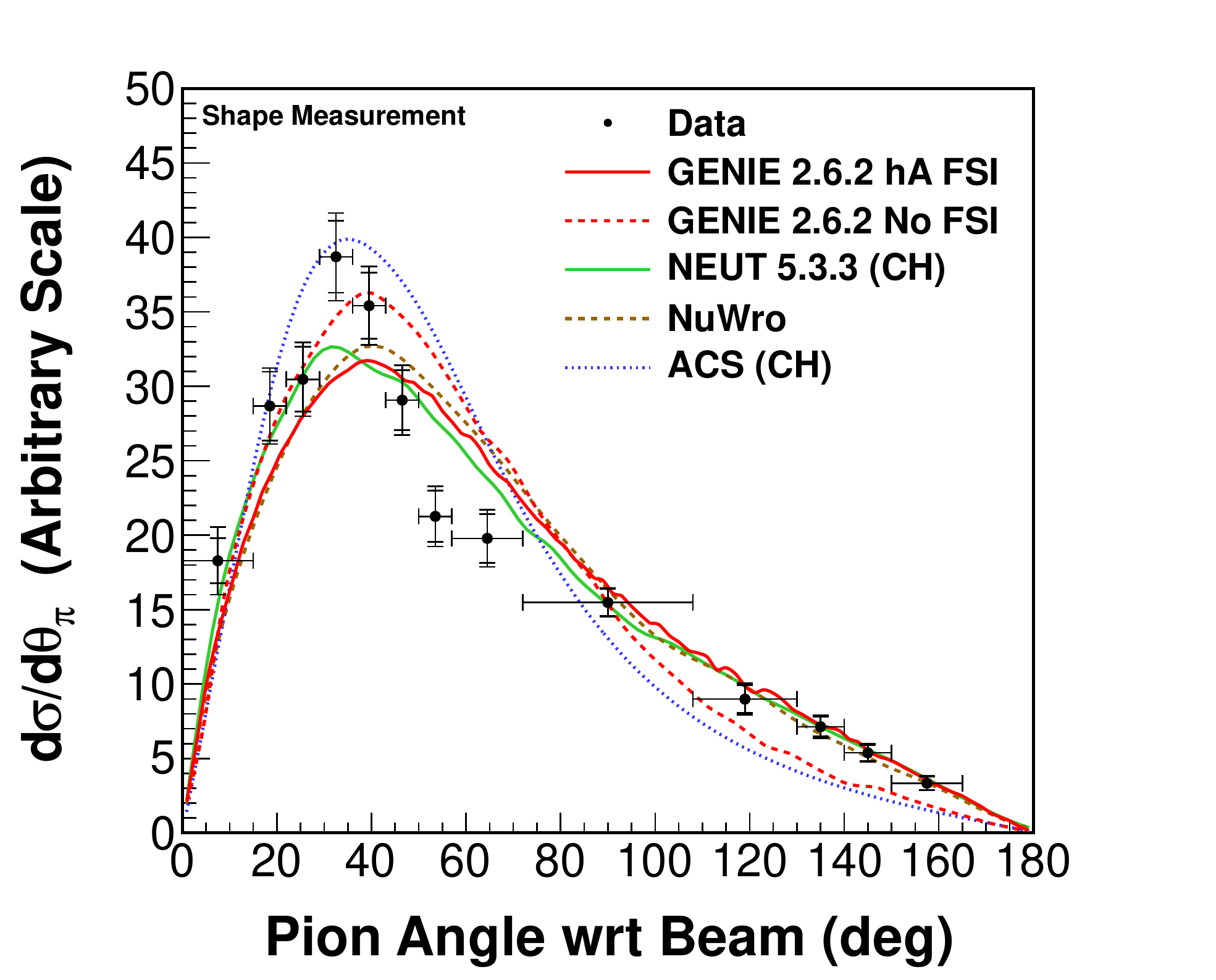}
}
%    \vspace{-7pt}
\caption{\cconepi $d\sigma / d\theta_\pi$ (top) and its shape (bottom) compared to the GENIE, ACS, NEUT, and NuWro models.
The shape predictions are normalized to the integral of the data.  The inner (outer) error bars correspond to the statistical (total) uncertainties.}
%    \vspace{-10pt}
\label{fig:piangle-1pi}
\end{figure}

The measured $d\sigma/d\theta_\pi$ for the \cconepi analysis
is shown in Fig.~\ref{fig:piangle-1pi}, along with predictions 
from the models discussed above.  The $\chi^2$ between the data and 
model predictions are listed in Table~\ref{tab:chi2}.  
The effect of FSI, shown in the comparison between 
the GENIE ``hA FSI'' and ``no FSI'' curves, is to deplete (increase) 
the forward (backward) angle cross section.  Both the absolute and 
shape measurements show a clear preference for models that implement FSI with a full or effective cascade model.  
In particular, the ``no FSI'' and ACS predictions do not 
describe the relative cross section for forward- and backward-going pions.

\begingroup
\squeezetable
\begin{table}[h]
\begin{tabular}{l|cc}
\hline\hline
Model & Absolute $\chi^{2}$ & Shape $\chi^{2}$ \\ 
\hline 
ACS (CH) & 78 & 89 \\ 
GENIE 2.6.2 hA FSI & 104 & 41 \\
GENIE 2.6.2 No FSI & 234 & 72 \\
NEUT 5.3.3 (CH) & 50 & 26 \\
NuWro & 67 & 46 \\
\hline\hline
ACS (CH) & 40 & 34  \\ 
GENIE 2.6.2 hA FSI & 21 & 7.4 \\
GENIE 2.6.2 No FSI & 105 & 23 \\
NEUT 5.3.3 (CH) & 26 & 13 \\
NuWro & 25 & 16 \\
\hline\hline
\end{tabular}
\caption{Top: Absolute (shape) $\chi^{2}$ with 13 (12) degrees of freedom between the \cconepi $d\sigma / d\theta_\pi$ measurement and various models. 
Bottom: Corresponding \cconepi $d\sigma / dT_\pi$ $\chi^{2}$ with 7 (6) degrees of freedom.}
\label{tab:chi2}
\end{table}
\endgroup
 
The shape of $d\sigma/d\theta_\pi$ could be 
sensitive to the $\Delta \rightarrow \pi$ decay angle distribution.  
GENIE and NuWro use an isotropic decay 
distribution while NEUT assumes the anisotropy in the original Rein-Sehgal 
model~\cite{Rein:1980wg}.  ACS calculates specific anisotropies 
for the $\Delta^{++}$ and the $\Delta^+$ separately.  
The larger effect, however, is the implementation of FSI.  

\begin{figure}[ht]
\centering
\subfloat[] {
    \includegraphics[width=\columnwidth]{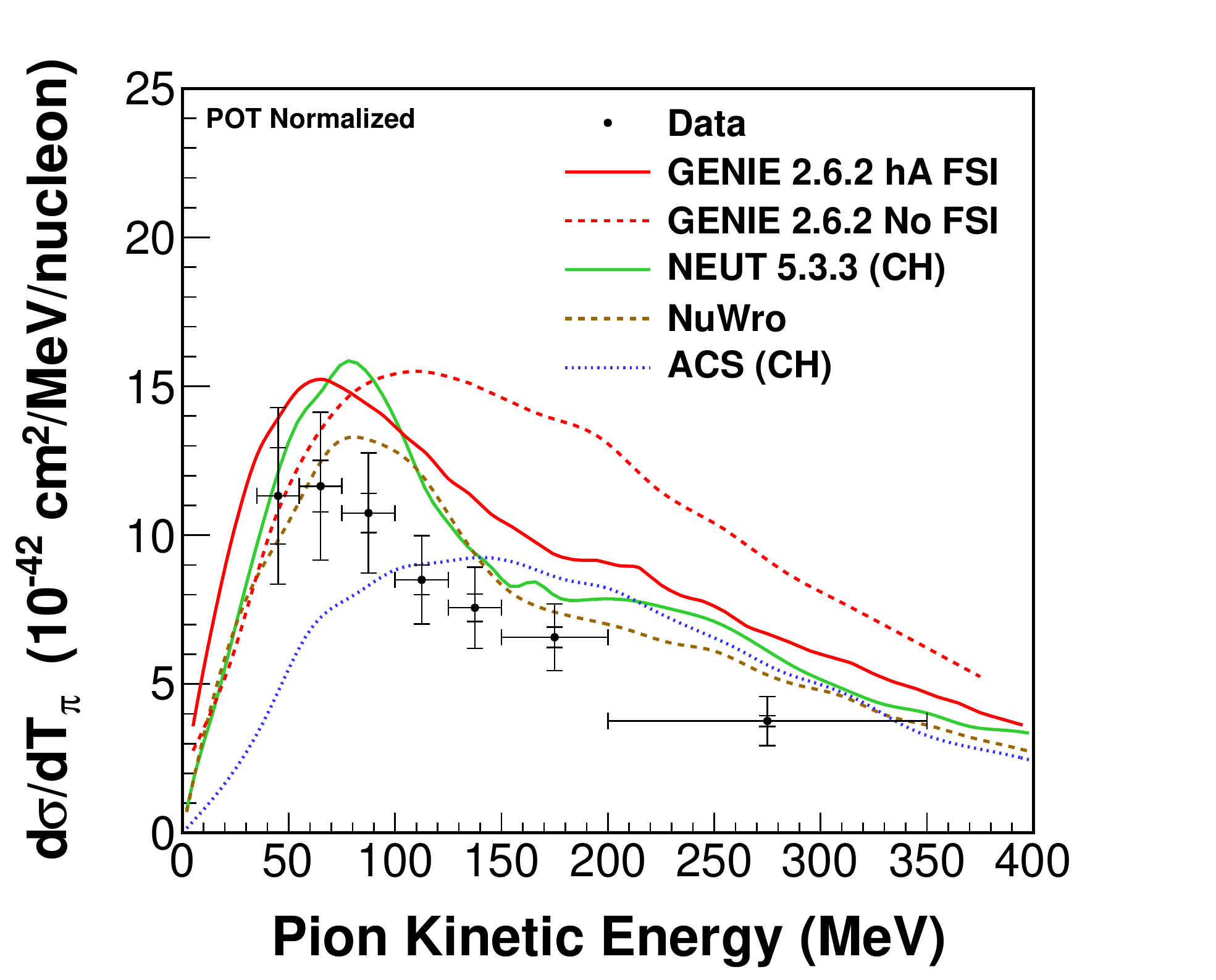}
}
\qquad
\subfloat[] {
    \includegraphics[width=\columnwidth]{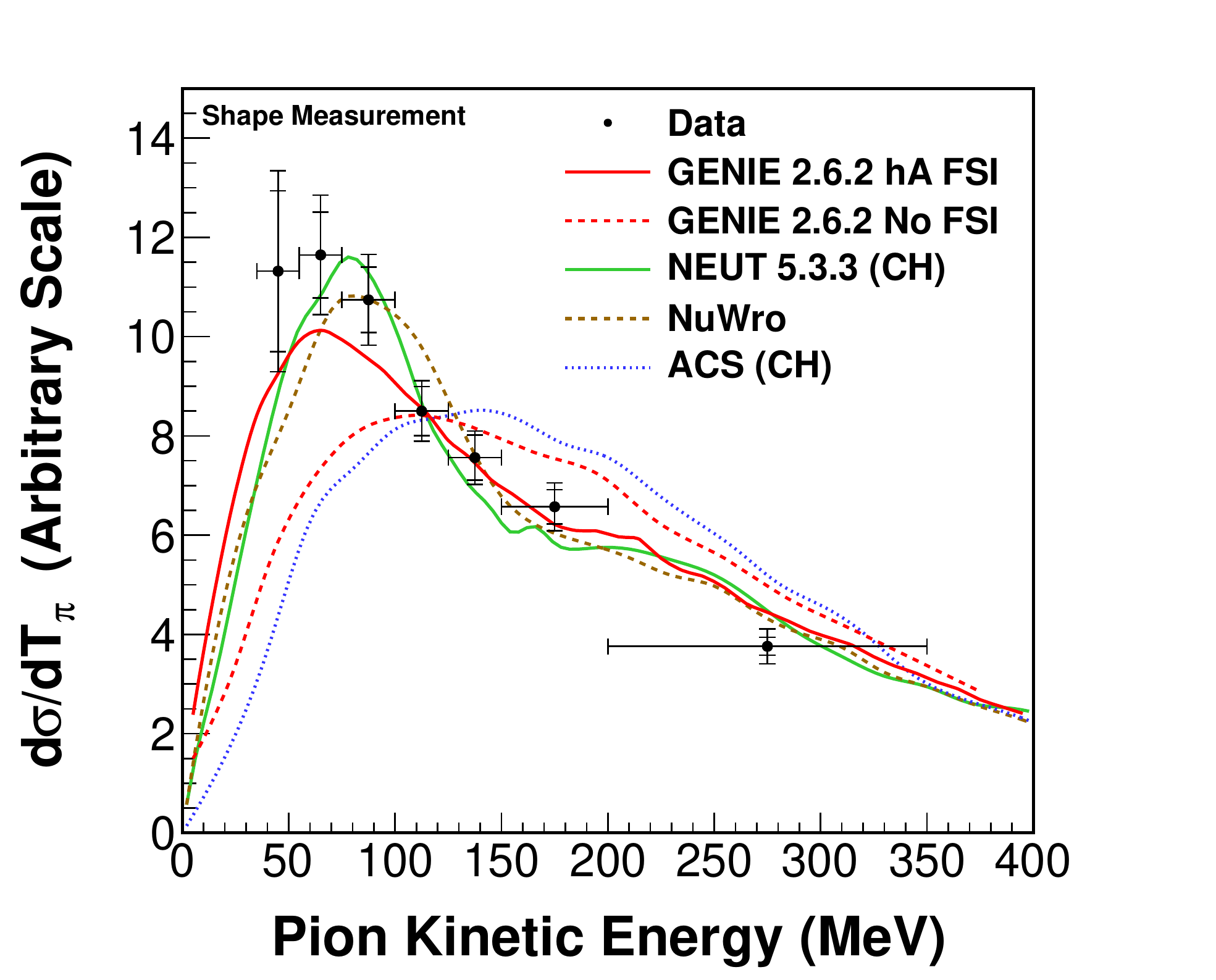}
}
%    \vspace{-7pt}
\caption{\cconepi $d\sigma / d\Tpi$ (top) and its shape (bottom) compared to the GENIE, ACS, NEUT, and NuWro models.
The shape predictions are normalized to the integral of the data.  The inner (outer) error bars correspond to the statistical (total) uncertainties.}
%    \vspace{-10pt}
\label{fig:pike-1pi}
\end{figure}

The measured \cconepi $d\sigma/dT_\pi$ is shown in  
Fig.~\ref{fig:pike-1pi} along with the model predictions.  The $\chi^2$ 
calculations are provided in Table~\ref{tab:chi2}.
FSI suppresses the charged pion production cross section through pion 
absorption and charge exchange, and migrates pions to lower energies 
through scattering.  These interactions are highly energy dependent, 
peaking between 100 and 220~MeV~\cite{Leitner:2008wx}, and significantly 
modify the shape of $d\sigma/dT_\pi$.  The significant reduction in the total cross section is seen
by comparing the solid and dashed GENIE predictions in the absolute measurement, while the energy dependence 
of the FSI can be clearly seen by comparing the GENIE predictions for the shape measurement.  
The data are in best agreement with models that implement full or effective particle cascade FSI algorithms.     

\begin{figure}[tp]
\centering
\includegraphics[width=\columnwidth]{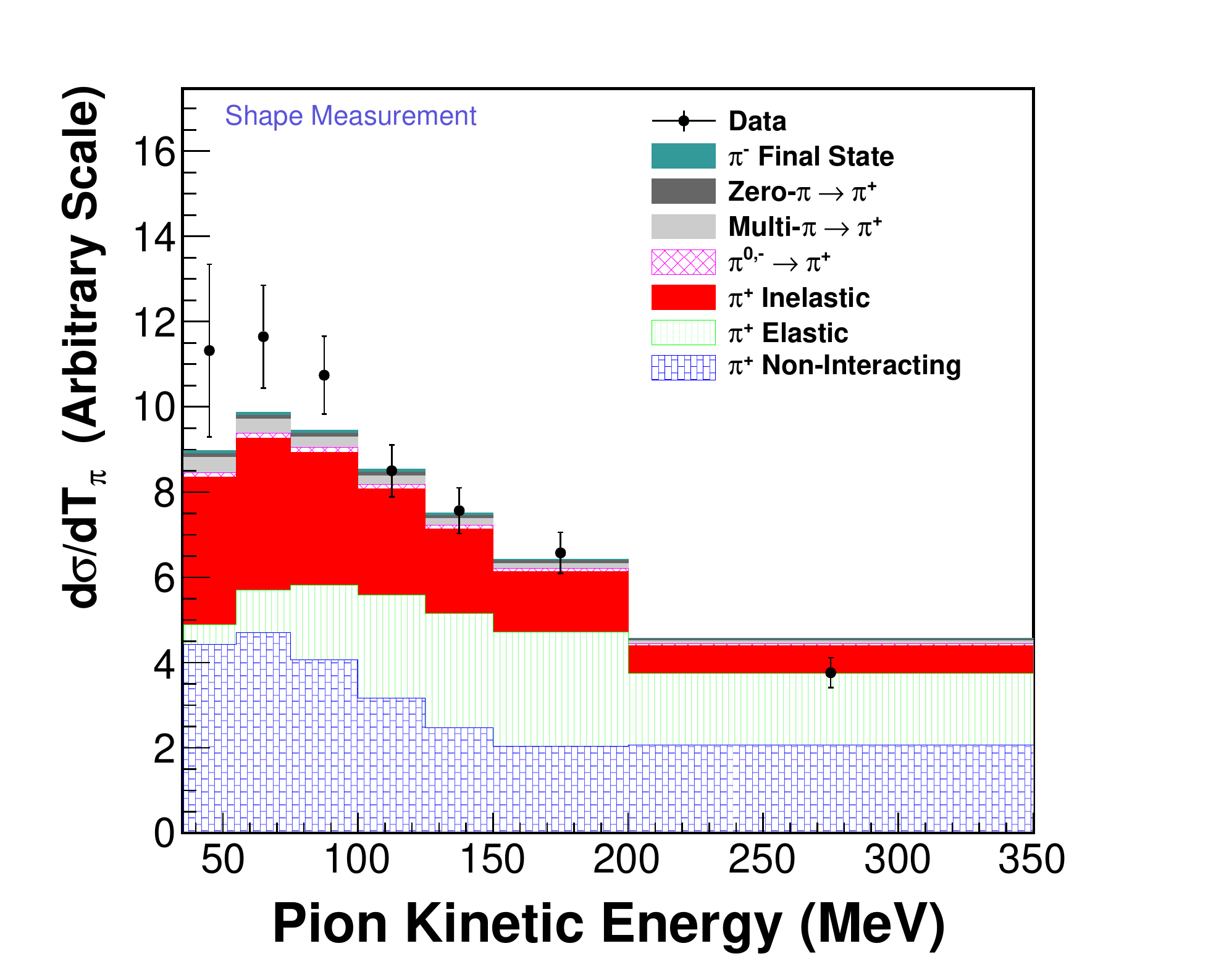} % shrink to fit PL 
%\vspace{-7pt}
\caption{The measured shape of $d\sigma / dT_\pi$ and a breakdown of the GENIE calculation
by FSI channel.  Coherent pion production and hydrogen interactions are included in the ``Non-Interacting'' category.  
The simulation is normalized to the integral of the data.}
    %\vspace{-10pt}
\label{fig:pike-1pi-AN-FSI}
\end{figure}

The shape of $d\sigma/d\Tpi$ is compared with the GENIE calculation subdivided by 
FSI channel in Fig.~\ref{fig:pike-1pi-AN-FSI}.  Effects of pion
absorption are significant but not directly seen because those
pions cannot be in the final state.  Inelastic scattering 
is the dominant contributor because the interaction probability is large
and the energy is significantly reduced.  Elastic scattering is also 
significant but does not affect the energy spectrum.  The calculation 
would agree with the data shape better if the inelastic
scattering contribution were increased within the estimated error
in the total pion inelastic cross section data ($\pm$40\%~\cite{Ashery:1981tq}).  

\begin{figure}[!htp]
\centering
  \includegraphics[width=0.9\columnwidth]{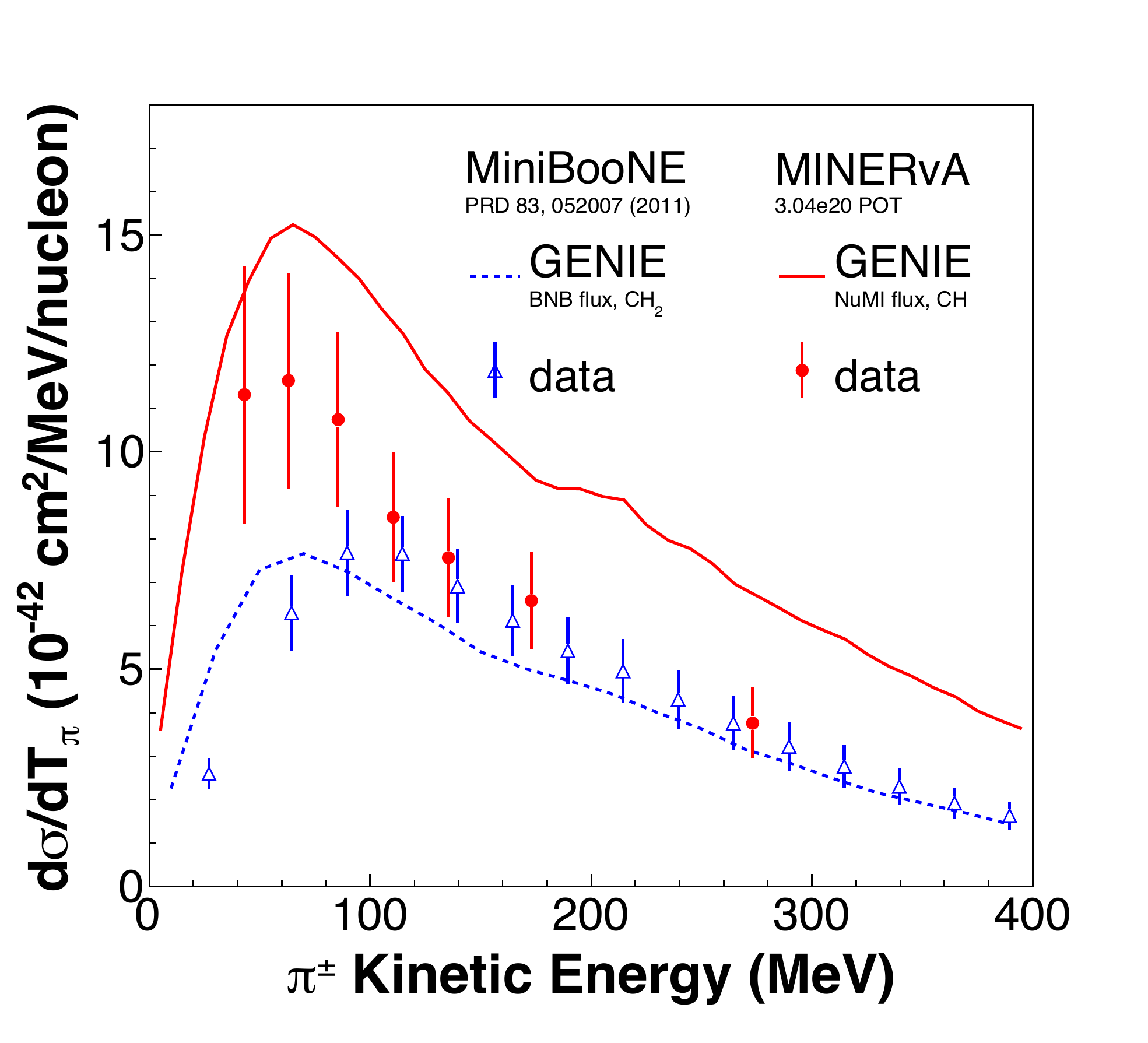}
    %\vspace{-7pt}
\caption{Comparison between the \minerva and \miniboone~\cite{miniboone_piprod} $d\sigma/d T_\pi$ data via the corresponding GENIE 2.6.2 ``hA FSI'' predictions.
Error bars indicate the total uncertainty.}
    %\vspace{-10pt}
\label{fig:mb-comp}
\end{figure} 
 
Since these data favor different calculations than the \miniboone 
data~\cite{Rodrigues:2014jfa}, a comparison of the two data sets is interesting.
Figure~\ref{fig:mb-comp} compares this measurement of \cconepi $d\sigma/dT_\pi$ with that 
of \miniboone along with the two corresponding GENIE predictions for 
the appropriate neutrino fluxes~\cite{miniboone_flux} and signal definitions.  
\minerva measures higher energy and higher 
$Q^2$ neutrino interactions than does \miniboone, but the $W$ and 
$T_\pi$ kinematic ranges overlap significantly.  \minerva reports
the cross section at $W<$ 1.4 GeV while \miniboone 
selects events with reconstructed $W<$ 1.35 GeV and uses the NUANCE event generator~\cite{NUANCE} 
to measure the cross section over the full $W$ range; 
GENIE predicts that 24\% of the \miniboone cross section result is at 
$W>$ 1.4 GeV. All these considerations lead to differences in the 
contributions due to $\Delta$ excitation and non-resonant 
backgrounds, but the key feature of attenuation due to pion FSI 
is expected to be similar. 

The \minerva and \miniboone results have a similar shape and magnitude above 
$T_\pi=100$~MeV in Fig.~\ref{fig:mb-comp}.  The shape agreement indicates some 
consistency in the pion absorption FSI process, while the agreement in 
magnitude is unexpected when considering the different $E_{\nu}$ and $W$ ranges 
of the measurements and is not presently described by any models.  
In fact, the \minerva cross section at higher \Tpi would nominally be
larger than the \miniboone result 
because the cross section for pion production
from nucleon targets rises with increasing $E_\nu$.
The shape disagreement below 100~MeV is also not explained 
by current models.  The GENIE model predicts the 
shape but overpredicts the level of the \minerva data (see Fig.~\ref{fig:pike-1pi} for shape), while it 
predicts the magnitude but not the shape of the MiniBooNE 
data.  
The same trend is seen with the GiBUU calculation, as shown in 
Ref.~\cite{gibuu-pi}.  

\begin{figure}[ht]
\centering
\subfloat[] {
    \includegraphics[width=\columnwidth]{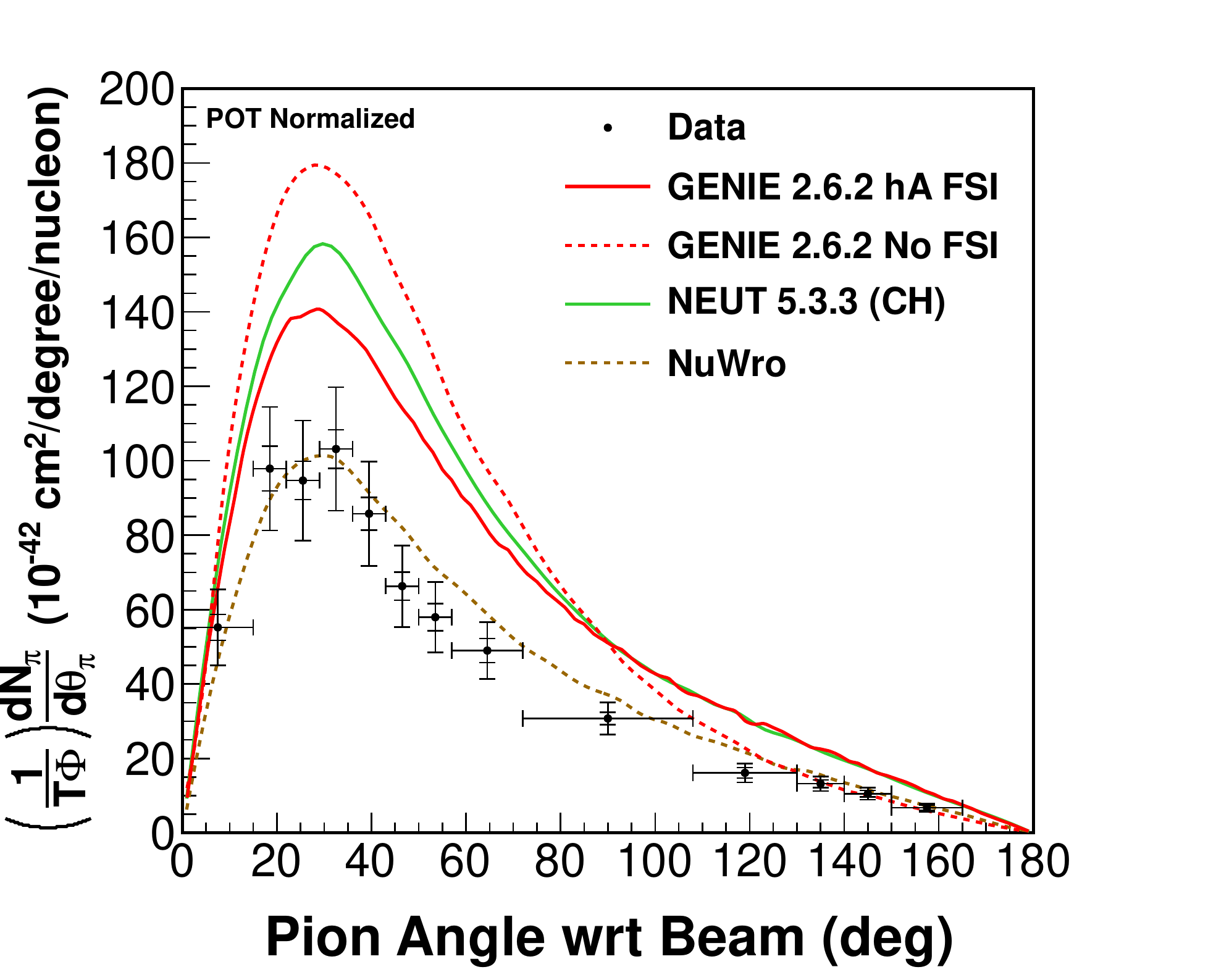}
}
\qquad
\subfloat[] {
    \includegraphics[width=\columnwidth]{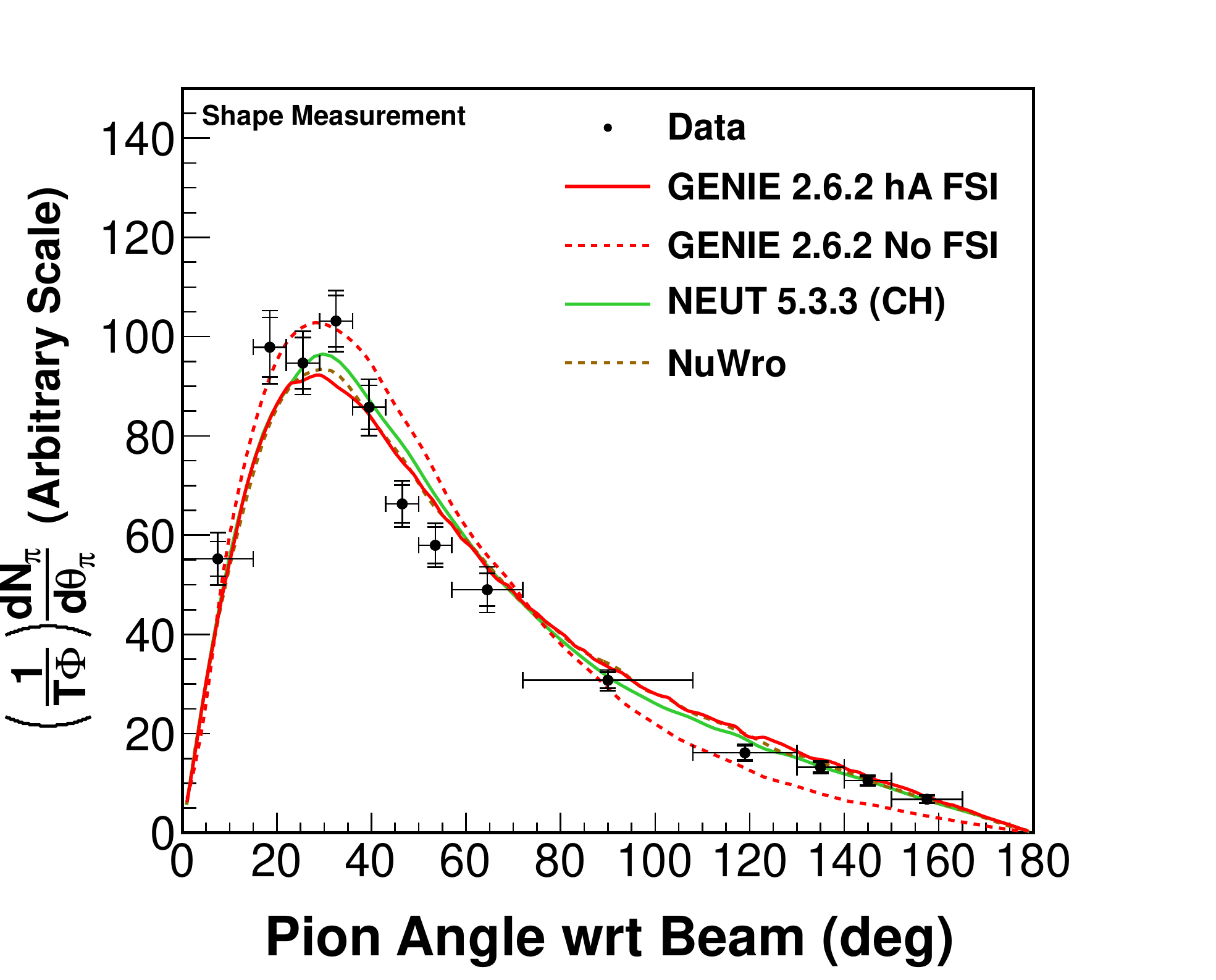}
}
%    \vspace{-7pt}
\caption{\ccpi $(1/T\Phi)(dN_{\pi} / d\theta_\pi)$ (top) and its shape (bottom) compared to the GENIE, NEUT, and NuWro models.
The inner (outer) error bars correspond to the statistical (total) uncertainties.}
%    \vspace{-10pt}
\label{fig:piangle-Npi-POT}
\end{figure}

\begin{figure}[ht]
\centering
\subfloat[] {
    \includegraphics[width=\columnwidth]{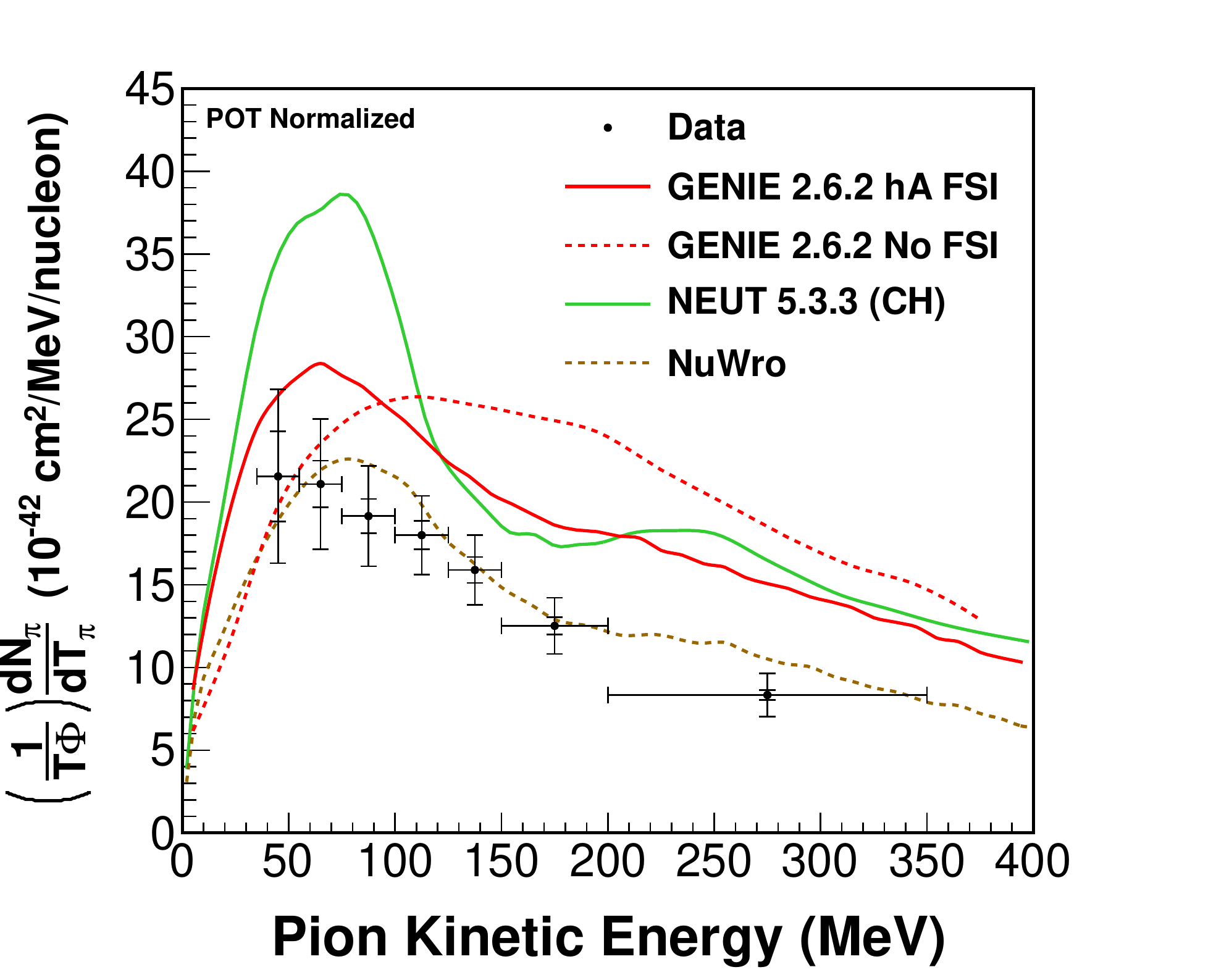}
}
\qquad
\subfloat[] {
    \includegraphics[width=\columnwidth]{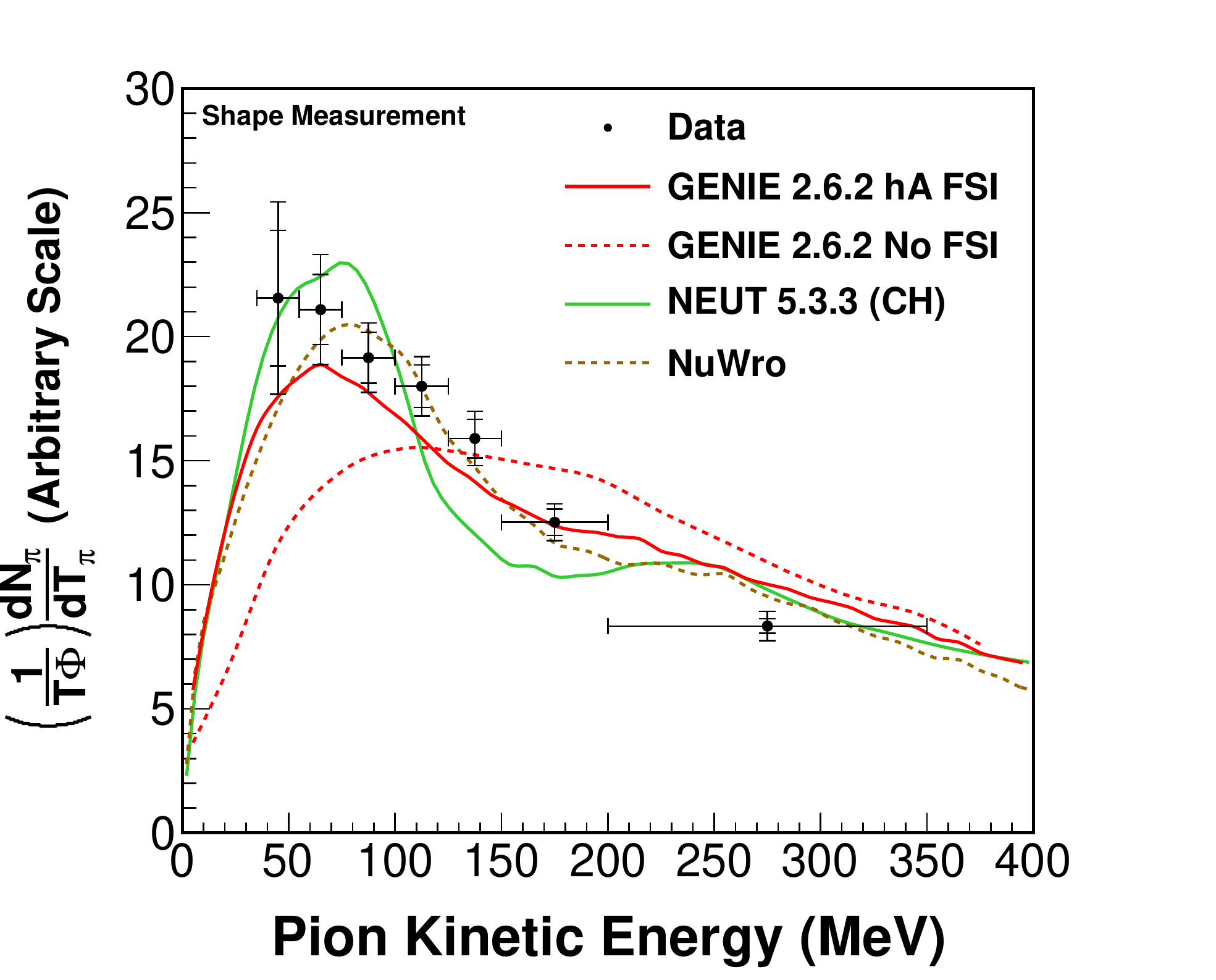}
}
%    \vspace{-7pt}
\caption{\ccpi $(1/T\Phi)(dN_{\pi} / dT_\pi)$ (top) and its shape (bottom) compared to the GENIE, NEUT, and NuWro models.
The inner (outer) error bars correspond to the statistical (total) uncertainties.}
%    \vspace{-10pt}
\label{fig:pike-Npi-POT}
\end{figure}

\subsection{\ccpi\ results}
Extension of the maximum $W$ from 1.4 to 1.8 GeV in the \ccpi
analysis includes additional nonresonant processes and $N^*$ resonances at high mass. For each event, 
more than one charged pion can be counted 
(see Section~\ref{sec:xs}), which causes these distributions to be sensitive to pion-producing FSI processes 
in higher-multiplicity events.  GENIE predicts that 19\% of the \ccpi charged pions come from two-pion events, and 
5\% come from events with three or more charged pions.

Figures~\ref{fig:piangle-Npi-POT} and~\ref{fig:pike-Npi-POT} show 
the results
of the \ccpi analysis as well as the GENIE (no FSI and with FSI),
NEUT, and NuWro predictions.  The data distributions are very
similar in shape to Figures~\ref{fig:piangle-1pi} and~\ref{fig:pike-1pi}
even though the total cross section is roughly 50\% larger, and the relative 
normalization of the GENIE prediction to the data does not change.  The NuWro 
prediction for the total \ccpi cross section improves slightly relative to its 
prediction for the \cconepi channel, while NEUT predicts a much larger increase 
in final state pions, particularly at \Tpi $<$ 100 MeV,  for the higher $W$ processes.  

\section{Summary}
\label{sec:summary}
This paper presents measurements of neutrino-induced pion production from 
a CH target and compares them to models with different FSI treatments and 
to the \miniboone measurement of the same process.  These data provide 
new information about the neutrino energy dependence of resonant pion production 
and can be used to place strong constraints on FSI. More generally, 
they provide an observational foundation for improving both the 
background and signal predictions needed for precise oscillation parameter 
measurements. 

Both the $d\sigma/d\theta_\pi$ and $d\sigma/dT_\pi$ distribution
shapes strongly favor models with FSI implemented as a full or effective cascade algorithm. 
For the \cconepi 
analysis, the calculations with FSI, NEUT and NuWro are in good 
agreement with the data while GENIE predicts cross sections
that are too large.  The distribution shape contains the most information
about FSI.  At $T_\pi$ greater than 100~MeV, 
where pion FSI effects largely deplete the yield, 
\minerva and \miniboone have similar shape.  However, the similarity
in magnitude is not expected.  There are also 
significant normalization and shape discrepancies between the two 
measurements below 100~MeV in comparison with the GENIE calculation.
A decomposition of the FSI channels in the GENIE calculation suggest 
that an increased inelastic contribution may improve agreement with the data.

For the \ccpi analysis, data results are similar to the \cconepi analysis.
However, differences among the models employed in the generators
produce significant changes with respect to the data.
The shape is strongly affected by FSI and the magnitude
disagreement can come from problems within the Monte Carlo models.

It is clear that the underlying pion production models and perhaps
other parts of the neutrino interaction will have to be 
modified to reproduce both data sets, which will in turn help improve 
predictions for oscillation experiments.

\ifnum\sizecheck=1
  \newpage
  {\Large Content after here does not count against size of PRL}
  \newpage
\fi
\begin{acknowledgments}

This work was supported by the Fermi National Accelerator Laboratory
under U.S. Department of Energy Contract
No.\@ DE-AC02-07CH11359 which included the \minerva construction project.
Construction support also
was granted by the United States National Science Foundation under
Grant No. PHY-0619727 and by the University of Rochester. Support for
participating scientists was provided by NSF and DOE (USA) by CAPES
and CNPq (Brazil), by CoNaCyT (Mexico), by CONICYT (Chile), by
CONCYTEC, DGI-PUCP and IDI/IGI-UNI (Peru), by Latin American Center for
Physics (CLAF), by the Swiss National Science Foundation, and by RAS and the Russian Ministry of Education and Science (Russia).  We
thank the MINOS Collaboration for use of its
near detector data. Finally, we thank the staff of
Fermilab for support of the beam line and detector.

\end{acknowledgments}

\bibliographystyle{apsrev4-1}
\bibliography{ChargedPionNeutrino.bib}

\begin{thebibliography}{61}%
\makeatletter
\providecommand \@ifxundefined [1]{%
 \@ifx{#1\undefined}
}%
\providecommand \@ifnum [1]{%
 \ifnum #1\expandafter \@firstoftwo
 \else \expandafter \@secondoftwo
 \fi
}%
\providecommand \@ifx [1]{%
 \ifx #1\expandafter \@firstoftwo
 \else \expandafter \@secondoftwo
 \fi
}%
\providecommand \natexlab [1]{#1}%
\providecommand \enquote  [1]{``#1''}%
\providecommand \bibnamefont  [1]{#1}%
\providecommand \bibfnamefont [1]{#1}%
\providecommand \citenamefont [1]{#1}%
\providecommand \href@noop [0]{\@secondoftwo}%
\providecommand \href [0]{\begingroup \@sanitize@url \@href}%
\providecommand \@href[1]{\@@startlink{#1}\@@href}%
\providecommand \@@href[1]{\endgroup#1\@@endlink}%
\providecommand \@sanitize@url [0]{\catcode `\\12\catcode `\$12\catcode
  `\&12\catcode `\#12\catcode `\^12\catcode `\_12\catcode `\%12\relax}%
\providecommand \@@startlink[1]{}%
\providecommand \@@endlink[0]{}%
\providecommand \url  [0]{\begingroup\@sanitize@url \@url }%
\providecommand \@url [1]{\endgroup\@href {#1}{\urlprefix }}%
\providecommand \urlprefix  [0]{URL }%
\providecommand \Eprint [0]{\href }%
\providecommand \doibase [0]{http://dx.doi.org/}%
\providecommand \selectlanguage [0]{\@gobble}%
\providecommand \bibinfo  [0]{\@secondoftwo}%
\providecommand \bibfield  [0]{\@secondoftwo}%
\providecommand \translation [1]{[#1]}%
\providecommand \BibitemOpen [0]{}%
\providecommand \bibitemStop [0]{}%
\providecommand \bibitemNoStop [0]{.\EOS\space}%
\providecommand \EOS [0]{\spacefactor3000\relax}%
\providecommand \BibitemShut  [1]{\csname bibitem#1\endcsname}%
\let\auto@bib@innerbib\@empty
%</preamble>
\bibitem [{\citenamefont {Abe}\ \emph {et~al.}(2011)\citenamefont {Abe} \emph
  {et~al.}}]{Abe:2011ks}%
  \BibitemOpen
  \bibfield  {author} {\bibinfo {author} {\bibfnamefont {K.}~\bibnamefont
  {Abe}} \emph {et~al.} (\bibinfo {collaboration} {T2K Collaboration}),\ }\href
  {\doibase http://dx.doi.org/10.1016/j.nima.2011.06.067} {\bibfield  {journal}
  {\bibinfo  {journal} {Nucl. Instrum. Methods Phys. Res., Sect. A}\ }\textbf
  {\bibinfo {volume} {659}},\ \bibinfo {pages} {106} (\bibinfo {year}
  {2011})}\BibitemShut {NoStop}%
%%CITATION = ARXIV:1106.1238;%%
\bibitem [{\citenamefont {Ayres}\ \emph {et~al.}(2004)\citenamefont {Ayres}
  \emph {et~al.}}]{Ayres:2004js}%
  \BibitemOpen
  \bibfield  {author} {\bibinfo {author} {\bibfnamefont {D.~S.}\ \bibnamefont
  {Ayres}} \emph {et~al.} (\bibinfo {collaboration} {NOvA Collaboration}),\
  }\href@noop {} {\  (\bibinfo {year} {2004})},\ \Eprint
  {http://arxiv.org/abs/hep-ex/0503053} {arXiv:hep-ex/0503053} \BibitemShut
  {NoStop}%
%%CITATION = HEP-EX/0503053;%%
\bibitem [{\citenamefont {Adams}\ \emph {et~al.}(2013)\citenamefont {Adams}
  \emph {et~al.}}]{LBNE}%
  \BibitemOpen
  \bibfield  {author} {\bibinfo {author} {\bibfnamefont {C.}~\bibnamefont
  {Adams}} \emph {et~al.} (\bibinfo {collaboration} {LBNE}),\ }\href@noop {} {\
   (\bibinfo {year} {2013})},\ \Eprint {http://arxiv.org/abs/1307.7335}
  {arXiv:1307.7335 [hep-ex]} \BibitemShut {NoStop}%
%%CITATION = ARXIV:1307.7335;%%
\bibitem [{\citenamefont {Fiorentini}\ \emph {et~al.}(2013)\citenamefont
  {Fiorentini} \emph {et~al.}}]{MINERVAnuprl}%
  \BibitemOpen
  \bibfield  {author} {\bibinfo {author} {\bibfnamefont {G.~A.}\ \bibnamefont
  {Fiorentini}} \emph {et~al.} (\bibinfo {collaboration} {MINERvA
  Collaboration}),\ }\href {\doibase 10.1103/PhysRevLett.111.022502} {\bibfield
   {journal} {\bibinfo  {journal} {Phys. Rev. Lett.}\ }\textbf {\bibinfo
  {volume} {111}},\ \bibinfo {pages} {022502} (\bibinfo {year}
  {2013})}\BibitemShut {NoStop}%
%%CITATION = ARXIV:1305.2243;%%
\bibitem [{\citenamefont {Fields}\ \emph {et~al.}(2013)\citenamefont {Fields}
  \emph {et~al.}}]{MINERVAnubarprl}%
  \BibitemOpen
  \bibfield  {author} {\bibinfo {author} {\bibfnamefont {L.}~\bibnamefont
  {Fields}} \emph {et~al.} (\bibinfo {collaboration} {MINERvA Collaboration}),\
  }\href {\doibase 10.1103/PhysRevLett.111.022501} {\bibfield  {journal}
  {\bibinfo  {journal} {Phys. Rev. Lett.}\ }\textbf {\bibinfo {volume} {111}},\
  \bibinfo {pages} {022501} (\bibinfo {year} {2013})}\BibitemShut {NoStop}%
%%CITATION = ARXIV:1305.2234;%%
\bibitem [{\citenamefont {Aguilar-Arevalo}\ \emph {et~al.}(2010)\citenamefont
  {Aguilar-Arevalo} \emph {et~al.}}]{mboone-ccqe}%
  \BibitemOpen
  \bibfield  {author} {\bibinfo {author} {\bibfnamefont {A.~A.}\ \bibnamefont
  {Aguilar-Arevalo}} \emph {et~al.},\ }\href {\doibase
  10.1103/PhysRevD.81.092005} {\bibfield  {journal} {\bibinfo  {journal} {Phys.
  Rev. D}\ }\textbf {\bibinfo {volume} {81}},\ \bibinfo {pages} {092005}
  (\bibinfo {year} {2010})}\BibitemShut {NoStop}%
\bibitem [{\citenamefont {Aguilar-Arevalo}\ \emph
  {et~al.}(2013{\natexlab{a}})\citenamefont {Aguilar-Arevalo} \emph
  {et~al.}}]{mboone-ccqe-nubar}%
  \BibitemOpen
  \bibfield  {author} {\bibinfo {author} {\bibfnamefont {A.~A.}\ \bibnamefont
  {Aguilar-Arevalo}} \emph {et~al.} (\bibinfo {collaboration} {MiniBooNE
  Collaboration}),\ }\href {\doibase 10.1103/PhysRevD.88.032001} {\bibfield
  {journal} {\bibinfo  {journal} {Phys. Rev. D}\ }\textbf {\bibinfo {volume}
  {88}},\ \bibinfo {pages} {032001} (\bibinfo {year}
  {2013}{\natexlab{a}})}\BibitemShut {NoStop}%
%%CITATION = ARXIV:1301.7067;%%
\bibitem [{\citenamefont {Abe}\ \emph {et~al.}(2014{\natexlab{a}})\citenamefont
  {Abe} \emph {et~al.}}]{t2k-theta13}%
  \BibitemOpen
  \bibfield  {author} {\bibinfo {author} {\bibfnamefont {K.}~\bibnamefont
  {Abe}} \emph {et~al.} (\bibinfo {collaboration} {T2K Collaboration}),\ }\href
  {\doibase 10.1103/PhysRevLett.112.061802} {\bibfield  {journal} {\bibinfo
  {journal} {Phys. Rev. Lett.}\ }\textbf {\bibinfo {volume} {112}},\ \bibinfo
  {pages} {061802} (\bibinfo {year} {2014}{\natexlab{a}})}\BibitemShut
  {NoStop}%
%%CITATION = ARXIV:1311.4750;%%
\bibitem [{\citenamefont {Aguilar-Arevalo}\ \emph
  {et~al.}(2013{\natexlab{b}})\citenamefont {Aguilar-Arevalo} \emph
  {et~al.}}]{AguilarArevalo:2010wv}%
  \BibitemOpen
  \bibfield  {author} {\bibinfo {author} {\bibfnamefont {A.~A.}\ \bibnamefont
  {Aguilar-Arevalo}} \emph {et~al.},\ }\href {\doibase
  10.1103/PhysRevLett.110.161801} {\bibfield  {journal} {\bibinfo  {journal}
  {Phys. Rev. Lett.}\ }\textbf {\bibinfo {volume} {110}},\ \bibinfo {pages}
  {161801} (\bibinfo {year} {2013}{\natexlab{b}})}\BibitemShut {NoStop}%
\bibitem [{\citenamefont {Lalakulich}\ \emph {et~al.}(2012)\citenamefont
  {Lalakulich}, \citenamefont {Mosel},\ and\ \citenamefont
  {Gallmeister}}]{Lalakulich:2012hs}%
  \BibitemOpen
  \bibfield  {author} {\bibinfo {author} {\bibfnamefont {O.}~\bibnamefont
  {Lalakulich}}, \bibinfo {author} {\bibfnamefont {U.}~\bibnamefont {Mosel}}, \
  and\ \bibinfo {author} {\bibfnamefont {K.}~\bibnamefont {Gallmeister}},\
  }\href {\doibase 10.1103/PhysRevC.86.054606} {\bibfield  {journal} {\bibinfo
  {journal} {Phys. Rev. C}\ }\textbf {\bibinfo {volume} {86}},\ \bibinfo
  {pages} {054606} (\bibinfo {year} {2012})}\BibitemShut {NoStop}%
%%CITATION = ARXIV:1208.3678;%%
\bibitem [{\citenamefont {Qian}\ \emph {et~al.}(2010)\citenamefont {Qian} \emph
  {et~al.}}]{Qian:2009aa}%
  \BibitemOpen
  \bibfield  {author} {\bibinfo {author} {\bibfnamefont {X.}~\bibnamefont
  {Qian}} \emph {et~al.},\ }\href {\doibase 10.1103/PhysRevC.81.055209}
  {\bibfield  {journal} {\bibinfo  {journal} {Phys. Rev. C}\ }\textbf {\bibinfo
  {volume} {81}},\ \bibinfo {pages} {055209} (\bibinfo {year}
  {2010})}\BibitemShut {NoStop}%
%%CITATION = ARXIV:0908.1616;%%
\bibitem [{\citenamefont {Bell}\ \emph {et~al.}(1978)\citenamefont {Bell} \emph
  {et~al.}}]{Bell:1978fnal}%
  \BibitemOpen
  \bibfield  {author} {\bibinfo {author} {\bibfnamefont {J.}~\bibnamefont
  {Bell}} \emph {et~al.},\ }\href {\doibase 10.1103/PhysRevLett.41.1008}
  {\bibfield  {journal} {\bibinfo  {journal} {Phys. Rev. Lett.}\ }\textbf
  {\bibinfo {volume} {41}},\ \bibinfo {pages} {1008} (\bibinfo {year}
  {1978})}\BibitemShut {NoStop}%
\bibitem [{\citenamefont {Radecky}\ \emph {et~al.}(1982)\citenamefont {Radecky}
  \emph {et~al.}}]{Radecky:1981fn}%
  \BibitemOpen
  \bibfield  {author} {\bibinfo {author} {\bibfnamefont {G.~M.}\ \bibnamefont
  {Radecky}} \emph {et~al.},\ }\href {\doibase 10.1103/PhysRevD.25.1161}
  {\bibfield  {journal} {\bibinfo  {journal} {Phys. Rev. D}\ }\textbf {\bibinfo
  {volume} {25}},\ \bibinfo {pages} {1161} (\bibinfo {year}
  {1982})}\BibitemShut {NoStop}%
%%CITATION = PHRVA,D25,1161;%%
\bibitem [{\citenamefont {Kitagaki}\ \emph {et~al.}(1986)\citenamefont
  {Kitagaki} \emph {et~al.}}]{Kitagaki:1986ct}%
  \BibitemOpen
  \bibfield  {author} {\bibinfo {author} {\bibfnamefont {T.}~\bibnamefont
  {Kitagaki}} \emph {et~al.},\ }\href {\doibase 10.1103/PhysRevD.34.2554}
  {\bibfield  {journal} {\bibinfo  {journal} {Phys. Rev. D}\ }\textbf {\bibinfo
  {volume} {34}},\ \bibinfo {pages} {2554} (\bibinfo {year}
  {1986})}\BibitemShut {NoStop}%
%%CITATION = PHRVA,D34,2554;%%
\bibitem [{\citenamefont {Allen}\ \emph {et~al.}(1986)\citenamefont {Allen}
  \emph {et~al.}}]{Allen:1986}%
  \BibitemOpen
  \bibfield  {author} {\bibinfo {author} {\bibfnamefont {P.}~\bibnamefont
  {Allen}} \emph {et~al.} (\bibinfo {collaboration}
  {Aachen-Birmingham-Bonn-CERN-London-Munich-Oxford Collaboration}),\ }\href
  {\doibase 10.1016/0550-3213(86)90480-3} {\bibfield  {journal} {\bibinfo
  {journal} {Nucl. Phys. B}\ }\textbf {\bibinfo {volume} {264}},\ \bibinfo
  {pages} {221 } (\bibinfo {year} {1986})}\BibitemShut {NoStop}%
\bibitem [{\citenamefont {Allasia}\ \emph {et~al.}(1990)\citenamefont {Allasia}
  \emph {et~al.}}]{WA25:1990}%
  \BibitemOpen
  \bibfield  {author} {\bibinfo {author} {\bibfnamefont {D.}~\bibnamefont
  {Allasia}} \emph {et~al.} (\bibinfo {collaboration}
  {Amsterdam-Bergen-Bologna-padova-Pisa-Saclay-Torino Collaboration}),\ }\href
  {\doibase 10.1016/0550-3213(90)90472-P} {\bibfield  {journal} {\bibinfo
  {journal} {Nucl. Phys. B}\ }\textbf {\bibinfo {volume} {343}},\ \bibinfo
  {pages} {285 } (\bibinfo {year} {1990})}\BibitemShut {NoStop}%
\bibitem [{\citenamefont {Block}\ \emph {et~al.}(1964)\citenamefont {Block}
  \emph {et~al.}}]{Block:1964}%
  \BibitemOpen
  \bibfield  {author} {\bibinfo {author} {\bibfnamefont {M.~M.}\ \bibnamefont
  {Block}} \emph {et~al.},\ }\href {\doibase 10.1016/0031-9163(64)91104-7}
  {\bibfield  {journal} {\bibinfo  {journal} {Phys. Lett.}\ }\textbf {\bibinfo
  {volume} {12}},\ \bibinfo {pages} {281 } (\bibinfo {year}
  {1964})}\BibitemShut {NoStop}%
\bibitem [{\citenamefont {Budagov}\ \emph {et~al.}(1969)\citenamefont {Budagov}
  \emph {et~al.}}]{Budagov:1969}%
  \BibitemOpen
  \bibfield  {author} {\bibinfo {author} {\bibfnamefont {I.}~\bibnamefont
  {Budagov}} \emph {et~al.},\ }\href {\doibase 10.1016/0370-2693(69)90041-0}
  {\bibfield  {journal} {\bibinfo  {journal} {Phys. Lett. B}\ }\textbf
  {\bibinfo {volume} {29}},\ \bibinfo {pages} {524 } (\bibinfo {year}
  {1969})}\BibitemShut {NoStop}%
\bibitem [{\citenamefont {Grabosch}\ \emph {et~al.}(1989)\citenamefont
  {Grabosch} \emph {et~al.}}]{SKAT:1989}%
  \BibitemOpen
  \bibfield  {author} {\bibinfo {author} {\bibfnamefont {H.~J.}\ \bibnamefont
  {Grabosch}} \emph {et~al.} (\bibinfo {collaboration} {SKAT Collaboration}),\
  }\href@noop {} {\bibfield  {journal} {\bibinfo  {journal} {Z. Phys. C}\
  }\textbf {\bibinfo {volume} {41}},\ \bibinfo {pages} {527 } (\bibinfo {year}
  {1989})}\BibitemShut {NoStop}%
\bibitem [{\citenamefont {Wilkinson}\ \emph {et~al.}(2014)\citenamefont
  {Wilkinson} \emph {et~al.}}]{newphil}%
  \BibitemOpen
  \bibfield  {author} {\bibinfo {author} {\bibfnamefont {C.}~\bibnamefont
  {Wilkinson}} \emph {et~al.},\ }\href {\doibase 10.1103/PhysRevD.90.112017}
  {\bibfield  {journal} {\bibinfo  {journal} {Phys.Rev.}\ }\textbf {\bibinfo
  {volume} {D90}},\ \bibinfo {pages} {112017} (\bibinfo {year}
  {2014})}\BibitemShut {NoStop}%
\bibitem [{\citenamefont {Rodriguez}\ \emph {et~al.}(2008)\citenamefont
  {Rodriguez} \emph {et~al.}}]{K2K_piprod}%
  \BibitemOpen
  \bibfield  {author} {\bibinfo {author} {\bibfnamefont {A.}~\bibnamefont
  {Rodriguez}} \emph {et~al.} (\bibinfo {collaboration} {K2K collaboration}),\
  }\href {\doibase 10.1103/PhysRevD.78.032003} {\bibfield  {journal} {\bibinfo
  {journal} {Phys. Rev. D}\ }\textbf {\bibinfo {volume} {78}},\ \bibinfo
  {pages} {032003} (\bibinfo {year} {2008})}\BibitemShut {NoStop}%
\bibitem [{\citenamefont {Aguilar-Arevalo}\ \emph
  {et~al.}(2009{\natexlab{a}})\citenamefont {Aguilar-Arevalo} \emph
  {et~al.}}]{miniboone_piratio}%
  \BibitemOpen
  \bibfield  {author} {\bibinfo {author} {\bibfnamefont {A.~A.}\ \bibnamefont
  {Aguilar-Arevalo}} \emph {et~al.},\ }\href {\doibase
  10.1103/PhysRevLett.103.081801} {\bibfield  {journal} {\bibinfo  {journal}
  {Phys. Rev. Lett.}\ }\textbf {\bibinfo {volume} {103}},\ \bibinfo {pages}
  {081801} (\bibinfo {year} {2009}{\natexlab{a}})}\BibitemShut {NoStop}%
\bibitem [{\citenamefont {Aguilar-Arevalo}\ \emph {et~al.}(2011)\citenamefont
  {Aguilar-Arevalo} \emph {et~al.}}]{miniboone_piprod}%
  \BibitemOpen
  \bibfield  {author} {\bibinfo {author} {\bibfnamefont {A.~A.}\ \bibnamefont
  {Aguilar-Arevalo}} \emph {et~al.} (\bibinfo {collaboration} {MiniBooNE
  Collaboration}),\ }\href {\doibase 10.1103/PhysRevD.83.052007} {\bibfield
  {journal} {\bibinfo  {journal} {Phys. Rev. D}\ }\textbf {\bibinfo {volume}
  {83}},\ \bibinfo {pages} {052007} (\bibinfo {year} {2011})}\BibitemShut
  {NoStop}%
\bibitem [{\citenamefont {{O.~Lalakulich and U.~Mosel}}(2013)}]{gibuu-pi}%
  \BibitemOpen
  \bibfield  {author} {\bibinfo {author} {\bibnamefont {{O.~Lalakulich and
  U.~Mosel}}},\ }\href {\doibase 10.1103/PhysRevC.87.014602} {\bibfield
  {journal} {\bibinfo  {journal} {Phys. Rev. C}\ }\textbf {\bibinfo {volume}
  {87}},\ \bibinfo {pages} {014602} (\bibinfo {year} {2013})}\BibitemShut
  {NoStop}%
\bibitem [{\citenamefont {{E.~Hern\'{a}ndez, J.~Nieves, and M.J.~Vicente
  Vacas}}(2013)}]{valencia-pi}%
  \BibitemOpen
  \bibfield  {author} {\bibinfo {author} {\bibnamefont {{E.~Hern\'{a}ndez,
  J.~Nieves, and M.J.~Vicente Vacas}}},\ }\href {\doibase
  10.1103/PhysRevD.87.113009} {\bibfield  {journal} {\bibinfo  {journal} {Phys.
  Rev. D}\ }\textbf {\bibinfo {volume} {87}},\ \bibinfo {pages} {113009}
  (\bibinfo {year} {2013})}\BibitemShut {NoStop}%
\bibitem [{\citenamefont {Rodrigues}(2015)}]{Rodrigues:2014jfa}%
  \BibitemOpen
  \bibfield  {author} {\bibinfo {author} {\bibfnamefont {P.~A.}\ \bibnamefont
  {Rodrigues}},\ }\href {\doibase 10.1063/1.4919470} {\bibfield  {journal}
  {\bibinfo  {journal} {AIP Conf. Proc.}\ }\textbf {\bibinfo {volume} {1663}},\
  \bibinfo {pages} {030006} (\bibinfo {year} {2015})}\BibitemShut {NoStop}%
\bibitem [{\citenamefont {Abe}\ \emph {et~al.}(2014{\natexlab{b}})\citenamefont
  {Abe} \emph {et~al.}}]{t2k_osc_long}%
  \BibitemOpen
  \bibfield  {author} {\bibinfo {author} {\bibfnamefont {K.}~\bibnamefont
  {Abe}} \emph {et~al.} (\bibinfo {collaboration} {T2K Collaboration}),\ }\href
  {\doibase 10.1103/PhysRevD.89.092003} {\bibfield  {journal} {\bibinfo
  {journal} {Phys. Rev. D}\ }\textbf {\bibinfo {volume} {89}},\ \bibinfo
  {pages} {092003} (\bibinfo {year} {2014}{\natexlab{b}})}\BibitemShut
  {NoStop}%
\bibitem [{\citenamefont {Anderson}\ \emph {et~al.}(1998)\citenamefont
  {Anderson}, \citenamefont {Bernstein}, \citenamefont {Boehnlein},
  \citenamefont {Bourkland}, \citenamefont {Childress} \emph
  {et~al.}}]{Anderson:1998zza}%
  \BibitemOpen
  \bibfield  {author} {\bibinfo {author} {\bibfnamefont {K.}~\bibnamefont
  {Anderson}}, \bibinfo {author} {\bibfnamefont {B.}~\bibnamefont {Bernstein}},
  \bibinfo {author} {\bibfnamefont {D.}~\bibnamefont {Boehnlein}}, \bibinfo
  {author} {\bibfnamefont {K.~R.}\ \bibnamefont {Bourkland}}, \bibinfo {author}
  {\bibfnamefont {S.}~\bibnamefont {Childress}},  \emph {et~al.},\ }\href@noop
  {} {\emph {\bibinfo {title} {{The NuMI Facility Technical Design Report}}}},\
  \bibinfo {type} {FERMILAB-DESIGN-1998-01}\ (\bibinfo {year}
  {1998})\BibitemShut {NoStop}%
%%CITATION = FERMILAB-DESIGN-1998-01 ETC.;%%
\bibitem [{\citenamefont {Michael}\ \emph {et~al.}(2008)\citenamefont {Michael}
  \emph {et~al.}}]{Michael:2008bc}%
  \BibitemOpen
  \bibfield  {author} {\bibinfo {author} {\bibfnamefont {D.~G.}\ \bibnamefont
  {Michael}} \emph {et~al.} (\bibinfo {collaboration} {MINOS Collaboration}),\
  }\href {\doibase 10.1016/j.nima.2008.08.003} {\bibfield  {journal} {\bibinfo
  {journal} {Nucl. Instrum. Methods Phys. Res., Sect. A}\ }\textbf {\bibinfo
  {volume} {596}},\ \bibinfo {pages} {190} (\bibinfo {year}
  {2008})}\BibitemShut {NoStop}%
%%CITATION = ARXIV:0805.3170;%%
\bibitem [{\citenamefont {Aliaga}\ \emph {et~al.}(2014)\citenamefont {Aliaga}
  \emph {et~al.}}]{minerva_nim}%
  \BibitemOpen
  \bibfield  {author} {\bibinfo {author} {\bibfnamefont {L.}~\bibnamefont
  {Aliaga}} \emph {et~al.} (\bibinfo {collaboration} {MINERvA Collaboration}),\
  }\href {\doibase 10.1016/j.nima.2013.12.053} {\bibfield  {journal} {\bibinfo
  {journal} {Nucl. Instrum. Methods in Phys. Res., Sect. A}\ }\textbf {\bibinfo
  {volume} {743}},\ \bibinfo {pages} {130} (\bibinfo {year}
  {2014})}\BibitemShut {NoStop}%
%%CITATION = ARXIV:1305.5199;%%
\bibitem [{\citenamefont {Agostinelli}\ \emph {et~al.}(2003)\citenamefont
  {Agostinelli} \emph {et~al.}}]{Agostinelli2003250}%
  \BibitemOpen
  \bibfield  {author} {\bibinfo {author} {\bibfnamefont {S.}~\bibnamefont
  {Agostinelli}} \emph {et~al.},\ }\href {\doibase
  10.1016/S0168-9002(03)01368-8} {\bibfield  {journal} {\bibinfo  {journal}
  {Nucl. Instrum. Methods Phys. Res., Sect. A}\ }\textbf {\bibinfo {volume}
  {506}},\ \bibinfo {pages} {250 } (\bibinfo {year} {2003})}\BibitemShut
  {NoStop}%
\bibitem [{\citenamefont {Allison}\ \emph {et~al.}(2006)\citenamefont {Allison}
  \emph {et~al.}}]{1610988}%
  \BibitemOpen
  \bibfield  {author} {\bibinfo {author} {\bibfnamefont {J.}~\bibnamefont
  {Allison}} \emph {et~al.},\ }\href {\doibase 10.1109/TNS.2006.869826}
  {\bibfield  {journal} {\bibinfo  {journal} {Nuclear Science, IEEE
  Transactions}\ }\textbf {\bibinfo {volume} {53}},\ \bibinfo {pages} {270}
  (\bibinfo {year} {2006})}\BibitemShut {NoStop}%
\bibitem [{\citenamefont {Alt}\ \emph {et~al.}(2007)\citenamefont {Alt} \emph
  {et~al.}}]{Alt:2006fr}%
  \BibitemOpen
  \bibfield  {author} {\bibinfo {author} {\bibfnamefont {C.}~\bibnamefont
  {Alt}} \emph {et~al.} (\bibinfo {collaboration} {NA49 Collaboration}),\
  }\href {\doibase 10.1140/epjc/s10052-006-0165-7} {\bibfield  {journal}
  {\bibinfo  {journal} {Eur. Phys. J. C}\ }\textbf {\bibinfo {volume} {49}},\
  \bibinfo {pages} {897} (\bibinfo {year} {2007})}\BibitemShut {NoStop}%
%%CITATION = HEP-EX/0606028;%%
\bibitem [{\citenamefont {Lebedev}(2007)}]{Lebedev:2007zz}%
  \BibitemOpen
  \bibfield  {author} {\bibinfo {author} {\bibfnamefont {A.~V.}\ \bibnamefont
  {Lebedev}},\ }\href@noop {} {\bibinfo {type} {Fermilab-thesis-2007-76}},\
  \bibinfo  {school} {Harvard University} (\bibinfo {year} {2007})\BibitemShut
  {NoStop}%
%%CITATION = FERMILAB-THESIS-2007-76 ETC.;%%
\bibitem [{\citenamefont {Pavlovic}(2008)}]{Pavlovic:2008zz}%
  \BibitemOpen
  \bibfield  {author} {\bibinfo {author} {\bibfnamefont {Z.}~\bibnamefont
  {Pavlovic}},\ }\href@noop {} {\bibinfo {type} {Fermilab-thesis-2008-59}},\
  \bibinfo  {school} {University of Texas} (\bibinfo {year} {2008})\BibitemShut
  {NoStop}%
%%CITATION = FERMILAB-THESIS-2008-59 ETC.;%%
\bibitem [{\citenamefont {Ashery}\ \emph {et~al.}(1981)\citenamefont {Ashery}
  \emph {et~al.}}]{Ashery:1981tq}%
  \BibitemOpen
  \bibfield  {author} {\bibinfo {author} {\bibfnamefont {D.}~\bibnamefont
  {Ashery}} \emph {et~al.},\ }\href {\doibase 10.1103/PhysRevC.23.2173}
  {\bibfield  {journal} {\bibinfo  {journal} {Phys. Rev. C}\ }\textbf {\bibinfo
  {volume} {23}},\ \bibinfo {pages} {2173} (\bibinfo {year}
  {1981})}\BibitemShut {NoStop}%
%%CITATION = PHRVA,C23,2173;%%
\bibitem [{\citenamefont {Allardyce}\ \emph {et~al.}(1973)\citenamefont
  {Allardyce} \emph {et~al.}}]{Allardyce:1973ce}%
  \BibitemOpen
  \bibfield  {author} {\bibinfo {author} {\bibfnamefont {B.~W.}\ \bibnamefont
  {Allardyce}} \emph {et~al.},\ }\href {\doibase 10.1016/0375-9474(73)90049-3}
  {\bibfield  {journal} {\bibinfo  {journal} {Nucl. Phys.}\ }\textbf {\bibinfo
  {volume} {A209}},\ \bibinfo {pages} {1} (\bibinfo {year} {1973})}\BibitemShut
  {NoStop}%
%%CITATION = NUPHA,A209,1;%%
\bibitem [{\citenamefont {Saunders}\ \emph {et~al.}(1996)\citenamefont
  {Saunders} \emph {et~al.}}]{Saunders:1996ic}%
  \BibitemOpen
  \bibfield  {author} {\bibinfo {author} {\bibfnamefont {A.}~\bibnamefont
  {Saunders}} \emph {et~al.},\ }\href {\doibase 10.1103/PhysRevC.53.1745}
  {\bibfield  {journal} {\bibinfo  {journal} {Phys. Rev. C}\ }\textbf {\bibinfo
  {volume} {53}},\ \bibinfo {pages} {1745} (\bibinfo {year}
  {1996})}\BibitemShut {NoStop}%
%%CITATION = PHRVA,C53,1745;%%
\bibitem [{\citenamefont {{T.S.H.~Lee and R.P.~Redwine}}(2002)}]{Lee:2002eq}%
  \BibitemOpen
  \bibfield  {author} {\bibinfo {author} {\bibnamefont {{T.S.H.~Lee and
  R.P.~Redwine}}},\ }\href {\doibase 10.1146/annurev.nucl.52.050102.090713}
  {\bibfield  {journal} {\bibinfo  {journal} {Ann. Rev. Nucl. Part. Sci.}\
  }\textbf {\bibinfo {volume} {52}},\ \bibinfo {pages} {23} (\bibinfo {year}
  {2002})}\BibitemShut {NoStop}%
%%CITATION = ARNUA,52,23;%%
\bibitem [{\citenamefont {Aliaga}\ \emph {et~al.}(2015)\citenamefont {Aliaga}
  \emph {et~al.}}]{mnv_testbeam}%
  \BibitemOpen
  \bibfield  {author} {\bibinfo {author} {\bibfnamefont {L.}~\bibnamefont
  {Aliaga}} \emph {et~al.} (\bibinfo {collaboration} {MINERvA}),\ }\href
  {\doibase 10.1016/j.nima.2015.04.003} {\bibfield  {journal} {\bibinfo
  {journal} {Nucl.Instrum.Meth.}\ }\textbf {\bibinfo {volume} {A789}},\
  \bibinfo {pages} {28} (\bibinfo {year} {2015})}\BibitemShut {NoStop}%
%%CITATION = ARXIV:1501.06431;%%
\bibitem [{\citenamefont {Andreopoulos}\ \emph {et~al.}(2010)\citenamefont
  {Andreopoulos} \emph {et~al.}}]{Andreopoulos201087}%
  \BibitemOpen
  \bibfield  {author} {\bibinfo {author} {\bibfnamefont {C.}~\bibnamefont
  {Andreopoulos}} \emph {et~al.},\ }\href {\doibase 10.1016/j.nima.2009.12.009}
  {\bibfield  {journal} {\bibinfo  {journal} {Nucl. Instrum. Methods Phys.
  Res., Sect. A}\ }\textbf {\bibinfo {volume} {A614}},\ \bibinfo {pages} {87 }
  (\bibinfo {year} {2010})},\ \bibinfo {note} {{P}rogram release 2.6.2 used
  here}\BibitemShut {NoStop}%
\bibitem [{\citenamefont {Rein}\ and\ \citenamefont
  {Sehgal}(1981)}]{Rein:1980wg}%
  \BibitemOpen
  \bibfield  {author} {\bibinfo {author} {\bibfnamefont {D.}~\bibnamefont
  {Rein}}\ and\ \bibinfo {author} {\bibfnamefont {L.~M.}\ \bibnamefont
  {Sehgal}},\ }\href {\doibase 10.1016/0003-4916(81)90242-6} {\bibfield
  {journal} {\bibinfo  {journal} {Ann. Phys.}\ }\textbf {\bibinfo {volume}
  {133}},\ \bibinfo {pages} {79} (\bibinfo {year} {1981})}\BibitemShut
  {NoStop}%
%%CITATION = APNYA,133,79;%%
\bibitem [{\citenamefont {Beringer}\ \emph {et~al.}(2012)\citenamefont
  {Beringer} \emph {et~al.}}]{pdg}%
  \BibitemOpen
  \bibfield  {author} {\bibinfo {author} {\bibfnamefont {J.}~\bibnamefont
  {Beringer}} \emph {et~al.} (\bibinfo {collaboration} {Particle Data Group}),\
  }\href {\doibase 10.1103/PhysRevD.86.010001} {\bibfield  {journal} {\bibinfo
  {journal} {Phys. Rev. D}\ }\textbf {\bibinfo {volume} {86}},\ \bibinfo
  {pages} {010001} (\bibinfo {year} {2012})}\BibitemShut {NoStop}%
%%CITATION = PHRVA,D86,010001;%%
\bibitem [{\citenamefont {Bodek}\ \emph {et~al.}(2005)\citenamefont {Bodek},
  \citenamefont {Park},\ and\ \citenamefont {Yang}}]{Bodek:2004pc}%
  \BibitemOpen
  \bibfield  {author} {\bibinfo {author} {\bibfnamefont {A.}~\bibnamefont
  {Bodek}}, \bibinfo {author} {\bibfnamefont {I.}~\bibnamefont {Park}}, \ and\
  \bibinfo {author} {\bibfnamefont {U.~K.}\ \bibnamefont {Yang}},\ }\href
  {\doibase 10.1016/j.nuclphysbps.2004.11.208} {\bibfield  {journal} {\bibinfo
  {journal} {Nucl. Phys. Proc. Suppl.}\ }\textbf {\bibinfo {volume} {139}},\
  \bibinfo {pages} {113} (\bibinfo {year} {2005})}\BibitemShut {NoStop}%
%%CITATION = HEP-PH/0411202;%%
\bibitem [{\citenamefont {{D.~Rein and L.M.~Sehgal}}(2007)}]{Rein:2007}%
  \BibitemOpen
  \bibfield  {author} {\bibinfo {author} {\bibnamefont {{D.~Rein and
  L.M.~Sehgal}}},\ }\href@noop {} {\bibfield  {journal} {\bibinfo  {journal}
  {Phys. Lett. B}\ }\textbf {\bibinfo {volume} {657}},\ \bibinfo {pages} {207 }
  (\bibinfo {year} {2007})}\BibitemShut {NoStop}%
\bibitem [{\citenamefont {Dytman}\ and\ \citenamefont
  {Meyer}(2011)}]{Dytman:2011zz}%
  \BibitemOpen
  \bibfield  {author} {\bibinfo {author} {\bibfnamefont {S.~A.}\ \bibnamefont
  {Dytman}}\ and\ \bibinfo {author} {\bibfnamefont {A.~S.}\ \bibnamefont
  {Meyer}},\ }\href {\doibase 10.1063/1.3661588} {\bibfield  {journal}
  {\bibinfo  {journal} {AIP Conf.Proc.}\ }\textbf {\bibinfo {volume} {1405}},\
  \bibinfo {pages} {213} (\bibinfo {year} {2011})}\BibitemShut {NoStop}%
%%CITATION = APCPC,1405,213;%%
\bibitem [{\citenamefont {Eberly}(2014)}]{pittir20853}%
  \BibitemOpen
  \bibfield  {author} {\bibinfo {author} {\bibfnamefont {B.}~\bibnamefont
  {Eberly}},\ }\href {http://d-scholarship.pitt.edu/20853/} {\bibinfo {type}
  {Fermilab-thesis-2014-18}},\ \bibinfo  {school} {Univ. of Pittsburgh}
  (\bibinfo {year} {2014})\BibitemShut {NoStop}%
\bibitem [{\citenamefont {D'Agostini}(1995)}]{D'Agostini:1994zf}%
  \BibitemOpen
  \bibfield  {author} {\bibinfo {author} {\bibfnamefont {G.}~\bibnamefont
  {D'Agostini}},\ }\href {\doibase 10.1016/0168-9002(95)00274-X} {\bibfield
  {journal} {\bibinfo  {journal} {Nucl. Instrum. Methods Phys. Res., Sect. A}\
  }\textbf {\bibinfo {volume} {362}},\ \bibinfo {pages} {487} (\bibinfo {year}
  {1995})}\BibitemShut {NoStop}%
%%CITATION = NUIMA,A362,487;%%
\bibitem [{\citenamefont {{M.S.~Athar, S.~Chauhan, and
  S.K.~Singh}}(2010)}]{Athar}%
  \BibitemOpen
  \bibfield  {author} {\bibinfo {author} {\bibnamefont {{M.S.~Athar,
  S.~Chauhan, and S.K.~Singh}}},\ }\href {\doibase 10.1140/epja/i2010-10908-0}
  {\bibfield  {journal} {\bibinfo  {journal} {Eur. Phys. J. A}\ }\textbf
  {\bibinfo {volume} {43}},\ \bibinfo {pages} {209} (\bibinfo {year}
  {2010})}\BibitemShut {NoStop}%
\bibitem [{\citenamefont {Hayato}(2009)}]{Hayato:2009zz}%
  \BibitemOpen
  \bibfield  {author} {\bibinfo {author} {\bibfnamefont {Y.}~\bibnamefont
  {Hayato}},\ }\href@noop {} {\bibfield  {journal} {\bibinfo  {journal} {Acta
  Phys. Polon. B}\ }\textbf {\bibinfo {volume} {40}},\ \bibinfo {pages} {2477}
  (\bibinfo {year} {2009})}\BibitemShut {NoStop}%
%%CITATION = APPOA,B40,2477;%%
\bibitem [{\citenamefont {Golan}\ \emph {et~al.}(2012)\citenamefont {Golan},
  \citenamefont {Juszczak},\ and\ \citenamefont {Sobczyk}}]{Golan:2012wx}%
  \BibitemOpen
  \bibfield  {author} {\bibinfo {author} {\bibfnamefont {T.}~\bibnamefont
  {Golan}}, \bibinfo {author} {\bibfnamefont {C.}~\bibnamefont {Juszczak}}, \
  and\ \bibinfo {author} {\bibfnamefont {J.~T.}\ \bibnamefont {Sobczyk}},\
  }\href {\doibase 10.1103/PhysRevC.86.015505} {\bibfield  {journal} {\bibinfo
  {journal} {Phys. Rev. C}\ }\textbf {\bibinfo {volume} {86}},\ \bibinfo
  {pages} {015505} (\bibinfo {year} {2012})}\BibitemShut {NoStop}%
%%CITATION = ARXIV:1202.4197;%%
\bibitem [{\citenamefont {Mosel}\ \emph {et~al.}(2014)\citenamefont {Mosel},
  \citenamefont {Lalakulich},\ and\ \citenamefont
  {Gallmeister}}]{gibuu-minerva}%
  \BibitemOpen
  \bibfield  {author} {\bibinfo {author} {\bibfnamefont {U.}~\bibnamefont
  {Mosel}}, \bibinfo {author} {\bibfnamefont {O.}~\bibnamefont {Lalakulich}}, \
  and\ \bibinfo {author} {\bibfnamefont {K.}~\bibnamefont {Gallmeister}},\
  }\href {\doibase 10.1103/PhysRevD.89.093003} {\bibfield  {journal} {\bibinfo
  {journal} {Phys. Rev. D}\ }\textbf {\bibinfo {volume} {89}},\ \bibinfo
  {pages} {093003} (\bibinfo {year} {2014})}\BibitemShut {NoStop}%
%%CITATION = ARXIV:1402.0297;%%
\bibitem [{\citenamefont {Mosel}(2015)}]{gibuu_mnv}%
  \BibitemOpen
  \bibfield  {author} {\bibinfo {author} {\bibfnamefont {U.}~\bibnamefont
  {Mosel}},\ }\href@noop {} {\bibfield  {journal} {\bibinfo  {journal} {Phys.
  Rev.}\ }\textbf {\bibinfo {volume} {C91}},\ \bibinfo {pages} {065501}
  (\bibinfo {year} {2015})}\BibitemShut {NoStop}%
%%CITATION = ARXIV:1502.08032;%%
\bibitem [{\citenamefont {{S.L.~Adler}}(1968)}]{nuwro_res}%
  \BibitemOpen
  \bibfield  {author} {\bibinfo {author} {\bibnamefont {{S.L.~Adler}}},\
  }\href@noop {} {\bibfield  {journal} {\bibinfo  {journal} {Ann. Phys.}\
  }\textbf {\bibinfo {volume} {50}},\ \bibinfo {pages} {189 } (\bibinfo {year}
  {1968})}\BibitemShut {NoStop}%
\bibitem [{\citenamefont {{S.L.~Adler}}(1975)}]{nuwro_res2}%
  \BibitemOpen
  \bibfield  {author} {\bibinfo {author} {\bibnamefont {{S.L.~Adler}}},\
  }\href@noop {} {\bibfield  {journal} {\bibinfo  {journal} {Phys. Rev. D}\
  }\textbf {\bibinfo {volume} {12}},\ \bibinfo {pages} {2644} (\bibinfo {year}
  {1975})}\BibitemShut {NoStop}%
\bibitem [{\citenamefont {{P.A.~Schreiner and F.~Von Hippel}}(1973)}]{acs_res}%
  \BibitemOpen
  \bibfield  {author} {\bibinfo {author} {\bibnamefont {{P.A.~Schreiner and
  F.~Von Hippel}}},\ }\href@noop {} {\bibfield  {journal} {\bibinfo  {journal}
  {Nucl. Phys. B}\ }\textbf {\bibinfo {volume} {58}},\ \bibinfo {pages} {333}
  (\bibinfo {year} {1973})}\BibitemShut {NoStop}%
\bibitem [{\citenamefont {{M.J.~Vicente-Vacas, M.K.~Khankhasayev, and
  S.G.~Mashnik}}(1994)}]{VicenteVacas:1994}%
  \BibitemOpen
  \bibfield  {author} {\bibinfo {author} {\bibnamefont {{M.J.~Vicente-Vacas,
  M.K.~Khankhasayev, and S.G.~Mashnik}}},\ }\href@noop {} {\  (\bibinfo {year}
  {1994})},\ \Eprint {http://arxiv.org/abs/nucl-th/9412023}
  {arXiv:nucl-th/9412023 [nucl-th]} \BibitemShut {NoStop}%
%%CITATION = NUCL-TH/9412023;%%
\bibitem [{\citenamefont {Salcedo}\ \emph {et~al.}(1988)\citenamefont
  {Salcedo}, \citenamefont {Oset}, \citenamefont {Vicente-Vacas},\ and\
  \citenamefont {Garcia-Recio}}]{Salcedo:1987md}%
  \BibitemOpen
  \bibfield  {author} {\bibinfo {author} {\bibfnamefont {L.~L.}\ \bibnamefont
  {Salcedo}}, \bibinfo {author} {\bibfnamefont {E.}~\bibnamefont {Oset}},
  \bibinfo {author} {\bibfnamefont {M.~J.}\ \bibnamefont {Vicente-Vacas}}, \
  and\ \bibinfo {author} {\bibfnamefont {C.}~\bibnamefont {Garcia-Recio}},\
  }\href {\doibase 10.1016/0375-9474(88)90310-7} {\bibfield  {journal}
  {\bibinfo  {journal} {Nucl. Phys.}\ }\textbf {\bibinfo {volume} {A484}},\
  \bibinfo {pages} {557} (\bibinfo {year} {1988})}\BibitemShut {NoStop}%
%%CITATION = NUPHA,A484,557;%%
\bibitem [{\citenamefont {Leitner}\ \emph {et~al.}(2009)\citenamefont
  {Leitner}, \citenamefont {Buss}, \citenamefont {Mosel},\ and\ \citenamefont
  {Alvarez-Ruso}}]{Leitner:2008wx}%
  \BibitemOpen
  \bibfield  {author} {\bibinfo {author} {\bibfnamefont {T.}~\bibnamefont
  {Leitner}}, \bibinfo {author} {\bibfnamefont {O.}~\bibnamefont {Buss}},
  \bibinfo {author} {\bibfnamefont {U.}~\bibnamefont {Mosel}}, \ and\ \bibinfo
  {author} {\bibfnamefont {L.}~\bibnamefont {Alvarez-Ruso}},\ }\href {\doibase
  10.1103/PhysRevC.79.038501} {\bibfield  {journal} {\bibinfo  {journal} {Phys.
  Rev. C}\ }\textbf {\bibinfo {volume} {79}},\ \bibinfo {pages} {038501}
  (\bibinfo {year} {2009})}\BibitemShut {NoStop}%
\bibitem [{\citenamefont {Aguilar-Arevalo}\ \emph
  {et~al.}(2009{\natexlab{b}})\citenamefont {Aguilar-Arevalo} \emph
  {et~al.}}]{miniboone_flux}%
  \BibitemOpen
  \bibfield  {author} {\bibinfo {author} {\bibfnamefont {A.~A.}\ \bibnamefont
  {Aguilar-Arevalo}} \emph {et~al.} (\bibinfo {collaboration} {MiniBooNE
  Collaboration}),\ }\href {\doibase 10.1103/PhysRevD.79.072002} {\bibfield
  {journal} {\bibinfo  {journal} {Phys. Rev. D}\ }\textbf {\bibinfo {volume}
  {79}},\ \bibinfo {pages} {072002} (\bibinfo {year}
  {2009}{\natexlab{b}})}\BibitemShut {NoStop}%
\bibitem [{\citenamefont {Casper}(2002)}]{NUANCE}%
  \BibitemOpen
  \bibfield  {author} {\bibinfo {author} {\bibfnamefont {D.}~\bibnamefont
  {Casper}},\ }\href {\doibase 10.1016/S0920-5632(02)01756-5} {\bibfield
  {journal} {\bibinfo  {journal} {Nucl. Phys. Proc. Suppl.}\ }\textbf {\bibinfo
  {volume} {112}},\ \bibinfo {pages} {161} (\bibinfo {year}
  {2002})}\BibitemShut {NoStop}%
\end{thebibliography}%

\ifnum\PRLsupp=0
  \clearpage
  \onecolumngrid
  \newcommand{\qsq}{\ensuremath{Q^2_{QE}}\xspace}
\renewcommand{\textfraction}{0.05}
\renewcommand{\topfraction}{0.95}
\renewcommand{\bottomfraction}{0.95}
\renewcommand{\floatpagefraction}{0.95}
\renewcommand{\dblfloatpagefraction}{0.95}
\renewcommand{\dbltopfraction}{0.95}
\setcounter{totalnumber}{5}
\setcounter{bottomnumber}{3}
\setcounter{topnumber}{3}
\setcounter{dbltopnumber}{3}

\renewcommand{\cconepi}{CC1\ensuremath{\pi^{\pm}}\xspace}
\renewcommand{\ccpi}{CCN\ensuremath{\pi^{\pm}}\xspace}
\renewcommand{\Tpi}{\ensuremath{T_{\pi}}\xspace}
\renewcommand{\thetapi}{\ensuremath{\theta_{\pi}}\xspace}

\ifdefined\PRLsupp
  \ifnum\PRLsupp=0
    \newcommand{\SuppName}{appendix}
  \else
    \newcommand{\SuppName}{supplemental material}
  \fi
\else
  \newcommand{\SuppName}{supplemental material}
\fi

\appendix*
\section{}
This\ \SuppName\ contains tables of measured cross sections, uncertainties, and bin correlations for the measurements presented in the paper.  
Additionally, these quantities are reported for alternative versions of the reported measurements in which the measured signal has the 
additional restriction that the muon angle with respect to the beam $\theta_\mu$ is less than 20$^\circ$.

%%%%%%%%%%%%%%%%%%%%%%%%%%%%%%%%%%%%%%%%%%%%%%%%%%%%%%%%%%%%%%%%%%%%%%%%%%%%
%full muon acceptance 1pi analysis - Theta
%%%%%%%%%%%%%%%%%%%%%%%%%%%%%%%%%%%%%%%%%%%%%%%%%%%%%%%%%%%%%%%%%%%%%%%%%%%%
\begingroup
\squeezetable
\begin{table}[h]
\begin{tabular}{l|ccccccccccccccc}
%Measured #theta_{#pi}, _1pi analysis
\hline\hline
$\theta_{\pi}$ (degree) Bins & 0 - 15 & 15 - 22 & 22 - 29 & 29 - 36 & 36 - 43 & 43 - 50 & 50 - 57 \\ 
\hline
Cross section in bin & 1.83 & 2.87 & 3.05 & 3.87 & 3.54 & 2.91 & 2.13  \\ 
$10^{-41}\mathrm{cm}^2/\mathrm{degree}/$nucleon &  $\pm$0.40 (0.23) &  $\pm$0.57 (0.26) &  $\pm$0.60 (0.25) &  $\pm$0.77 (0.29) &  $\pm$0.74 (0.26) &  $\pm$0.61 (0.23) &  $\pm$0.45 (0.20) \\ 
\hline
$\theta_{\pi}$ (degree) Bins & 57 - 72 & 72 - 108 & 108 - 130 & 130 - 140 & 140 - 150 & 150 - 165 \\ 
\hline
Cross section in bin & 1.98 & 1.55 & 0.90 & 0.71 & 0.54 & 0.33 \\ 
$10^{-41}\mathrm{cm}^2/\mathrm{degree}/$nucleon &  $\pm$0.40 (0.19) &  $\pm$0.29 (0.10) &  $\pm$0.19 (0.11) &  $\pm$0.14 (0.08) &  $\pm$0.11 (0.06) &  $\pm$0.07 (0.05) \\ 
\hline\hline
\end{tabular}
\caption{Measured \cconepi $d\sigma / d\theta_\pi$ and total uncertainties.  The absolute uncertainties are followed by shape uncertainties in parentheses.}
\end{table}
\endgroup

\begingroup
\squeezetable
\begin{table}[ht]
\begin{tabular}{l|ccccccccccccccc}
%Full Correlation matrix for #theta_{#pi}, _1pi analysis
\hline\hline
Bins (degree) & 0 - 15 & 15 - 22 & 22 - 29 & 29 - 36 & 36 - 43 & 43 - 50 & 50 - 57 & 57 - 72 & 72 - 108 & 108 - 130 & 130 - 140 & 140 - 150 & 150 - 165 \\ 
\hline
0 - 15 & 1 & 0.78 & 0.71 & 0.71 & 0.73 & 0.71 & 0.66 & 0.65 & 0.78 & 0.73 & 0.72 & 0.69 & 0.60 \\ 
15 - 22 &  & 1 & 0.82 & 0.82 & 0.83 & 0.82 & 0.78 & 0.76 & 0.84 & 0.74 & 0.74 & 0.73 & 0.64 \\ 
22 - 29 &  &  & 1 & 0.87 & 0.86 & 0.86 & 0.83 & 0.81 & 0.85 & 0.72 & 0.73 & 0.73 & 0.65 \\ 
29 - 36 &  &  &  & 1 & 0.89 & 0.88 & 0.85 & 0.83 & 0.85 & 0.73 & 0.75 & 0.74 & 0.66 \\ 
36 - 43 &  &  &  &  & 1 & 0.90 & 0.86 & 0.83 & 0.87 & 0.75 & 0.76 & 0.76 & 0.67 \\ 
43 - 50 &  &  &  &  &  & 1 & 0.86 & 0.83 & 0.86 & 0.74 & 0.75 & 0.75 & 0.67 \\ 
50 - 57 &  &  &  &  &  &  & 1 & 0.82 & 0.84 & 0.72 & 0.73 & 0.73 & 0.66 \\ 
57 - 72 &  &  &  &  &  &  &  & 1 & 0.82 & 0.72 & 0.73 & 0.72 & 0.65 \\ 
72 - 108 &  &  &  &  &  &  &  &  & 1 & 0.80 & 0.80 & 0.79 & 0.71 \\ 
108 - 130 &  &  &  &  &  &  &  &  &  & 1 & 0.75 & 0.73 & 0.66 \\ 
130 - 140 &  &  &  &  &  &  &  &  &  &  & 1 & 0.76 & 0.67 \\ 
140 - 150 &  &  &  &  &  &  &  &  &  &  &  & 1 & 0.66 \\ 
150 - 165 &  &  &  &  &  &  &  &  &  &  &  &  & 1 \\ 
\hline
0 - 15 & 1 & 0.19 & -0.16 & -0.24 & -0.20 & -0.23 & -0.31 & -0.26 & 0.02 & 0.13 & 0.08 & 0.04 & -0.03 \\ 
15 - 22 &  & 1 & 0.03 & -0.02 & -0.04 & -0.07 & -0.13 & -0.13 & -0.03 & -0.02 & -0.04 & -0.04 & -0.06 \\ 
22 - 29 &  &  & 1 & 0.16 & 0.06 & 0.07 & 0.08 & 0.07 & -0.06 & -0.16 & -0.13 & -0.11 & -0.07 \\ 
29 - 36 &  &  &  & 1 & 0.19 & 0.18 & 0.13 & 0.09 & -0.12 & -0.21 & -0.14 & -0.12 & -0.11 \\ 
36 - 43 &  &  &  &  & 1 & 0.24 & 0.16 & 0.02 & -0.10 & -0.14 & -0.16 & -0.12 & -0.11 \\ 
43 - 50 &  &  &  &  &  & 1 & 0.17 & 0.07 & -0.09 & -0.14 & -0.14 & -0.11 & -0.09 \\ 
50 - 57 &  &  &  &  &  &  & 1 & 0.13 & -0.03 & -0.11 & -0.11 & -0.07 & -0.02 \\ 
57 - 72 &  &  &  &  &  &  &  & 1 & -0.04 & -0.05 & -0.01 & -0.02 & 0.02 \\ 
72 - 108 &  &  &  &  &  &  &  &  & 1 & 0.07 & 0.04 & 0.05 & 0.06 \\ 
108 - 130 &  &  &  &  &  &  &  &  &  & 1 & 0.17 & 0.13 & 0.11 \\ 
130 - 140 &  &  &  &  &  &  &  &  &  &  & 1 & 0.23 & 0.16 \\ 
140 - 150 &  &  &  &  &  &  &  &  &  &  &  & 1 & 0.17 \\ 
150 - 165 &  &  &  &  &  &  &  &  &  &  &  &  & 1 \\ 
\hline\hline
\end{tabular}
\caption{Top: Full correlation matrix for the \cconepi $d\sigma / d\theta_\pi$ uncertainties.  Bottom: Corresponding shape correlation matrix.}
\end{table}
\endgroup

\clearpage
%%%%%%%%%%%%%%%%%%%%%%%%%%%%%%%%%%%%%%%%%%%%%%%%%%%%%%%%%%%%%%%%%%%%%%%%%%%%
%full muon acceptance 1pi analysis - KE
%%%%%%%%%%%%%%%%%%%%%%%%%%%%%%%%%%%%%%%%%%%%%%%%%%%%%%%%%%%%%%%%%%%%%%%%%%%%
\begingroup
\squeezetable
\begin{table}[h]
\begin{tabular}{l|ccccccc}
%Measured T_{#pi}, _1pi analysis
\hline\hline
$T_{\pi}$ (MeV) Bins & 35 - 55 & 55 - 75 & 75 - 100 & 100 - 125 & 125 - 150 & 150 - 200 & 200 - 350 \\ 
\hline
Cross section in bin & 1.13 & 1.16 & 1.07 & 0.85 & 0.76 & 0.66 & 0.38 \\ 
$10^{-41}\mathrm{cm}^2/\mathrm{MeV}/$nucleon &  $\pm$0.30 (0.20) &  $\pm$0.25 (0.12) &  $\pm$0.20 (0.09) &  $\pm$0.15 (0.06) &  $\pm$0.14 (0.05) &  $\pm$0.11 (0.05) &  $\pm$0.08 (0.04) \\ 
\hline\hline
\end{tabular}
\caption{Measured \cconepi $d\sigma / dT_\pi$ and total uncertainties.  The absolute uncertainties are followed by shape uncertainties in parentheses.}
\end{table}
\endgroup

\begingroup
\squeezetable
\begin{table}[h]
\begin{tabular}{l|ccccccc}
%Full Correlation matrix for T_{#pi}, _1pi analysis
\hline\hline
Bins (MeV) & 35 - 55 & 55 - 75 & 75 - 100 & 100 - 125 & 125 - 150 & 150 - 200 & 200 - 350 \\ 
\hline
35 - 55 & 1 & 0.74 & 0.72 & 0.68 & 0.68 & 0.59 & 0.56 \\ 
55 - 75 &  & 1 & 0.87 & 0.82 & 0.81 & 0.72 & 0.70 \\ 
75 - 100 &  &  & 1 & 0.85 & 0.84 & 0.76 & 0.71 \\ 
100 - 125 &  &  &  & 1 & 0.88 & 0.83 & 0.79 \\ 
125 - 150 &  &  &  &  & 1 & 0.84 & 0.81 \\ 
150 - 200 &  &  &  &  &  & 1 & 0.89 \\ 
200 - 350 &  &  &  &  &  &  & 1 \\ 
\hline
35 - 55 & 1 & 0.29 & 0.20 & 0.01 & -0.02 & -0.30 & -0.36 \\ 
55 - 75 &  & 1 & 0.39 & 0.09 & 0.02 & -0.40 & -0.47 \\ 
75 - 100 &  &  & 1 & 0.21 & 0.13 & -0.22 & -0.53 \\ 
100 - 125 &  &  &  & 1 & 0.25 & 0.01 & -0.31 \\ 
125 - 150 &  &  &  &  & 1 & 0.05 & -0.21 \\ 
150 - 200 &  &  &  &  &  & 1 & 0.27 \\ 
200 - 350 &  &  &  &  &  &  & 1 \\ 
\hline\hline
\end{tabular}
\caption{Top: Full correlation matrix for the \cconepi $d\sigma / dT_\pi$ uncertainties.  Bottom: Corresponding shape correlation matrix.}
\end{table}
\endgroup

\clearpage
%%%%%%%%%%%%%%%%%%%%%%%%%%%%%%%%%%%%%%%%%%%%%%%%%%%%%%%%%%%%%%%%%%%%%%%%%%%%
%limited muon acceptance 1pi analysis - Theta
%%%%%%%%%%%%%%%%%%%%%%%%%%%%%%%%%%%%%%%%%%%%%%%%%%%%%%%%%%%%%%%%%%%%%%%%%%%%
\begingroup
\squeezetable
\begin{table}[h]
\begin{tabular}{l|ccccccccccccccc}
%Measured #theta_{#pi}, _1pi analysis
\hline\hline
$\theta_{\pi}$ (degree) Bins & 0 - 15 & 15 - 22 & 22 - 29 & 29 - 36 & 36 - 43 & 43 - 50 & 50 - 57 \\ 
\hline
Cross section in bin & 1.15 & 1.66 & 1.76 & 2.26 & 2.16 & 1.87 & 1.44  \\ 
$10^{-41}\mathrm{cm}^2/\mathrm{degree}/$nucleon &  $\pm$0.25 (0.17) &  $\pm$0.29 (0.16) &  $\pm$0.29 (0.15) &  $\pm$0.37 (0.17) &  $\pm$0.38 (0.16) &  $\pm$0.34 (0.15) &  $\pm$0.27 (0.14) \\ 
\hline
$\theta_{\pi}$ (degree) Bins & 57 - 72 & 72 - 108 & 108 - 130 & 130 - 140 & 140 - 150 & 150 - 165 \\ 
\hline
Cross section in bin & 1.41 & 1.18 & 0.69 & 0.55 & 0.414 & 0.256 \\ 
$10^{-41}\mathrm{cm}^2/\mathrm{degree}/$nucleon &  $\pm$0.26 (0.14) &  $\pm$0.21 (0.08) &  $\pm$0.15 (0.08) &  $\pm$0.11 (0.06) &  $\pm$0.081 (0.047) &  $\pm$0.054 (0.037) \\ 
\hline\hline
\end{tabular}
\caption{Measured $d\sigma / d\theta_\pi$ and total uncertainties for the \cconepi analysis with the additional signal requirement of $\theta_\mu < 20^\circ$.  The absolute uncertainties 
are followed by shape uncertainties in parentheses.}
\end{table}
\endgroup

\begingroup
\squeezetable
\begin{table}[h]
\begin{tabular}{l|ccccccccccccccc}
%Full Correlation matrix for #theta_{#pi}, _1pi analysis
\hline\hline
Bins (degree) & 0 - 15 & 15 - 22 & 22 - 29 & 29 - 36 & 36 - 43 & 43 - 50 & 50 - 57 & 57 - 72 & 72 - 108 & 108 - 130 & 130 - 140 & 140 - 150 & 150 - 165 \\ 
\hline
0 - 15 & 1 & 0.77 & 0.65 & 0.61 & 0.61 & 0.58 & 0.51 & 0.53 & 0.68 & 0.67 & 0.65 & 0.62 & 0.54 \\ 
15 - 22 &  & 1 & 0.74 & 0.73 & 0.72 & 0.70 & 0.65 & 0.65 & 0.76 & 0.71 & 0.70 & 0.67 & 0.61 \\ 
22 - 29 &  &  & 1 & 0.80 & 0.79 & 0.78 & 0.75 & 0.75 & 0.79 & 0.70 & 0.69 & 0.67 & 0.63 \\ 
29 - 36 &  &  &  & 1 & 0.84 & 0.83 & 0.80 & 0.79 & 0.82 & 0.72 & 0.73 & 0.71 & 0.65 \\ 
36 - 43 &  &  &  &  & 1 & 0.86 & 0.82 & 0.79 & 0.84 & 0.74 & 0.74 & 0.73 & 0.66 \\ 
43 - 50 &  &  &  &  &  & 1 & 0.82 & 0.79 & 0.83 & 0.72 & 0.72 & 0.70 & 0.64 \\ 
50 - 57 &  &  &  &  &  &  & 1 & 0.78 & 0.80 & 0.68 & 0.68 & 0.67 & 0.62 \\ 
57 - 72 &  &  &  &  &  &  &  & 1 & 0.77 & 0.67 & 0.67 & 0.65 & 0.59 \\ 
72 - 108 &  &  &  &  &  &  &  &  & 1 & 0.78 & 0.77 & 0.75 & 0.67 \\ 
108 - 130 &  &  &  &  &  &  &  &  &  & 1 & 0.75 & 0.73 & 0.65 \\ 
130 - 140 &  &  &  &  &  &  &  &  &  &  & 1 & 0.76 & 0.67 \\ 
140 - 150 &  &  &  &  &  &  &  &  &  &  &  & 1 & 0.66 \\ 
150 - 165 &  &  &  &  &  &  &  &  &  &  &  &  & 1 \\ 
\hline
0 - 15 & 1 & 0.36 & -0.04 & -0.24 & -0.30 & -0.32 & -0.39 & -0.30 & -0.10 & 0.11 & 0.06 & 0.01 & -0.03 \\ 
15 - 22 &  & 1 & 0.09 & -0.07 & -0.16 & -0.19 & -0.24 & -0.17 & -0.11 & 0.01 & 0.001 & -0.01 & 0.001 \\ 
22 - 29 &  &  & 1 & 0.14 & -0.01 & 0.01 & 0.03 & 0.09 & -0.10 & -0.14 & -0.11 & -0.10 & -0.02 \\ 
29 - 36 &  &  &  & 1 & 0.14 & 0.15 & 0.13 & 0.17 & -0.09 & -0.16 & -0.07 & -0.06 & -0.03 \\ 
36 - 43 &  &  &  &  & 1 & 0.25 & 0.19 & 0.13 & -0.02 & -0.07 & -0.07 & -0.04 & -0.01 \\ 
43 - 50 &  &  &  &  &  & 1 & 0.22 & 0.16 & 0.01 & -0.12 & -0.11 & -0.08 & -0.05 \\  
50 - 57 &  &  &  &  &  &  & 1 & 0.20 & 0.05 & -0.12 & -0.11 & -0.07 & -0.01 \\ 
57 - 72 &  &  &  &  &  &  &  & 1 & -0.04 & -0.14 & -0.11 & -0.11 & -0.05 \\ 
72 - 108 &  &  &  &  &  &  &  &  & 1 & 0.02 & -0.04 & -0.01 & -0.04 \\ 
108 - 130 &  &  &  &  &  &  &  &  &  & 1 & 0.18 & 0.15 & 0.10 \\ 
130 - 140 &  &  &  &  &  &  &  &  &  &  & 1 & 0.25 & 0.14 \\ 
140 - 150 &  &  &  &  &  &  &  &  &  &  &  & 1 & 0.16 \\ 
150 - 165 &  &  &  &  &  &  &  &  &  &  &  &  & 1 \\ 
\hline\hline
\end{tabular}
\caption{Top:  Full correlation matrix for the $d\sigma / d\theta_\pi$ uncertainties in the \cconepi analysis with the additional signal requirement of $\theta_\mu < 20^\circ$.  Bottom: Corresponding shape correlation matrix.} 
\end{table}
\endgroup

\clearpage
%%%%%%%%%%%%%%%%%%%%%%%%%%%%%%%%%%%%%%%%%%%%%%%%%%%%%%%%%%%%%%%%%%%%%%%%%%%%
%limited muon acceptance 1pi analysis - KE
%%%%%%%%%%%%%%%%%%%%%%%%%%%%%%%%%%%%%%%%%%%%%%%%%%%%%%%%%%%%%%%%%%%%%%%%%%%%
\begingroup
\squeezetable
\begin{table}[h]
\begin{tabular}{l|ccccccc}
%Measured T_{#pi}, _1pi analysis
\hline\hline
$T_{\pi}$ (MeV) Bins & 35 - 55 & 55 - 75 & 75 - 100 & 100 - 125 & 125 - 150 & 150 - 200 & 200 - 350 \\ 
\hline
Cross section in bin & 0.85 & 0.90 & 0.84 & 0.66 & 0.59 & 0.496 & 0.257 \\ 
$10^{-41}\mathrm{cm}^2/\mathrm{MeV}/$nucleon &  $\pm$0.21 (0.15) &  $\pm$0.18 (0.09) &  $\pm$0.15 (0.07) &  $\pm$0.11 (0.05) &  $\pm$0.10 (0.04) &  $\pm$0.079 (0.036) &  $\pm$0.053 (0.024) \\ 
\hline\hline
\end{tabular}
\caption{Measured $d\sigma / dT_\pi$ and total uncertainties for the \cconepi analysis with the additional signal requirement of $\theta_\mu < 20^\circ$.  The absolute uncertainties are followed by shape uncertainties in parentheses.}
\end{table}
\endgroup

\begingroup
\squeezetable
\begin{table}[h]
\begin{tabular}{l|ccccccc}
%Full Correlation matrix for T_{#pi}, _1pi analysis
\hline\hline
Bins (MeV) & 35 - 55 & 55 - 75 & 75 - 100 & 100 - 125 & 125 - 150 & 150 - 200 & 200 - 350 \\ 
\hline
35 - 55 & 1 & 0.71 & 0.69 & 0.65 & 0.65 & 0.56 & 0.52 \\ 
55 - 75 &  & 1 & 0.86 & 0.81 & 0.80 & 0.70 & 0.67 \\ 
75 - 100 &  &  & 1 & 0.84 & 0.83 & 0.74 & 0.68 \\ 
100 - 125 &  &  &  & 1 & 0.87 & 0.81 & 0.77 \\ 
125 - 150 &  &  &  &  & 1 & 0.83 & 0.79 \\ 
150 - 200 &  &  &  &  &  & 1 & 0.88 \\ 
200 - 350 &  &  &  &  &  &  & 1 \\
\hline
35 - 55 & 1 & 0.26 & 0.17 & -0.02 & -0.05 & -0.29 & -0.34 \\ 
55 - 75 &  & 1 & 0.36 & 0.05 & -0.01 & -0.39 & -0.44 \\ 
75 - 100 &  &  & 1 & 0.18 & 0.10 & -0.22 & -0.52 \\ 
100 - 125 &  &  &  & 1 & 0.22 & 0.01 & -0.27 \\ 
125 - 150 &  &  &  &  & 1 & 0.04 & -0.17 \\ 
150 - 200 &  &  &  &  &  & 1 & 0.29 \\ 
200 - 350 &  &  &  &  &  &  & 1 \\ 
\hline\hline
\end{tabular}
\caption{Top: Full correlation matrix for the $d\sigma / dT_\pi$ uncertainties in the \cconepi analysis with the additional signal requirement of $\theta_\mu < 20^\circ$.  Bottom: Corresponding shape correlation matrix.}
\end{table}
\endgroup

\clearpage
%%%%%%%%%%%%%%%%%%%%%%%%%%%%%%%%%%%%%%%%%%%%%%%%%%%%%%%%%%%%%%%%%%%%%%%%%%%%
%full muon acceptance Npi analysis - Theta
%%%%%%%%%%%%%%%%%%%%%%%%%%%%%%%%%%%%%%%%%%%%%%%%%%%%%%%%%%%%%%%%%%%%%%%%%%%%
\begingroup
\squeezetable
\begin{table}[h]
\begin{tabular}{l|ccccccccccccc}
%Measured #theta_{#pi}, _Npi analysis
\hline\hline
$\theta_{\pi}$ (degree) Bins & 0 - 15 & 15 - 22 & 22 - 29 & 29 - 36 & 36 - 43 & 43 - 50 & 50 - 57 \\ 
\hline
Measurement in bin & 5.5 & 9.8 & 9.5 & 10.3 & 8.9 & 6.6 & 5.80 \\ 
$10^{-41}\mathrm{cm}^2/\mathrm{degree}/$nucleon &  $\pm$1.0 (0.5) &  $\pm$1.7 (0.7) &  $\pm$1.6 (0.6) &  $\pm$1.7 (0.6) &  $\pm$1.4 (0.6) &  $\pm$1.1 (0.5) &  $\pm$0.94 (0.44) \\ 
\hline
$\theta_{\pi}$ (degree) Bins & 57 - 72 & 72 - 108 & 108 - 130 & 130 - 140 & 140 - 150 & 150 - 165 \\ 
\hline
Measurement in bin & 4.90 & 3.08 & 1.61 & 1.32 & 1.05 & 0.68 \\ 
$10^{-41}\mathrm{cm}^2/\mathrm{degree}/$nucleon & $\pm$0.77 (0.46) &  $\pm$0.43 (0.21) &  $\pm$0.25 (0.17) &  $\pm$0.20 (0.13) &  $\pm$0.16 (0.11) &  $\pm$0.11 (0.09) \\ 
\hline\hline
\end{tabular}
\caption{Measured \ccpi $(1/T\Phi)(dN_{\pi} / d\theta_\pi)$ and total uncertainties.  The absolute uncertainties are followed by shape uncertainties in parentheses.}
\end{table}
\endgroup

\begingroup
\squeezetable
\begin{table}[h]
\begin{tabular}{l|ccccccccccccc}
%Full Correlation matrix for #theta_{#pi}, _Npi analysis
\hline\hline
Bins (degree) & 0 - 15 & 15 - 22 & 22 - 29 & 29 - 36 & 36 - 43 & 43 - 50 & 50 - 57 & 57 - 72 & 72 - 108 & 108 - 130 & 130 - 140 & 140 - 150 & 150 - 165 \\ 
\hline
0 - 15 & 1 & 0.85 & 0.85 & 0.84 & 0.78 & 0.75 & 0.69 & 0.56 & 0.69 & 0.59 & 0.60 & 0.58 & 0.52 \\ 
15 - 22 &  & 1 & 0.86 & 0.85 & 0.82 & 0.79 & 0.75 & 0.64 & 0.75 & 0.65 & 0.65 & 0.63 & 0.57 \\ 
22 - 29 &  &  & 1 & 0.88 & 0.86 & 0.84 & 0.79 & 0.68 & 0.78 & 0.65 & 0.66 & 0.65 & 0.59 \\ 
29 - 36 &  &  &  & 1 & 0.87 & 0.85 & 0.81 & 0.70 & 0.79 & 0.65 & 0.68 & 0.66 & 0.59 \\ 
36 - 43 &  &  &  &  & 1 & 0.88 & 0.84 & 0.71 & 0.77 & 0.63 & 0.65 & 0.63 & 0.58 \\ 
43 - 50 &  &  &  &  &  & 1 & 0.84 & 0.74 & 0.79 & 0.64 & 0.66 & 0.64 & 0.59 \\ 
50 - 57 &  &  &  &  &  &  & 1 & 0.78 & 0.79 & 0.67 & 0.67 & 0.65 & 0.59 \\ 
57 - 72 &  &  &  &  &  &  &  & 1 & 0.78 & 0.69 & 0.71 & 0.68 & 0.61 \\ 
72 - 108 &  &  &  &  &  &  &  &  & 1 & 0.74 & 0.75 & 0.74 & 0.67 \\ 
108 - 130 &  &  &  &  &  &  &  &  &  & 1 & 0.68 & 0.66 & 0.60 \\ 
130 - 140 &  &  &  &  &  &  &  &  &  &  & 1 & 0.71 & 0.63 \\ 
140 - 150 &  &  &  &  &  &  &  &  &  &  &  & 1 & 0.63 \\ 
150 - 165 &  &  &  &  &  &  &  &  &  &  &  &  & 1 \\ 
\hline
0 - 15 & 1 & 0.37 & 0.30 & 0.20 & -0.01 & -0.13 & -0.29 & -0.45 & -0.33 & -0.22 & -0.24 & -0.23 & -0.20 \\ 
15 - 22 &  & 1 & 0.20 & 0.13 & -0.01 & -0.09 & -0.22 & -0.33 & -0.23 & -0.13 & -0.18 & -0.17 & -0.15 \\ 
22 - 29 &  &  & 1 & 0.19 & 0.10 & 0.03 & -0.14 & -0.32 & -0.26 & -0.22 & -0.26 & -0.23 & -0.19 \\ 
29 - 36 &  &  &  & 1 & 0.16 & 0.07 & -0.05 & -0.23 & -0.24 & -0.24 & -0.20 & -0.18 & -0.19 \\ 
36 - 43 &  &  &  &  & 1 & 0.30 & 0.16 & -0.11 & -0.21 & -0.26 & -0.25 & -0.23 & -0.16 \\ 
43 - 50 &  &  &  &  &  & 1 & 0.21 & 0.01 & -0.10 & -0.18 & -0.19 & -0.17 & -0.11 \\ 
50 - 57 &  &  &  &  &  &  & 1 & 0.24 & 0.06 & -0.01 & -0.02 & -0.03 & -0.02 \\ 
57 - 72 &  &  &  &  &  &  &  & 1 & 0.27 & 0.25 & 0.28 & 0.23 & 0.19 \\ 
72 - 108 &  &  &  &  &  &  &  &  & 1 & 0.28 & 0.29 & 0.29 & 0.25 \\ 
108 - 130 &  &  &  &  &  &  &  &  &  & 1 & 0.29 & 0.26 & 0.22 \\ 
130 - 140 &  &  &  &  &  &  &  &  &  &  & 1 & 0.37 & 0.28 \\ 
140 - 150 &  &  &  &  &  &  &  &  &  &  &  & 1 & 0.29 \\ 
150 - 165 &  &  &  &  &  &  &  &  &  &  &  &  & 1 \\ 
\hline\hline
\end{tabular}
\caption{Top: Full correlation matrix for the \ccpi $(1/T\Phi)(dN_{\pi} / d\theta_\pi)$ uncertainties.  Bottom: Corresponding shape correlation matrix.}
\end{table}
\endgroup

\clearpage
%%%%%%%%%%%%%%%%%%%%%%%%%%%%%%%%%%%%%%%%%%%%%%%%%%%%%%%%%%%%%%%%%%%%%%%%%%%%
%full muon acceptance Npi analysis - KE
%%%%%%%%%%%%%%%%%%%%%%%%%%%%%%%%%%%%%%%%%%%%%%%%%%%%%%%%%%%%%%%%%%%%%%%%%%%%
\begingroup
\squeezetable
\begin{table}[h]
\begin{tabular}{l|ccccccc}
%Measured T_{#pi}, _Npi analysis
\hline\hline
$T_{\pi}$ (MeV) Bins & 35 - 55 & 55 - 75 & 75 - 100 & 100 - 125 & 125 - 150 & 150 - 200 & 200 - 350 \\ 
\hline
Measurement in bin & 2.16 & 2.11 & 1.92 & 1.80 & 1.59 & 1.25 & 0.83 \\ 
$10^{-41}\mathrm{cm}^2/\mathrm{MeV}/$nucleon &  $\pm$0.53 (0.39) &  $\pm$0.39 (0.22) &  $\pm$0.30 (0.14) &  $\pm$0.24 (0.12) &  $\pm$0.21 (0.11) &  $\pm$0.17 (0.07) &  $\pm$0.13 (0.06) \\ 
\hline\hline
\end{tabular}
\caption{Measured \ccpi $(1/T\Phi)(dN_{\pi} / dT_\pi)$ and total uncertainties.  The absolute uncertainties are followed by shape uncertainties in parentheses.}
\end{table}
\endgroup

\begingroup
\squeezetable
\begin{table}[h]
\begin{tabular}{l|cccccccc}
%Full Correlation matrix for T_{#pi}, _Npi analysis
\hline\hline
Bins (MeV) & 35 - 55 & 55 - 75 & 75 - 100 & 100 - 125 & 125 - 150 & 150 - 200 & 200 - 350 \\ 
\hline
35 - 55 & 1 & 0.74 & 0.69 & 0.59 & 0.52 & 0.51 & 0.46 \\ 
55 - 75 &  & 1 & 0.84 & 0.72 & 0.66 & 0.64 & 0.61 \\ 
75 - 100 &  &  & 1 & 0.80 & 0.76 & 0.74 & 0.70 \\ 
100 - 125 &  &  &  & 1 & 0.85 & 0.81 & 0.73 \\ 
125 - 150 &  &  &  &  & 1 & 0.84 & 0.78 \\ 
150 - 200 &  &  &  &  &  & 1 & 0.89 \\ 
200 - 350 &  &  &  &  &  &  & 1 \\ 
\hline
35 - 55 & 1 & 0.44 & 0.25 & -0.12 & -0.28 & -0.42 & -0.46 \\ 
55 - 75 &  & 1 & 0.41 & -0.11 & -0.28 & -0.49 & -0.51 \\ 
75 - 100 &  &  & 1 & 0.08 & -0.09 & -0.36 & -0.45 \\ 
100 - 125 &  &  &  & 1 & 0.42 & 0.15 & -0.23 \\ 
125 - 150 &  &  &  &  & 1 & 0.32 & -0.01 \\ 
150 - 200 &  &  &  &  &  & 1 & 0.40 \\ 
200 - 350 &  &  &  &  &  &  & 1 \\ 
\hline\hline
\end{tabular}
\caption{Top: Full correlation matrix for the \ccpi $(1/T\Phi)(dN_{\pi} / dT_\pi)$ uncertainties.  Bottom: Corresponding shape correlation matrix.}
\end{table}
\endgroup

\clearpage
%%%%%%%%%%%%%%%%%%%%%%%%%%%%%%%%%%%%%%%%%%%%%%%%%%%%%%%%%%%%%%%%%%%%%%%%%%%%
%limited muon acceptance Npi analysis - Theta
%%%%%%%%%%%%%%%%%%%%%%%%%%%%%%%%%%%%%%%%%%%%%%%%%%%%%%%%%%%%%%%%%%%%%%%%%%%%
\begingroup
\squeezetable
\begin{table}[h]
\begin{tabular}{l|ccccccccccccccc}
%Measured #theta_{#pi}, _Npi analysis
\hline\hline
$\theta_{\pi}$ (degree) Bins & 0 - 15 & 15 - 22 & 22 - 29 & 29 - 36 & 36 - 43 & 43 - 50 & 50 - 57 \\ 
\hline
Measurement in bin & 2.30 & 4.23 & 4.42 & 5.18 & 4.61 & 3.79 & 3.48 \\ 
$10^{-41}\mathrm{cm}^2/\mathrm{degree}/$nucleon &  $\pm$0.41 (0.22) &  $\pm$0.68 (0.31) &  $\pm$0.72 (0.30) &  $\pm$0.80 (0.32) &  $\pm$0.74 (0.32) &  $\pm$0.61 (0.27) &  $\pm$0.53 (0.25) \\ 
\hline
$\theta_{\pi}$ (degree) Bins & 57 - 72 & 72 - 108 & 108 - 130 & 130 - 140 & 140 - 150 & 150 - 165 \\ 
\hline
Measurement in bin & 3.12 & 2.06 & 1.08 & 0.88 & 0.70 & 0.452 \\ 
$10^{-41}\mathrm{cm}^2/\mathrm{degree}/$nucleon &  $\pm$0.45 (0.27) &  $\pm$0.28 (0.13) &  $\pm$0.17 (0.11) &  $\pm$0.13 (0.08) &  $\pm$0.10 (0.07) &  $\pm$0.073 (0.053) \\ 
\hline\hline
\end{tabular}
\caption{Measured $(1/T\Phi)(dN_{\pi} / d\theta_\pi)$ and total uncertainties in the \ccpi analysis with the additional signal requirement of $\theta_\mu < 20^\circ$.  The absolute uncertainties are followed by shape uncertainties in parentheses.}
\end{table}
\endgroup

\begingroup
\squeezetable
\begin{table}[h]
\begin{tabular}{l|ccccccccccccccc}
%Full Correlation matrix for #theta_{#pi}, _Npi analysis
\hline\hline
Bins (degree) & 0 - 15 & 15 - 22 & 22 - 29 & 29 - 36 & 36 - 43 & 43 - 50 & 50 - 57 & 57 - 72 & 72 - 108 & 108 - 130 & 130 - 140 & 140 - 150 & 150 - 165 \\ 
\hline
0 - 15 & 1 & 0.84 & 0.83 & 0.83 & 0.78 & 0.76 & 0.71 & 0.59 & 0.72 & 0.61 & 0.62 & 0.60 & 0.57 \\ 
15 - 22 &  & 1 & 0.85 & 0.84 & 0.82 & 0.80 & 0.77 & 0.66 & 0.76 & 0.66 & 0.66 & 0.64 & 0.60 \\ 
22 - 29 &  &  & 1 & 0.87 & 0.86 & 0.84 & 0.80 & 0.69 & 0.78 & 0.66 & 0.67 & 0.66 & 0.62 \\ 
29 - 36 &  &  &  & 1 & 0.87 & 0.85 & 0.81 & 0.71 & 0.78 & 0.65 & 0.69 & 0.67 & 0.62 \\ 
36 - 43 &  &  &  &  & 1 & 0.87 & 0.83 & 0.70 & 0.75 & 0.62 & 0.65 & 0.63 & 0.60 \\ 
43 - 50 &  &  &  &  &  & 1 & 0.83 & 0.72 & 0.76 & 0.63 & 0.66 & 0.64 & 0.61 \\ 
50 - 57 &  &  &  &  &  &  & 1 & 0.75 & 0.77 & 0.66 & 0.68 & 0.66 & 0.61 \\ 
57 - 72 &  &  &  &  &  &  &  & 1 & 0.74 & 0.66 & 0.69 & 0.66 & 0.60 \\ 
72 - 108 &  &  &  &  &  &  &  &  & 1 & 0.73 & 0.73 & 0.72 & 0.65 \\ 
108 - 130 &  &  &  &  &  &  &  &  &  & 1 & 0.66 & 0.64 & 0.58 \\ 
130 - 140 &  &  &  &  &  &  &  &  &  &  & 1 & 0.69 & 0.60 \\ 
140 - 150 &  &  &  &  &  &  &  &  &  &  &  & 1 & 0.60 \\ 
150 - 165 &  &  &  &  &  &  &  &  &  &  &  &  & 1 \\ 
\hline
0 - 15 & 1 & 0.32 & 0.26 & 0.20 & 0.06 & -0.02 & -0.18 & -0.37 & -0.21 & -0.16 & -0.19 & -0.19 & -0.13 \\ 
15 - 22 &  & 1 & 0.17 & 0.13 & 0.06 & 0.01 & -0.12 & -0.24 & -0.15 & -0.10 & -0.16 & -0.16 & -0.11 \\ 
22 - 29 &  &  & 1 & 0.21 & 0.18 & 0.13 & -0.06 & -0.26 & -0.21 & -0.19 & -0.22 & -0.20 & -0.14 \\ 
29 - 36 &  &  &  & 1 & 0.23 & 0.14 & -0.002 & -0.20 & -0.24 & -0.22 & -0.17 & -0.15 & -0.14 \\ 
36 - 43 &  &  &  &  & 1 & 0.33 & 0.15 & -0.15 & -0.29 & -0.27 & -0.25 & -0.22 & -0.13 \\ 
43 - 50 &  &  &  &  &  & 1 & 0.16 & -0.08 & -0.20 & -0.21 & -0.21 & -0.18 & -0.10 \\ 
50 - 57 &  &  &  &  &  &  & 1 & 0.12 & -0.07 & -0.05 & -0.05 & -0.06 & -0.03 \\ 
57 - 72 &  &  &  &  &  &  &  & 1 & 0.12 & 0.15 & 0.20 & 0.15 & 0.12 \\ 
72 - 108 &  &  &  &  &  &  &  &  & 1 & 0.22 & 0.20 & 0.20 & 0.13 \\ 
108 - 130 &  &  &  &  &  &  &  &  &  & 1 & 0.21 & 0.19 & 0.12 \\ 
130 - 140 &  &  &  &  &  &  &  &  &  &  & 1 & 0.28 & 0.16 \\ 
140 - 150 &  &  &  &  &  &  &  &  &  &  &  & 1 & 0.19 \\ 
150 - 165 &  &  &  &  &  &  &  &  &  &  &  &  & 1 \\
\hline\hline
\end{tabular}
\caption{Top: Full correlation matrix for the $(1/T\Phi)(dN_{\pi} / d\theta_\pi)$ uncertainties in the \ccpi analysis with the additional signal requirement of $\theta_\mu < 20^\circ$.  Bottom: Corresponding shape correlation matrix.}
\end{table}
\endgroup

\clearpage
%%%%%%%%%%%%%%%%%%%%%%%%%%%%%%%%%%%%%%%%%%%%%%%%%%%%%%%%%%%%%%%%%%%%%%%%%%%%
%limited muon acceptance Npi analysis - KE
%%%%%%%%%%%%%%%%%%%%%%%%%%%%%%%%%%%%%%%%%%%%%%%%%%%%%%%%%%%%%%%%%%%%%%%%%%%%
\begingroup
\squeezetable
\begin{table}[h]
\begin{tabular}{l|ccccccc}
%Measured T_{#pi}, _Npi analysis
\hline\hline
$T_{\pi}$ (MeV) Bins & 35 - 55 & 55 - 75 & 75 - 100 & 100 - 125 & 125 - 150 & 150 - 200 & 200 - 350 \\ 
\hline
Measurement in bin & 1.42 & 1.42 & 1.29 & 1.21 & 1.06 & 0.82 & 0.501 \\ 
$10^{-41}\mathrm{cm}^2/\mathrm{MeV}/$nucleon &  $\pm$0.34 (0.25) &  $\pm$0.26 (0.14) &  $\pm$0.20 (0.09) &  $\pm$0.15 (0.08) &  $\pm$0.14 (0.07) &  $\pm$0.11 (0.05) &  $\pm$0.076 (0.036) \\ 
\hline\hline
\end{tabular}
\caption{Measured $(1/T\Phi)(dN_{\pi} / dT_\pi)$ and total uncertainties in the \ccpi analysis with the additional signal requirement of $\theta_\mu < 20^\circ$.  The absolute uncertainties are followed by shape uncertainties in parentheses.}
\end{table}
\endgroup

\begingroup
\squeezetable
\begin{table}[h]
\begin{tabular}{l|ccccccc}
%Full Correlation matrix for T_{#pi}, _Npi analysis
\hline\hline
Bins (MeV) & 35 - 55 & 55 - 75 & 75 - 100 & 100 - 125 & 125 - 150 & 150 - 200 & 200 - 350 \\ 
\hline
35 - 55 & 1 & 0.73 & 0.68 & 0.58 & 0.52 & 0.51 & 0.46 \\ 
55 - 75 &  & 1 & 0.84 & 0.72 & 0.66 & 0.65 & 0.62 \\ 
75 - 100 &  &  & 1 & 0.80 & 0.75 & 0.73 & 0.69 \\ 
100 - 125 &  &  &  & 1 & 0.84 & 0.80 & 0.72 \\ 
125 - 150 &  &  &  &  & 1 & 0.83 & 0.77 \\ 
150 - 200 &  &  &  &  &  & 1 & 0.88 \\ 
200 - 350 &  &  &  &  &  &  & 1 \\ 
\hline
35 - 55 & 1 & 0.41 & 0.22 & -0.13 & -0.28 & -0.41 & -0.44 \\ 
55 - 75 &  & 1 & 0.38 & -0.13 & -0.29 & -0.48 & -0.48 \\ 
75 - 100 &  &  & 1 & 0.07 & -0.10 & -0.35 & -0.42 \\ 
100 - 125 &  &  &  & 1 & 0.41 & 0.14 & -0.21 \\ 
125 - 150 &  &  &  &  & 1 & 0.31 & 0.01 \\ 
150 - 200 &  &  &  &  &  & 1 & 0.41 \\ 
200 - 350 &  &  &  &  &  &  & 1 \\ 
\hline\hline
\end{tabular}
\caption{Top: Full correlation matrix for the $(1/T\Phi)(dN_{\pi} / dT_\pi)$ uncertainties in the \ccpi analysis with the additional signal requirement of $\theta_\mu < 20^\circ$.  Bottom: Corresponding shape correlation matrix.}
\end{table}
\endgroup

%\clearpage

\fi

\end{document}